\documentclass[cleveref,autoref,pdfa,english]{oasics-v2021}
\pdfoutput=1

\usepackage{booktabs} %
\usepackage{graphicx}   %
\usepackage{mathtools}
\usepackage{amsmath}
\usepackage{color, colortbl}
\usepackage[nolist]{acronym}
\usepackage{multirow}
\usepackage{adjustbox}
\usepackage{array}
\usepackage{appendix}
\usepackage[tableposition=bottom]{caption}
\usepackage{placeins}
\usepackage[pagecolor=none]{pagecolor}

        \makeatletter
        \let\c@author\relax
        \makeatother
        \usepackage[backend=biber]{biblatex}
\usepackage{hyperref}
\def\Snospace~{\S{}}

\hypersetup{
    colorlinks,%
    citecolor=[rgb]{0.7,0,0},
    filecolor=[rgb]{0.2,0.2,0.2},
    linkcolor=[rgb]{0,0,0.7},
    urlcolor=[rgb]{0.7,0,0},
}

\definecolor{Gray}{gray}{0.9}
\newcommand{\Vs}[1]{\text{\textit{#1}}}  %

\newcolumntype{C}{>{\centering\arraybackslash}m{4.5em}}
\newcolumntype{D}{>{\centering\arraybackslash}m{3.5em}}
\newcolumntype{M}[1]{>{\centering\arraybackslash}m{#1}}
\newcolumntype{P}[1]{>{\centering\arraybackslash}p{#1}}
\newcolumntype{N}{@{}m{0pt}@{}}

        \hideOASIcs
        \nolinenumbers %

\makeatletter
  \newcommand\EatSpacesHack{\@bsphack\@esphack}
\makeatother

\acrodefplural{AS}[ASes]{Autonomous Systems}
\acrodefplural{VP}[VPs]{Vantage Points}
\acrodef{NAT}[NAT]{Network Address Translation}
\acrodef{CG-NAT}[CG-NAT]{Carrier-Grade NAT}
\acrodefplural{CDN}[CDNs]{Content Delivery Networks}
\acrodef{RIR}[RIR]{Regional Internet Registry}
\acrodefplural{RIR}[RIRs]{Regional Internet Registries}

\author{Guillermo Baltra}
  {USC/ISI, Marina del Rey, California, USA}
  {baltra@ant.isi.edu}
  {https://orcid.org/0009-0005-6312-5403}{}

\author{Tarang Saluja}
  {Swarthmore College, Swarthmore, Pennsylvania, USA}
  {tsaluja@ant.isi.edu}
  {}{}

\author{Yuri Pradkin}
  {USC/ISI, Marina del Rey, California, USA}
  {yuri@isi.edu}
  {https://orcid.org/0009-0008-7966-7030}{}

\author{John Heidemann}
  {USC/ISI, Marina del Rey, California, USA}
  {johnh@isi.edu}
  {https://orcid.org/0000-0002-1225-7562}{}

\authorrunning{Baltra, Saluja, Pradkin, Heidemann}

\ccsdesc{Networks~Network measurement}
\ccsdesc{Networks~Public Internet}
\ccsdesc{Networks~Network economics}

\Copyright{Guillermo Baltra, Tarang Saluja, Yuri Pradkin, John Heidemann}

\acknowledgements{
The authors would like to thank John Wroclawski,
  Wes Hardaker, Ramakrishna Padmanabhan,
  Ramesh Govindan,
  Eddie Kohler,
  and the Internet Architecture Board for their input on
  on an early version of this paper.
We would also like to thank the NINeS chairs and steering committee for
  building a venue that is receptive for this kind of paper,
  and to our shepherd, Michael Schapira, and the reviewers for their input.
}

\funding{The work is
  supported in part by
   the National Science Foundation,
   CISE Directorate, award CNS-2007106 %
  NSF-2028279, %
  OAC-2530698, %
and DARPA Contract HR001124C0403. %
Any opinions,
findings and conclusions or recommendations expressed in this material are
those of the author(s) and do not necessarily reflect the views of
NSF or DARPA.
The U.S.~Government is authorized to reproduce and distribute
reprints for Governmental purposes notwithstanding any copyright
notation thereom.}

\title{Understanding Partial Reachability in the Internet Core}

  \subtitle{Technical Report: arXiv:2601.12196}
  \EventLongTitle{arXiv}
  \EventShortTitle{arXiv}
  \EventLogo{}
  \ArticleNo{1}

\keywords{Internet, Internet reliability, Network outages, Active measurements}

\begin{acronym}[AS]
\acro{AS}{Autonomous System}
\acro{RDNS}{Reverse DNS}
\acro{VP}{Vantage Point}
\end{acronym}

\begin{document}

\maketitle

\begin{abstract}
Routing strives to connect all the Internet,
  but compete: %
political pressure threatens routing fragmentation;
  architectural changes such as private clouds, carrier-grade NAT, and firewalls
    make connectivity conditional;
  and commercial disputes create partial reachability for days or years.
This paper suggests \emph{persistent, partial reachability is fundamental to the Internet}
  and an underexplored problem.
We first
  \emph{derive a conceptual definition of the Internet core}
  based on connectivity, not authority.
We identify \emph{peninsulas}: persistent, partial connectivity;
  and \emph{islands}: when computers are partitioned
    from the Internet core.
Second, we develop algorithms to observe each across the Internet,
  and apply them to
  two existing measurement systems:
  Trinocular, where 6 locations observe 5M networks frequently,
  and RIPE Atlas, where 13k locations scan the DNS roots frequently.
Cross-validation shows our findings are stable over \emph{three years of data},
  and consistent with as few as 3 geographically-distributed observers.
We validate peninsulas and islands against
  CAIDA Ark, showing
  good recall (0.94)
  and bounding precision between 0.42 and 0.82.
Finally, our work has broad practical impact:
  we show that
  \emph{peninsulas are more common than Internet outages}.
Factoring out peninsulas and islands as noise can
  \emph{improve existing measurement systems};
  their ``noise'' is $5\times$ to $9.7\times$ larger
  than the operational events
  in RIPE's DNSmon.
We show that
  most peninsula events are routing transients (45\%),
  but most peninsula-time (90\%) is due to a few (7\%) long-lived events.
Our work helps inform Internet policy
  and governance,
  with our neutral definition showing
  no single country or organization can unilaterally control the Internet core.
\end{abstract}

\vspace*{-1ex}

\section{Introduction}
	\label{sec:introduction}

The Internet was created
  to allow disparate networks to communicate~\cite{Cerf74a,Postel80b,clark1988design},
  making \emph{network partition} its nemesis.
Routing is designed to heal partitions,
  so that ``communication must continue despite loss of networks or gateways''~\cite{clark1988design}.
Yet the reality of partitions prompts leadership-election algorithms such as Paxos~\cite{Lamport98a}.

Worse than complete network partition is \emph{long-lived partial reachability}.
Although transient reachability problems are well known (for example,~\cite{Wang06b}),
  and human errors occur~\cite{Mahajan02a},
  \emph{policy choices} can cause persistent partial connectivity.
Economic differences result in peering disputes~\cite{ipv6peeringdisputes,google_cogent,cloudflare_he};
  while political choices can limit access~\cite{Reuters22a},
  or emphasize sovereignty~\cite{BBC19a,RBC21a,Reuters21a}.
Research~\cite{andersen2001resilient,katz2008studying,katz2012lifeguard}
  and production~\cite{Schlinker17a,Yap17a} work around persistent unreachability.

\textbf{Challenges:}
But today \emph{universal reachability in the Internet core is often challenged}:
\emph{Political} pressure
  may Balkanize the Internet along national borders.
Examples include
  Russia's 2019 sovereign-Internet law~\cite{BBC19a,RBC21a,Reuters21a}
and
  national ``Internet kill switches''
  that are debated in U.S.~\cite{GovTrack20a} and the U.K.,
  and deployed elsewhere~\cite{Cowie11a,Coca18a,Griffiths19a,Taye19a}. %
These pressures prompted policy discussions about fragmentation~\cite{Drake16a,Drake22a}.
We suggest that \emph{technical methods can help inform policy discussions}
  and that threats such as de-peering
  place the global Internet at risk.
We will show that no single country
  can unilaterally control the Internet core today  (\autoref{sec:other_applications}),
  and that de-peering \emph{can} fragment the Internet core
  into pieces (\autoref{sec:internet_partition}).

\emph{Architecturally}, 25 years of evolution
  have segmented the Internet core:
  many services live in clouds;
  users are usually second-class clients due to \ac{NAT};
  firewalls interrupt connectivity;
  and Internet has both IPv4 and IPv6.
Politics can influence architecture,
  with
  China's Great Firewall~\cite{Anonymous12a,Anonymous14a},
  and
  a proposed ``new Internet''~\cite{huawei2020}.
We suggest that technical methods
  help us \emph{reason about changes to Internet architecture},
  to understand implications of partial reachability
  and evaluate IPv6 deployment.

\emph{Operationally}, even when ISP peering is mature,
  disputes can cause long-term  partial unreachability~\cite{ipv6peeringdisputes}.
Such unreachability detected experimentally~\cite{Dhamdhere18a},
  and systems built to mitigate partial
  reachability~\cite{andersen2001resilient,katz2008studying,katz2012lifeguard}.
We show several operational uses of our work.
We show that \emph{accounting for partial reachability can make existing measurement
  systems more sensitive}.
By applying these results to widely used RIPE DNSmon (\autoref{sec:dnsmon}),
   we show that its observations of
   persistent high query loss (5--8\% to the DNS Root~\cite{RootServers16a})
   are mostly measurement error and
   persistent partial connectivity.
These factors are $5\times$ and $9.7\times$ (IPv4 and v6)
  larger than operationally important signals.
Our analysis also helps resolve uncertainty in
  Internet outage detection (\autoref{sec:outage_detection_short}),
  clarifying ``corner cases''
  due to conflicting observations~\cite{Schulman11a,quan2013trinocular,Shah17a,richter2018advancing,guillot2019internet}.
We show partial reachability is a common cause,
   and it occurs at least as often as complete outages  (\autoref{sec:peninsula_frequency}).
Finally, our work helps quantify the applicability of
  route-failure mitigation~\cite{andersen2001resilient,katz2008studying,katz2012lifeguard},
  and
  of cloud egress selection~\cite{Schlinker17a}.

\textbf{Contributions:}
Our first contribution is to
  \emph{recognize that partial reachability is a fundamental part of the Internet},
   and addressing it requires a
   \emph{rigorous definition of what \emph{is} the Internet's core} (\autoref{sec:problem}).
In 1982, the Internet was 83 hosts~\cite{Smallberg82a} globally reachable with TCP/IP~\cite{Postel80b}.
In 1995,
  the Federal Networking Council defined ``Internet''
  as (i) a global address space,
  (ii) supporting TCP/IP and its follow-ons,
  that (iii) provides services~\cite{nitrd}.
Later work added DNS~\cite{IAB20a} and IPv6.
But today's Internet is much changed:
Both users on PCs
  and the majority of users on mobile devices
  access the Internet indirectly through
  \ac{NAT}~\cite{Tsuchiya93a}
  and \ac{CG-NAT}~\cite{Richter16c}.
Many public services operate from the cloud,
  visible through rented or imported IP addresses,
  backed
  by network virtualization~\cite{Greenberg09a}.
Media is replicated in \acp{CDN}.
Access is mediated by firewalls.
Yet users find Internet services so seamless
  that technology recedes
  and the web, Facebook, and phone apps are their ``Internet''.

\emph{We define Internet core as the strongly-connected component of more than
  50\% of active, public IP addresses that can bidirectionally route to each other}
  (\autoref{sec:definition}).
This definition has several unique characteristics.
First, %
  captures the uniform, \emph{peer-to-peer nature
  of the Internet core}
  necessary for first-class services.
Second, it defines \emph{one, unique} Internet core
  by requiring reachability of more than 50\%---there can be only one
  since multiple majorities
  are impossible.
Finally, unlike prior work, this \emph{conceptual} definition
  avoids dependence on any specific measurement system,
  nor does it depend on historical precedent, special locations, or central authorities.
Although an operational measurements will
  reflect observation error,
  the conceptual Internet core
  defines an asymptote against which our current and future measurements can compare,
  unlike prior definitions from specific systems~\cite{andersen2001resilient,katz2008studying,katz2012lifeguard}.

Our second contribution is to use this definition to
  identify two classes of persistent unreachability (\autoref{sec:internet_landscape}),
  and develop algorithms to quantify each (\autoref{sec:design}).
We define  \emph{peninsulas} as when a network sees persistent, partial connectivity
  to part of Internet core.
We present the \emph{Taitao} algorithm to detect peninsulas
  that often result from peering disputes or long-term firewalls.
We define
  \emph{islands} as
  when one or more computers are partitioned from the main Internet core
  as detected by \emph{Chiloe}, our second algorithm.

\begin{table}[]
        \centering
\mbox{
\begin{tabular}{@{}lllll@{}}
\toprule
  \multicolumn{1}{c}{\multirow{2}{*}{\textbf{data}}} & \multicolumn{1}{c}{\multirow{2}{*}{\textbf{num.}}} & \multicolumn{3}{c}{\textbf{measurement}}  \\ \cmidrule(l){3-5}
  \multicolumn{1}{c}{\textbf{source}}                & \multicolumn{1}{c}{\textbf{VPs}} & \multicolumn{1}{c}{\textbf{freq.}} & \multicolumn{1}{c}{\textbf{targets}} & \multicolumn{1}{c}{\textbf{duration}} \\ \midrule
Trinocular~\cite{quan2013trinocular} & 6$^a$ & 11~min. & 5M /24s & 4 years \\
RIPE Atlas~\cite{Ripe15c} & 12,086$^b$ & 5~min. & 13 RSOs & 3 years \\
CAIDA Ark~\cite{ark_data} & 171$^c$ & \,24~hrs. & all IPv4 & selected \\
Routeviews~\cite{routeviews} & 55$^d$ & 1~hour & all IPv4 & selected \\ \bottomrule
\multicolumn{5}{l}{{\footnotesize a: In 2017 and 2019.  b: On 2024-01-30.  c: On 2017-12-01. d: In 2024-01. }} \\
\end{tabular}
}
\vspace*{-2ex}
\caption{Types of data sources used in this paper.}
	\label{tab:data_types}
\end{table}

We apply these algorithms to data from two operational %
  systems (\autoref{tab:data_types}):
  Trinocular, with frequent measurements
  of 5M networks
    from six \acp{VP}~\cite{quan2013trinocular},
  and RIPE Atlas,
    with frequent measurements of
  the DNS root~\cite{RootServers16a} from 13k \acp{VP}~\cite{Ripe15c}.
By applying new algorithms to existing, publicly available, multi-year data
  we are able to provide longitudinal analysis with some results covering more than three years.
These two systems demonstrate our approach works on active
  probes covering millions of networks (although from few observers)
  and also from more than 13k \acp{VP} (although probing only limited destinations),
  strongly suggesting the results generalize,
    since no practical system can cover the $O(n^2)$ cost of all destinations from all sources.

In addition varying \acp{VP} and destinations across the design space,
  we evaluate the accuracy of our systems with rigorous measurements (\autoref{sec:validation}).
We quantify the independence of the Trinocular sites (\autoref{sec:site_correlation})
  with cross-validation.
Our analysis shows that combinations of any three independent \acp{VP}
  provide a result that is statistically indistinguishable from the asymptote
  \autoref{sec:peninsula_frequency}.
We show our results are stable over more than three years
  with samples from Trinocular (\autoref{sec:chiloe_validation})
  and continuous results from RIPE Atlas (\autoref{sec:dnsmon}).
Finally, we validate both algorithms against a third measurement system,
  CAIDA Archipelago,
  where 171 \acp{VP} scan millions of networks, daily~\cite{CAIDA07b}.
Although comparing very different systems is challenging,
  these results provide strong bounds on accuracy (\autoref{sec:taitao_validation}),
  with very good recall (0.94) and reasonable precision
  (lower and upper bounds from 0.42 and to 0.82).

Our final contribution
 \emph{uses these algorithms
  to address current operational questions}.
We show that partial reachability is a \emph{pervasive problem}
  today, meriting attention.
We prove that peninsulas
  occur \emph{more} often than outages, as subject of wide attention~\cite{Schulman11a,Dainotti11a,quan2013trinocular,Shah17a,Dainotti17a,wan2020origin}.
We bring technical light to policy choices
  around national networks (\autoref{sec:other_applications}) and de-peering (\autoref{sec:internet_partition}).
We improve sensitivity of RIPE Atlas' DNSmon~\cite{Amin15a} (\autoref{sec:dnsmon}),
  resolve corner cases in outage detection (\autoref{sec:outage_detection_short}),
  and quantify opportunities for route detouring (\autoref{sec:peninsula_frequency}).

These contributions range from
  a theoretical definition,
  to experimental measurements,
  and their practical application.
Each depends on the other---the definition enables the algorithms,
  which are then applied to show utility.

\textbf{Artifacts and ethics:}
Data used  (\autoref{tab:data_types})
  and
    created~\cite{ANT22b}
  in this paper
  is available at no cost.
Our work poses no ethical concerns
  (\autoref{sec:research_ethics})
  by not identifying individuals
  and avoiding additional traffic by reanalysis
  with new algorithms.
IRB review says it is non-human subjects research
  (USC IRB IIR00001648).

An earlier of this work was released as a technical report was released in July 2021~\cite{Baltra21a_v1}.
 In May 2022 it was updated with several additions:
 for
 a more careful definitions in \autoref{sec:definition} and \autoref{sec:internet_landscape},
new information about island durations \autoref{sec:islands_duration}
  and sizes \autoref{sec:islands_sizes},
expanded applications in \autoref{sec:applications}
  and \autoref{sec:other_applications} and \autoref{sec:internet_partition},
considerable additional details and supporting data in appendices, and
many writing improvements.
In October 2023 we updated it to address typos and missing references.
In January 2026 it was revised with extensive editing throughout,
  including a more careful definition
  and additional and rearragned appendicies~\cite{Baltra26b};
  this version adds several appendices to but otherwise reflects the conference
  version published at ACM NINeS~\cite{Baltra26a}.

\section{Problem: Partial Reachability}
	\label{sec:problem}

Understanding partial reachability requires a rigorous definition
  of \emph{what} is being reached.
We next define \emph{the Internet core}
  to which we connect,
  to answer the political, architectural, and operational questions
  from \autoref{sec:introduction}.

We suggest a definition must be both
  \emph{conceptual} and \emph{operational}~\cite{scientific_methods}.
Our conceptual definition (\autoref{sec:definition})
  articulates what the Internet \emph{is and is not}.
it provides a goal which our implementation (\autoref{sec:design}) approximates,
  and we apply it improve real-world, operational systems (\autoref{sec:dnsmon}).
Prior definitions~\cite{Cerf74a,Postel80b,nitrd} are too vague to operationalize.

Second, a definition must give both sufficient \emph{and}
  necessary conditions to be part of the Internet core.
Prior work gave
  properties the core must have (sufficient conditions, like supporting TCP).
We add \emph{necessary} conditions to  define
  when networks \emph{leave} the Internet core (\autoref{sec:internet_partition}).

\subsection{The Internet: A Conceptual Definition}
	\label{sec:definition}

We define the Internet core as \emph{all active IP addresses
  that can Bidirectionally Route to more than
  50\% of the public, Potentially Reachable Internet}.
We define these key terms next,
  and
  expand their motivation and implications
  later (\autoref{sec:definition_motivations}).

Two addresses are \emph{Bidirectionally Routable} when
  each can initiate a connection to the other.
In our realization we measure connectivity
  with either ICMP echo-request
  or with DNS queries and replies,
  considering alternatives in \autoref{sec:definition_motivations}.

The
  \emph{Potentially Reachable Internet}
  is all IP addresses
  in a graph-theoretic strongly-connected component,
  with graph edges defined by Bidirectional Routability.
This definition means
  any node in the set can reach any other,
  either directly or perhaps through one or more hops.

\subsection{Motivation for \emph{This} Definition}
	\label{sec:definition_motivations}

We define the potentially reachable Internet
  via observation,
  so it depends only on testable, shared information,
  and not a central authority such as ICANN.
Defining the Potentially Reachable Internet
  as active addresses
  also implies that the vast parts of unallocated IPv6 do
  not change our conclusions.

\textbf{Why both bidirectional routability and potential reachability?}
\emph{Bidirectional Routability} is connectivity in
        the networking sense, so each address must have a routing table
        entry that covers the other, and there must be some BGP-level
        reachability between them.
\emph{Graph-Theoretic Reachability}
  shows transitive connectivity, even when disputes mean some pairs
  cannot reach each other.

Bidirectional Routability is required to capture the idea of
        IP routing from prior  definitions~\cite{Cerf74a,Postel80b,nitrd},
        where all hosts should be able to communicate directly.
It excludes private, NAT'ed addresses~\cite{Rekhter96a},
  which, although useful clients,
  require rendezvous protocols (STUN~\cite{Rosenberg03a},
  UPnP~\cite{Miller01a}, or PMP~\cite{Cheshire13d})
  to partially link to the core,
  and also non-public cloud addresses hidden behind load balancers~\cite{Greenberg09a}.
However, cloud VMs with fully-reachable public addresses are part of the core,
  including cloud-hosted services using public IP addresses from the cloud operator
  or their own (BYOIP).

Graph-Theoretic Reachability is required
  to define what ``100\%'' is,
  so  we guarantee one (or no) Internets
  by looking for a non-overlapping majority,
  even in the face of conflicting claims (\autoref{sec:half_proof}).
The combination of terms help us resolve such conflicts
  as different peninsulas sharing a common Internet core (although perhaps
  requiring relay through a third party).

\textbf{Why more than 50\%?}
We take as an %
  axiom that there should be \emph{one Internet core}
  per address space (IPv4 and IPv6),
  or a reason why that Internet core no longer exists.
Thus we require a definition to unambiguously identify ``the'' Internet core
  given conflicting claims;
  any larger value is excessive,
  and anything smaller would allow multiple viable claims.
(In practice, \autoref{fig:atlas_revisited}
          we see 98.5--99.5\% agreement on the core,
          so values at the 50\% threshold are unlikely.)

Requiring a majority of active addresses
  ensures that there can be only one Internet core,
  since any two majorities must overlap.
Any smaller fraction could allow two groups to make
  valid claims.
We discuss how to identify the core in the face of conflicting
  claims in \autoref{sec:half_proof}.

The definition of the Internet core should not require a central authority.
``Majority'' supports assessment
  independent of any authority.
Any computer to prove it is in the Internet core
  by reaching half of active addresses,
  as defined by
  multiple, independent, long-term evaluations~\cite{Heidemann08c,Zander14b,Dainotti16a}.
It also avoids identification of ``tier-1'' ISPs,
  an imprecise term determined only by private business agreements.

Finally, a majority defines \emph{an Internet core that can end:}
  fragmentation occurs
  should the current Internet core break into three or more disconnected components
  where none retains a majority of active addresses.
If a large enough organization or group
  chose to secede, or are expelled,
  the Internet core could become several no-longer internets (\autoref{sec:internet_partition}).

\textbf{Why all addresses?}
In each of IPv4 and IPv6
  we consider all addresses equally. %
Public Internet addresses are global,
  and the Internet core was intentionally designed without a hierarchy~\cite{clark1988design}.
Consistent with
  decentralization trends~\cite{dinrg},
  a definition should not create hierarchy,
  nor designate special addresses by age or importance.

These definitions are relatively apolitical
  and reduce first-mover bias, discussed in \autoref{sec:internet_partition}.
Addresses are an Internet-centric metric,
  unlike population or countries.
Requiring activity
  reduces the influence of
  large allocated, but unused, space,
  such as in legacy IPv4 /8s
  and new IPv6 allocations.

\textbf{Reachability, Protocols and Firewalls:}
End-to-end reachability avoids
  difficult discovery of router-level topology.

Our conceptual definition allows different definitions of reachability.
Reachability may be measured by protocols
  such as ICMP echo-request (pings),
  DNS or HTTP queries,
  or by data-plane reachability with BGP\@.
Any specific test will provide an operational realization
  of our conceptual definition.
(Measurement must tolerate transient failures,
  perhaps with multiple targets (Trinocular)
  or retransmissions (Atlas).)
\autoref{sec:peninsula_frequency} examines how well
  using ICMP-based measures converge,
  and \autoref{sec:dnsmon} shows DNS stability over years.

Firewalls complicate observing reachability
  and can make it conditional.
We accept that the results of specific observations
  may vary with different protocols or observation times;
  experiments show results are stable (\autoref{sec:peninsula_frequency}).
Measurement allows us to evaluate
  policy-driven unreachability
  (see \autoref{sec:country_peninsulas}).

We have two implementations of peninsula and island detection;
  both use publicly-available data from existing measurement systems.
One uses
  Trinocular~\cite{quan2013trinocular},
  because of its frequent, Internet-wide
  ICMP echo requests (11-minutes to 5M IPv4 /24s).
Prior work has shown ICMP provides the most response~\cite{Bartlett07d,quan2013trinocular,durumeric2014internet},
  and can avoid rate limiting~\cite{Guo18a},
  other other protocol options are possible.
Our second uses RIPE Atlas because of its use in DNS (\autoref{sec:dnsmon}).

\textbf{Why reachability and not applications?}
Users care about applications, and a user-centric view
  might emphasize reachability of HTTP or to Facebook
  rather than at the IP layer.
Our second realization uses public data from RIPE Atlas,
  with DNS as the application, as described
  in \autoref{sec:dnsmon}.
Many large outages are failures of applications
  such as DNS~\cite{Prince25a}; their study would require a different evaluator
  than IP reachability.
Future work may look at other, more user-centric applications.
However, we
  suggest reachability at the IP layer
  is a more fundamental concept.
IP has changed only twice since 1969 with IPv4 and IPv6,
  but applications wax and wane,
  and some (like e-mail) extend beyond the Internet.

\subsection{Cases of Partial Reachability}
  \label{sec:internet_landscape}

\begin{figure}
\begin{center}
    \mbox{\includegraphics[width=.4\linewidth,trim=210 300 220 145,clip]{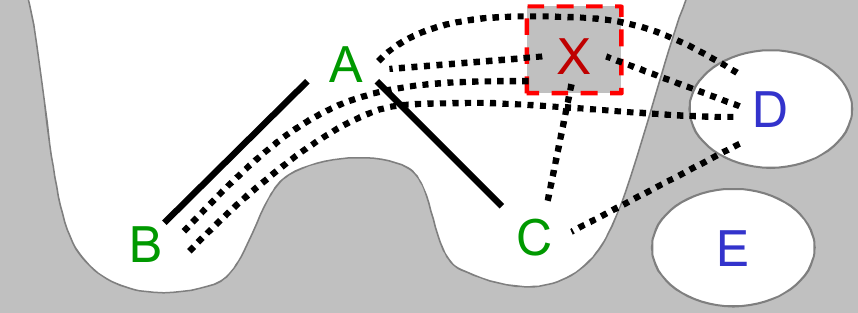}}
\end{center}
    \caption{$A$, $B$ and $C$ are the connected core,
      $B$ and $C$ peninsulas,
      $D$ and $E$ islands,
      $X$ is out.}
      \label{fig:term_concept}
\end{figure}

We use our definition of the Internet core
  to consider three types of partial reachability,
  shown in \autoref{fig:term_concept}.
Here long-term and current routability are dotted and solid lines,
  and white regions show current data-plane reachability.
All address blocks but $E$ form the \emph{core}.
Blocks $B$ and $C$ are on \emph{peninsulas}
  because they do not route to each other,
  although data could relay through $A$.
Block $X$ has an \emph{outage};
  its routes are temporarily down.
Blocks $D$ and $E$ are \emph{islands}:
$D$ usually can route to the core,
  but not currently.
$E$ uses public addresses, but has never announced routes publicly.

\subsubsection{Outages}
\label{sec:outages}

A number of groups have examined Internet outages~\cite{Schulman11a,quan2013trinocular,richter2018advancing,guillot2019internet}.
These systems observe the public IPv4 Internet and identify networks
  that are no longer reachable---they have left the Internet.
Often these systems define outages operationally
  (network $X$ is out since none of our \acp{VP} can reach it).
In this paper, we define an outage as when all computers
  in a block are off,
  perhaps due to power loss.
We next define islands,
  when the computers are on but cannot reach the Internet core.

\subsubsection{Islands: Isolated Networks}
	\label{sec:island}

An \emph{island} is a group of public IP addresses
  partitioned from the Internet core,
  but still able to communicate among themselves.
Operationally, outages (X in \autoref{fig:term_concept}) and islands (like D and E)
  are both unreachable from external \acp{VP} and appear identical,
  but computers in an island are on and can reach each other.

Islands occur when an organization is
  no longer connected to the Internet core.
A business with one ISP becomes an island
  when its router upstream connection fails,
  even though computers in the business can reach each other.
An \emph{address island} is when
  a computer can reach only itself.

\textbf{Example Islands:}
Islands are common in RIPE Atlas~\cite{Amin15a} when a \ac{VP}
  has an IPv6 address on the LAN, but lacks routes to the public IPv6 Internet.
In \autoref{sec:dnsmon} we show that this kind of misconfiguration
  accounts for
  $5\times$ more IPv6 unreachability than other, more meaningful problems.

We also see islands in reanalysis of data from Trinocular outage detection~\cite{quan2013trinocular}.
Over three years, from 2017 to 2020,
  we saw 14 cases where one of the 6 Trinocular \acp{VP}
  was active and could reach its LAN, but could not reach the rest of the Internet.
Network operators confirm local routing failures in several of these cases.
  \autoref{sec:additional_details_island_example} shows one example.

\subsubsection{Peninsulas: Partial Connectivity}
	\label{sec:peninsula_definition}

Link and power failures create islands,
  \emph{peninsulas} are
  \emph{partial} connectivity,
  when a group of public IP addresses can reach some destinations,
  but not others.
    (In a geographic peninsula, the mainland may be visible over water, but reachable only with a detour; similarly,
    in \autoref{fig:term_concept},
    $B$ can reach $A$, but not $C$.)
Peninsulas occur when
  an upstream provider of a multi-homed network
  accepts traffic it cannot deliver or forward,
  when Tier-1 ISPs refuse to peer,
  or when firewalls block traffic.
Experimental overlay networks route around peninsulas~\cite{andersen2001resilient,katz2008studying,katz2012lifeguard}.

\textbf{Peninsulas in IPv6:}
An long-term peninsula follows from
  the IPv6 peering dispute between Hurricane Electric (HE) and Cogent.
These ISPs decline to peer in IPv6 (IPv4 is fine),
  nor do they forward their IPv6 through
  another party.
HE and Cogent customers could not reach each other in 2009~\cite{ipv6peeringdisputes},
  and this problem persists through 2025, as we show in DNSmon (\autoref{sec:dnsmon}).
We further confirm unreachability
  between HE and Cogent users in IPv6 with traceroutes
  from looking glasses~\cite{he_looking_glass,cogent_looking_glass}
  (HE at 2001:470:20::2 and Cogent at 2001:550:1:a::d):
  neither can reach their neighbor's server,
  but both reach their own.
Other IPv6 disputes include Cogent and Google~\cite{google_cogent}, and
 Cloudflare and Hurricane Electric~\cite{cloudflare_he}.
Disputes can arise from an inability to  agree to %
  settlement-free or paid peering.

\textbf{Peninsulas in IPv4:}
We observed a peninsula lasting 3 hours starting 2017-10-23t22:02Z,
  where
  five Polish \acp{AS}
  had 1716 /24 blocks that were always reachable one Los Angeles,
  but not from four other \acp{VP}
  (as seen in public data from  Trinocular~\cite{LANDER17a}).
Before the peninsula, these blocks
  received service through Multimedia Polska (\emph{MP}, AS21021),
  via Cogent (AS174), or through Tata (AS6453).
When the peninsula occurred, traffic to all blocks continued through Cogent
  but was blackholed; it did not shift to Tata. %
The successful \ac{VP} could reach MP
  through  Tata for the entire event,
  proving MP was connected.
After 3 hours, we see a burst of 23k BGP updates
  and MP is again reachable from all VPs.
  A graph showing reachability to this peninsula is in \autoref{sec:additional_details_peninsula_example}.

\section{Detecting Partial Connectivity}
	\label{sec:design}

We now introduce the
  \emph{Taitao} algorithm to detects peninsulas,
  and \emph{Chiloe}, islands
  (names from Patagonian geography).

\subsection{Taitao: a Peninsula Detector}
\label{sec:disagreements}

Peninsulas occur when portions of the Internet core
  are reachable from some locations and not others.
They can be seen by two \acp{VP} disagreeing on reachability.

Detecting peninsulas presents three challenges.
Without \acp{VP} everywhere,
  when all \acp{VP} are on the same ``side'' of a peninsula ($A$ and $C$ in \autoref{fig:term_concept}),
  their reachability agrees even though \acp{VP} may disagree (like $B$).
Second, asynchronous observations test reachability at different times:
  observations in Trinocular spread over 11~minutes,
  and in Atlas, 5~minutes.
Observations at times before and after a network change
  will disagree, but both are true---a difference
  due to weak synchronization,
  and not a peninsula.
Third, connectivity problems near the observer (or when an observer is an island)
  should not reflect on the intended destination.

We identify peninsulas by detecting disagreements in
  block state by comparing
  successful \ac{VP} observations that occur at about the same time.
Since probing rounds occur asynchronously,
  we compare measurements within the measurement system's window
  (11 or 5 minutes for Trinocular and  Atlas).
This approach sees peninsulas lasting longer than one window duration,
  but may miss briefer ones,
  or when \acp{VP} are not on ``both sides''.

Formally, $O_{i,b}$ is the set of observers with valid observations
about block $b$ at round $i$.
We look for disagreements in $O_{i,b}$,
  defining $O_{i,b}^\Vs{up} \subset O_{i,b}$ as the
set of observers that measure block $b$ as up at round $i$.
We detect a peninsula when:
\begin{align}
  0 < |O_{i,b}^\Vs{up}| < |O_{i,b}|
\end{align}

When only one \ac{VP} reaches a block,
  we must classify it as a peninsula or an island,
  as described next.

\subsection{Chiloe: an Island Detector}
	\label{sec:chiloe}

According \autoref{sec:island}, islands occur
when the Internet core is partitioned, and the component
  with fewer than half the active addresses
  is the island.
Typical islands are much smaller.

We can find islands by looking for networks that
  are only reachable from less than half of the Internet core.
However, to classify such networks as an island
  and not merely a peninsula, we need to show that it is partitioned,
  which requires global knowledge.
In addition, if islands are partitioned from all \acp{VP},
  we cannot tell an island,
  with active but disconnected computers,
  from an outage, where they are off.

For these reasons, %
  we must look for islands that include \acp{VP} in their partition.
Because we know the VP is active and scanning
  we can determine how much of the Internet core is in its partition,
  ruling out an outage.
We also can confirm the Internet core is not reachable,
  to rule out a peninsula.

Formally, we say that $B$ is the set of blocks
  in the Internet core.
$B^\Vs{up}_{i,o} \subseteq B$ are blocks reachable from observer
$o$ at round $i$, while
$B^\Vs{dn}_{i,o} \subseteq B$ is its complement.
We detect that observer $o$ is in an island when
  it thinks half or more of the
  observable Internet core
  is down:
\begin{align}
  0 \leq |B^\Vs{up}_{i,o}| < |B^\Vs{dn}_{i,o}|
\end{align}

This method is independent of measurement systems,
  but is limited to detecting islands that contain \acp{VP},
  so \emph{any deployment will certainly undercount islands}.
We evaluate islands in Trinocular and Atlas
  (\autoref{sec:how_common_are_islands}),
  confirming more \acp{VP} see more islands,
  but that \emph{nearly all reported islands are correct}.

Finally, because observations are not instantaneous,
  we must avoid confusing short-lived islands with long-lived peninsulas.
For islands lasting longer than 11-minutes,
  we also require
 $|B^\Vs{up}_{i,o}| \rightarrow 0$.
With $|B^\Vs{up}_{i,o}| = 0$, it is an address island.

\subsection{Deployment with Existing Systems}
    \label{sec:deployment_status}

We have deployed our algorithms as extensions to two systems:
  Trinocular and RIPE Atlas.
In both cases, each system provides data to us via
  existing APIs and
  we then apply Taitao and Chiloe and share results back.
Processing time for both is modest, with DNSmon running in minutes
  and Trinocular taking less time than Trinocular outage detection.

For DNSmon, we provide daily outages and peninsulas since 2022-01-01
  on a public
  website~\cite{Saluja22b}.
We have also discussed these results with RIPE and the root operators;
  RIPE currently identifies islands manually,
  and one root operator is using our results to guide operations.
We provide 3.5 years Trinocular analysis at our
  website~\cite{ANT17c},
  and are working with Trinocular operators to operationalize our algorithms.

\section{Validating our approach}
	\label{sec:validation}

We next validate our algorithms with three data sources.

\begin{table*}
  \begin{minipage}[b]{.49\linewidth}
  \footnotesize
  \resizebox{\textwidth}{!}{
  \begin{tabular}{c c c | c c c}
    & & & \multicolumn{3}{c}{\normalsize \textbf{Ark}} \\
    & & Sites Up & Conflicting & All Down & All Up \\
	\cline{3-6}
    \multirow{7}{2pt}{\rotatebox[origin=l]{90}{\parbox{1.6cm}{\normalsize \textbf{Trinocular}}}}
    & \multirow{5}{2pt}{\rotatebox[origin=l]{90}{\parbox{35pt}{Conflicting}}}
      & 1 & \cellcolor[HTML]{99ee77}20  & \cellcolor[HTML]{99ee77}6  & \cellcolor[HTML]{FFF9C4}\emph{15}  \\
    & & 2 & \cellcolor[HTML]{99ee77}13  & \cellcolor[HTML]{99ee77}5  & \cellcolor[HTML]{FFF9C4}\emph{11}  \\
    & & 3 & \cellcolor[HTML]{99ee77}13  & \cellcolor[HTML]{99ee77}1  & \cellcolor[HTML]{FFF9C4}\emph{5}   \\
    & & 4 & \cellcolor[HTML]{99ee77}26  & \cellcolor[HTML]{99ee77}4  & \cellcolor[HTML]{FFF9C4}\emph{19}  \\
    & & 5 & \cellcolor[HTML]{99ee77}83  & \cellcolor[HTML]{99ee77}13 & \cellcolor[HTML]{FFF9C4}\emph{201} \\
	\cline{3-6}
    & \multirow{2}{2pt}{\rotatebox[origin=l]{90}{\parbox{21pt}{Agree}}}
    & 0 & \cellcolor[HTML]{F0ABAB}\textbf{6} & \cellcolor[HTML]{CCFF99}97 & \cellcolor[HTML]{F0ABAB}\textbf{6}     \\
    & & 6 & \cellcolor[HTML]{CCFF99}491,120 & \cellcolor[HTML]{CCFF99}90
    & \cellcolor[HTML]{CCFF99}1,485,394 \\
\end{tabular}}
  \caption{Trinocular and Ark agreement table. Dataset A30, 2017q4.}
  \label{tab:taitao_validation_table}
\end{minipage}
\hspace{3mm}
\begin{minipage}[b]{.49\linewidth}
  \centering
  \footnotesize
  \resizebox{\textwidth}{!}{
  \begin{tabular}{c P{40pt} P{40pt} c c}
    & & \multicolumn{3}{c}{\normalsize \textbf{Ark}} \\
    & & Peninsula & \multicolumn{2}{c}{Non Peninsula} \\
    \multirow{2}{2pt}{\rotatebox[origin=l]{90}{\parbox{30pt}{\normalsize \textbf{Taitao}}}}
    & Peninsula & \cellcolor[HTML]{99ee77} 184 & \cellcolor[HTML]{FFF9C4}
    \emph{251 (strict)} & \parbox[30pt][20pt][c]{30pt}{\cellcolor[HTML]{FDD835} \emph{40 (loose)}} \\
    & \parbox[40pt][20pt][c]{40pt}{\centering Non Peninsula} & \cellcolor[HTML]{F0ABAB} \textbf{12} &
    \multicolumn{2}{c}{\cellcolor[HTML]{CCFF99} 1,976,701} \\
  \end{tabular}
}
  \vspace{10pt}
  \caption{Taitao confusion matrix.  Dataset: A30, 2017q4.}
  \label{tab:taitao_confusion_matrix}
\end{minipage}
\end{table*}

\subsection{Can Taitao Detect Peninsulas?}
	\label{sec:taitao_validation}

We compare Taitao detections
  from 6 \acp{VP}
 to independent observations taken from more than 100 VPs in CAIDA's Ark~\cite{ark_data}.
This comparison is challenging,
  because both Taitao and Ark are imperfect operational systems
  that differ in probing frequency, targets, and method.
Neither defines perfect ground truth,
  but agreement suggests likely truth.

We believe this complexity is warranted because
  Ark provides a more diverse perspective (with 171 locations),
  if we can account for its much sparser frequency.
Ark traceroutes also allow us to assess \emph{where} peninsulas begin.
We expect to see a strong correlation between Taitao peninsulas and Ark observations.
(We considered RIPE Atlas as another external dataset,
  but its coverage is sparse, while Ark covers all /24s.)

\textbf{Identifying comparable blocks:}
We study 21 days of Ark observations from 2017-10-10 to -31.
Ark covers all networks with two strategies.
With team probing in 2017,
  a 40 VP ``team'' traceroutes to
  all routed /24 about once per day.
For prefix probing,
  about 35 VPs each traceroute to .1 addresses of all routed /24s every day.
We use both types of data: the three Ark teams
  and all available prefix probing VPs.
We group results by /24 blocks,%
  considering /24s instead of ASes to be sensitive to intra-AS peninsulas.

Ark differs from Taitao's Trinocular input in three ways:
  the target is a random address or the .1 address in each block;
  it uses traceroute, not ping;
  and it probes blocks daily, not every 11 minutes.
Sometimes these differences cause Ark traceroutes to fail
  when a simple ping succeeds.
First, Trinocular's targets respond more often because
  it uses a curated hitlist~\cite{Fan10a}
  while Ark does not.
Second, Ark's traceroutes can terminate due to path
  \emph{loops}
  or \emph{gaps} in the path,
  (in addition to succeeding or reporting target unreachable).
We do not consider results with gaps, %
  so problems on the path do not
  bias results for endpoints reachable by direct pings.

To correct for differences in target addresses,
  we must avoid
  misinterpreting a block as unreachable
  when the block is online but Ark's target address is not,
  we discard traces sent to never-active addresses
  (those not observed in 3 years of complete IPv4 scans), %
  and blocks
  for which Ark did not get a single successful response.%
Since
  dynamic addressing~\cite{Padmanabhan16a} means Ark often
  fails with an unreachable last hop,
  we see conflicting observations in Ark, implying false peninsulas.
We therefore trust Ark confirmation of outages and full reachability,
  but question Ark-only peninsulas.

To correct for Ark's less frequent probing,
  we compare \emph{long-lived} Trinocular down-events (5 hours or more).
Ark measurements are infrequent (once every 24 hours) compared to Trinocular's 11-minute reports,
  so short Trinocular events are often unobserved by Ark.
(Since outage durations are heavy-tailed, 5\,h gives
  Ark some time to confirm without discarding too many events.)
To confirm agreements or conflicting reports from Ark,
  we require at least 3 Ark observations within the peninsula's span of time.
Varying these parameters is potential future work;
  with small quantitative changes likely,
  but changes to overall bounds unlikely.

We filter out blocks with frequent transient changes
  or signs of network-level filtering, as prior work~\cite{quan2013trinocular,Shah17a,richter2018advancing}.
We define the ``reliable'' blocks suitable for comparison
  as those responsive for at least 85\% of the quarter
  from each of the 6 Trinocular VPs.
(This threshold avoids diurnal blocks or blocks with long outages;
  values of 90\% or less have similar results.)
We also discard flaky blocks whose responses are frequently inconsistent across \acp{VP}.
(We consider more than 10 combinations of \ac{VP} as frequently inconsistent.)
For the 21 days, we find 4M unique Trinocular /24 blocks,
  and 11M Ark /24 blocks,
  making 2M blocks in both available for study.

\textbf{Results:}
\autoref{tab:taitao_confusion_matrix} shows outcomes,
   treating Taitao as prediction and Ark as truth,
  with details in \autoref{tab:taitao_validation_table}.
Dark green indicates true positives (TP):
  when (a) either both Taitao and Ark show mixed results, both indicating a peninsula,
  or when (b) Taitao indicates a peninsula (1 to 5 sites up but at least one down),
Ark shows all-down during the event and up before and after.
We treat Ark in case (b) as positive
  because the infrequency of Ark probing
  (one probe per team every 24 hours) %
  means we cannot guarantee
  VPs in the peninsula will probe responsive targets in time.
Since peninsulas are not common, so too are true positives,
  but we see 184 TPs.

We show \emph{true negatives} as light green and neither bold nor italic.
In almost all of these cases (1.4M)
  both Taitao and Ark reach the block, agreeing.
The vast majority of these are an artifact of our use of
  Ark as ``ground truth'', when it is not designed
  to accurately measure partitions.
The challenge of an Ark claim of peninsula is that
  about  5/6ths of Ark probes fail in the last hop because it probes a single
  random address (see \cite{quan2013trinocular} figure 6).
As a result, while positive Ark results support non-partitions,
  negative Ark results are most likely a missed target and not
  an unreachable block;
  we expand on this analysis in
  \autoref{sec:false_negative_details}.
We therefore treat this second most-common result (491k cases) as a true negative.
For the same reason, we include the small number (97) of cases where
  both Ark and Taitao report all-down,
  assuming Ark terminates at an empty address.
We include in this category the 90 events where Ark is all-down and Trinocular
is all-up.
We attribute Ark's failure to reach its targets to infrequent probing.

We mark \emph{false negatives} as red and bold.
For these few cases (only 12),
  all Trinocular \acp{VP} are down, but
  Ark reports all or some responding.
We believe these cases indicate blocks that have chosen to drop Trinocular traffic.

Finally, yellow italics shows when Taitao's peninsulas
  are \emph{false positives}, since all Ark probes reached the target block.
This case occurs when either traffic from some Trinocular \acp{VP} is
  filtered, or all Ark VPs are ``inside'' the peninsula.
Light yellow (strict) shows all the 251 cases that Taitao detects.
For most of these cases (201),
  five Trinocular \acp{VP} responding and one does not,
  suggesting network problems are near one of the Trinocular VPs
  (since five of six independent VPs have working paths).
Discarding these cases we get 40 (orange); still conservative but a \emph{looser} estimate.

The strict scenario sees precision 0.42, recall 0.94, and $F_1$ score
  0.58,
  and in the loose scenario, precision improves to 0.82 and
  $F_1$ score to 0.88.
We consider these results a strong lower bound on the size of problem,
  and confirmation that the peninsulas detected by Taitao are correct.

Of course custom measurement could align with our analysis
  and should close this bound,
  but the need to build in long-term, existing data,
  motivates these early, rough bounds.
We expect future work to tighten these bounds.

\subsection{Can Chiloe Detect Islands?}
\label{sec:chiloe_validation}

Chiloe (\autoref{sec:chiloe}) detects islands when a \ac{VP} within the island
  can reach less than half the rest of the world.

\textbf{Trinocular:}
To validate Chiloe's correctness,
  we compare when a single \ac{VP} believes to be in an island,
  against what the rest of the world believes about that \ac{VP}.
We begin with Trinocular, where we have strong evidence for a few \acp{VP},
  then we summarize Atlas with 13k \acp{VP}.

Islands are unreachable, like $D$ in \autoref{fig:term_concept}.
We measure blocks,
  so if any address in block $D$ can reach another, it is an island.
If no external \acp{VP} can reach $D$'s block,
  Chiloe confirms an island,
  but some \ac{VP} reaching $D$'s block implies a peninsula.
In \autoref{sec:site_correlation} we show that Trinocular \acp{VP} are independent,
  and therefore no two \acp{VP} live within the same island.
We believe this definition is the best possible ground truth,
  since perfect classification requires instant,
  global knowledge and cannot be measured in practice.

We take 3 years worth of data from all six
  Trinocular \acp{VP}.
From Trinocular's pacing, we analyze 11-minute bins.

  \begin{table}
\centering
	\resizebox{0.38\columnwidth}{!}{
\subfloat[Chiloe confusion matrix \label{tab:chiloe_validation}]{%
    \centering
    \begin{tabular}{c c P{30pt} P{35pt} @{}m{0pt}@{}}
      & & \multicolumn{2}{c}{\normalsize \textbf{Chiloe}} \\
      & & \parbox{40pt}{\centering Island} & \parbox{40pt}{\centering Peninsula} \\
      \multirow{3}{10pt}{\rotatebox[origin=l]{90}{\parbox{50pt}{\normalsize \textbf{Trinocular}}}} &
      \parbox[10pt][20pt][c]{60pt}{Block Island} &
      \cellcolor[HTML]{99ee77} 2 & \cellcolor[HTML]{F0ABAB} \textbf{0} \\
      & \parbox[10pt][20pt][c]{60pt}{Addr Island} &
      \cellcolor[HTML]{99ee77} 19 & \cellcolor[HTML]{F0ABAB} \textbf{8} \\
      & \parbox[10pt][20pt][c]{40pt}{Peninsula}
      & \cellcolor[HTML]{FFF9C4} \emph{2} & \cellcolor[HTML]{CCFF99} 566
    \end{tabular}

	  }}%
\hspace{20mm}
	\resizebox{0.34\columnwidth}{!}{
\subfloat[Detected islands\label{tab:island_summary}]{%
    \centering
  \begin{tabular}{c c c}
   Sites	& Events	& Per Year \\
   \midrule
   W	    & 5		    & 1.67 \\
   C	    & 11  		& 3.67 \\
   J	    & 1		    & 0.33 \\
   G	    & 1		    & 0.33 \\
   E	    & 3		    & 1.00 \\
   N	    & 2		    & 0.67 \\
    \hline
    All (norm.)     & 23        & 7.67 (1.28)  \\
  \end{tabular}

	  }}
    \caption{(a) Chiloe confusion matrix, events between 2017-01-04 and 2020-03-31, datasets A28 through A39.
(b) Islands detected from 2017q2 to 2020q1.
	  }
\vspace*{-2ex}
\end{table}

In \Cref{tab:chiloe_validation} we show that Chiloe detects 23 islands
across three years.
In 2 of these events, the block is unreachable from other \acp{VP},
  confirming the island with our validation methodology.
Manual inspection confirms that the remaining
19 events are islands too, but at the address level---the \ac{VP}
  was unable to reach anything but did not lose power,
  and other addresses in its block were reachable from \acp{VP} at other locations.
These observations suggest a VP-specific problem making it an island.
Finally, for 2 events, the prober's block was reachable during the event by
every site including the prober itself which suggests partial connectivity
(a peninsula), and therefore a false positive.

In the 566 non-island events (true negatives),
  a single \ac{VP} cannot reach more than 5\% but less than 50\% of
  the Internet core.
In each of these cases, one or more
other \acp{VP} were able to reach the affected \ac{VP}'s block,
  showing they were not an island (although perhaps a peninsula).
The table omits the frequent events when less than 5\% of the network is unavailable from the \ac{VP},
  although they too are true negatives.

Bold red shows 8 false negatives. These are events that last about 2 Trinocular
rounds or less (22 min), often not enough time for Trinocular to change its
belief on block state.

\textbf{Atlas:}
With 13k VPs, RIPE Atlas provides a broader view of islands.
We find 188 (v4) and 388 (v6) Atlas VPs are islands (\autoref{sec:dnsmon}),
  accounting for \emph{the majority of DNS unreachable events}.
RIPE operators confirmed these are often misconfigurations.

\textbf{Operators:}
Beyond this quantitative comparison, we discussed
  islands with Trinocular and RIPE Atlas operators.
They confirm our examples %
  and trends (\autoref{fig:tsaluja_longitudinal_partial_outages}).

\subsection{Are the Sites Independent?}
	\label{sec:site_correlation}

Our evaluation assumes \acp{VP} do not share common network paths.
\acp{VP} improve path diversity
  by network diversity and physical distance,
  particularly with today's ``flatter'' Internet~\cite{Labovitz10c}.
We next quantify and validate this assumption.

We measure similarity of observations
  between pairs of VPs.
We examine only cases where one of the pair disagrees with some other VP,
  since when all agree, we have no new information.
If the pair agrees with each other, but not with the majority,
  the pair shows similarity.
If they disagree with each other, they are dissimilar.
We quantify similarity $S_P$ for a pair of sites $P$ as
	${S_P = (P_1 + P_0)/(P_1 + P_0 + D_*)}$,
where $P_s$ indicates the pair agrees on the network having state $s$ of
  up (1) or down (0) and disagrees with the others,
  and for $D_*$, the pair disagrees with each other.
$S_P$ ranges from 1, where the pair always agrees,
  to 0, where they always disagree.

\autoref{tab:overall_correlation} shows similarities for each pair
of the 6 Trinocular VPs (as half of the symmetric matrix).
No two sites have a similarity more than 0.14,
  and most pairs are under 0.08.
This result shows that no two sites are particularly correlated.

\begin{table}
\centering
\begin{minipage}[b]{.38\textwidth}
  \resizebox{\textwidth}{!}{
	\begin{tabular}{c|c@{\hspace{0.7ex}}c@{\hspace{0.7ex}}c@{\hspace{0.7ex}}c@{\hspace{0.7ex}}c}
       & C      & J      & G      & E      & N      \\
	\hline
	W  & 0.017  & 0.031  & 0.019  & 0.035  & 0.020  \\
    C  &        & 0.077  & 0.143  & 0.067  & 0.049  \\
    J  &        &        & 0.044  & 0.036  & 0.046  \\
    G  &        &        &        & 0.050  & 0.100  \\
    E  &        &        &        &        & 0.058  \\
    \end{tabular}
    }
    \captionsetup{type=table}
    \caption{Similarities all VPs. Dataset: A30, 2017q4.}
    \label{tab:overall_correlation}
\end{minipage}
\hspace{10mm}
\begin{minipage}[b]{.53\linewidth}
    \resizebox{1\textwidth}{!}{
    \begin{tabular}{l r r r r r r}
      & \multicolumn{4}{c}{\textbf{IPv4 Addresses}} &
      \multicolumn{2}{c}{\textbf{IPv6 Addresses}} \\
      \textbf{RIR}       &
      \multicolumn{2}{c}{\textbf{Active}}  &
      \multicolumn{2}{c}{\textbf{Allocated}}  &
      \multicolumn{2}{c}{\textbf{Allocated}} \\
      \midrule
      AFRINIC   & 15M    &  2\%   & 121M   & 3.3\%      & 9,661   & 3\%       \\
      APNIC     & 223M   & \cellcolor[HTML]{99ee77}33\%   & 892M  & 24.0\%      & 88,614  & 27.8\%    \\
      \rowcolor[HTML]{DCDCDC}
      \hspace{1mm} \emph{China}   & 112M & 17\% &  345M   &    9.3\%     & 54,849  & \cellcolor[HTML]{FFF9C4}17.2\%    \\
      ARIN      & 150M   & 22\%   & 1,673M & \cellcolor[HTML]{99ee77}45.2\%      & 56,172  & 17.6\%    \\
      \rowcolor[HTML]{DCDCDC}
      \hspace{1mm} \emph{U.S.}    & 140M & \cellcolor[HTML]{FFF9C4}21\% & 1,617M  & \cellcolor[HTML]{FFF9C4}43.7\%        & 55,026  & \cellcolor[HTML]{FFF9C4}17.3\%  \\
      LACNIC    & 82M  & 12\%     & 191M  & 5.2\%      & 15,298  & 4.8\%     \\
      RIPE NCC  & 206M & 30\%     & 826M  & 22.3\%      & 148,881 & \cellcolor[HTML]{99ee77}46.7\%    \\
      \rowcolor[HTML]{DCDCDC}
      \hspace{1mm} \emph{Germany} & 40M & 6\% & 124M  &    3.3\%     & 22,075  & 6.9\%     \\
      \midrule
      Total &  676M &  100\% & 3,703M & 100\%      & 318,626 & 100\%     \\
    \end{tabular}}
    \captionof{table}{RIR IPv4 hosts and IPv6 /32 allocation  \cite{iana_v4, iana_v6}. }
    \label{tab:rir_allocation}
\end{minipage}
\end{table}

\subsection{Stability Across Time}

We confirm our results are not time-dependent
  by repeating key results in multiple years,
  including operational result from 2022 to 2025
  (\autoref{fig:tsaluja_longitudinal_partial_outages} in \autoref{sec:dnsmon}),
  and confirm all results with multiple sources and dates
   (see \autoref{sec:2020}).
We expect these results to apply today
  since partial reachability has persisted since 2001~\cite{andersen2001resilient},
  with some events lasting years~\cite{cloudflare_he},
  as our results document (\autoref{fig:tsaluja_longitudinal_partial_outages}).
We use older data in some examples
  to avoid limitations of measurement deployments.
During 2017q4, Trinocular had six active VPs
  and Ark had three teams,
  providing strong statements from many perspectives.
Trinocular had fewer VPs in 2019 and early 2020,
  and Ark has fewer teams today,
  but 2020 gives quantitatively similar results
   (see \autoref{sec:2020}).
\autoref{sec:peninsula_locations} uses 2020q3 data
  because Ark observed a very large number of loops in 2017q4.

\subsection{Varying Parameters and Geography}

Our algorithms are influenced by the parameters
  in our data sources, including
  how often and where they probe,
  where they are placed,
  and how many \acp{VP} they employ, and how much data we analyze.
We vary \emph{all of these parameters} across our datasets
  (see \autoref{tab:data_types}),
  but the requirement for Internet-wide data
  spanning months and years
  means we depend on existing deployed infrastructure.
Systematically varying \ac{VP} frequency and location
  is challenging future work.

We believe these diverse data sources \emph{confirm our results apply
  over a range of geographic locations}.
We study locations quantitatively in \autoref{sec:site_correlation})
  and confirm stable results with Atlas across 3k ASes and 12k locations
  in \autoref{sec:dnsmon}.
Thus, while we certainly greatly \emph{undercount} the absolute
  numbers of peninsulas and islands observed from Trinocular's 6 locations
  (\autoref{sec:evaluation}),
  Atlas confirms these trends apply with 12k \acp{VP}.

\textbf{IPv6:}
Given data, our algorithms apply to both IPv4 and IPv6.
We provide results for both v4 and v6
  with RIPE Atlas and DNSmon (\autoref{sec:dnsmon}),
  and for Internet-wide v4 with Trinocular.
Internet-wide IPv6 results depend on v6 outage detection,
  an area of active and future research.

\section{Internet Islands and Peninsulas}
	\label{sec:evaluation}

We now examine islands and peninsulas in the Internet core.

\subsection{How Common Are Peninsulas?}
	\label{sec:peninsula_frequency}

We estimate how often peninsulas occur  in the Internet core
  in three ways.
First, we directly measure the visibility of peninsulas %
  by summing the duration of peninsulas as seen from six VPs.
Second, we confirm the accuracy of this estimate
  by evaluating its convergence as we vary the number of VPs---more VPs
  show more peninsula-time, but a result that converges
  suggests it is
  approaching the limit.
Third, we compare peninsula-time to outage-time,
  showing that, in the limit, observers see both for about the same
  duration.
Outages correspond to service downtime~\cite{down_time_cost},
  and are a recognized problem in academia and industry.
Our results show that \emph{peninsulas are as common as outages},
  suggesting peninsulas are an important new problem deserving attention.

\textbf{Peninsula-time:}
We estimate the duration an observer can see a peninsula
  by considering three types of events: \emph{all up}, \emph{all down}, and
  \emph{disagreement} between six VPs.
Disagreement, the last case, suggests a peninsula,
  while agreement (all up or down),
  suggests no problem or an outage.
We compute peninsula-time by summing the time each target /24
  has disagreeing observations from Trinocular VPs.

\begin{figure*}
\adjustbox{valign=b}{\begin{minipage}[b]{.30\linewidth}
    \includegraphics[width=1\linewidth]{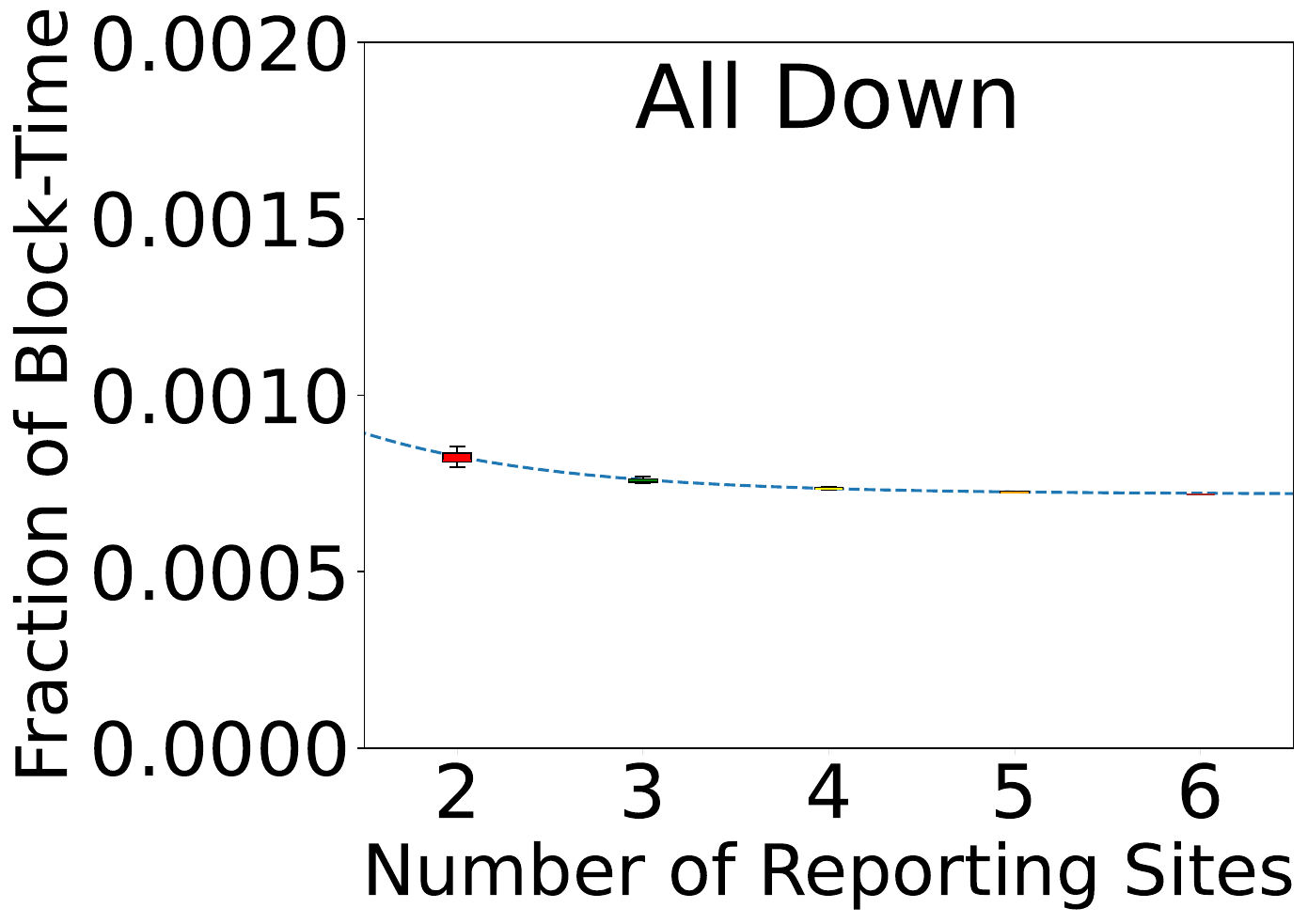}
\end{minipage}}\quad
\adjustbox{valign=b}{\begin{minipage}[b]{.30\linewidth}
    \includegraphics[width=1\linewidth]{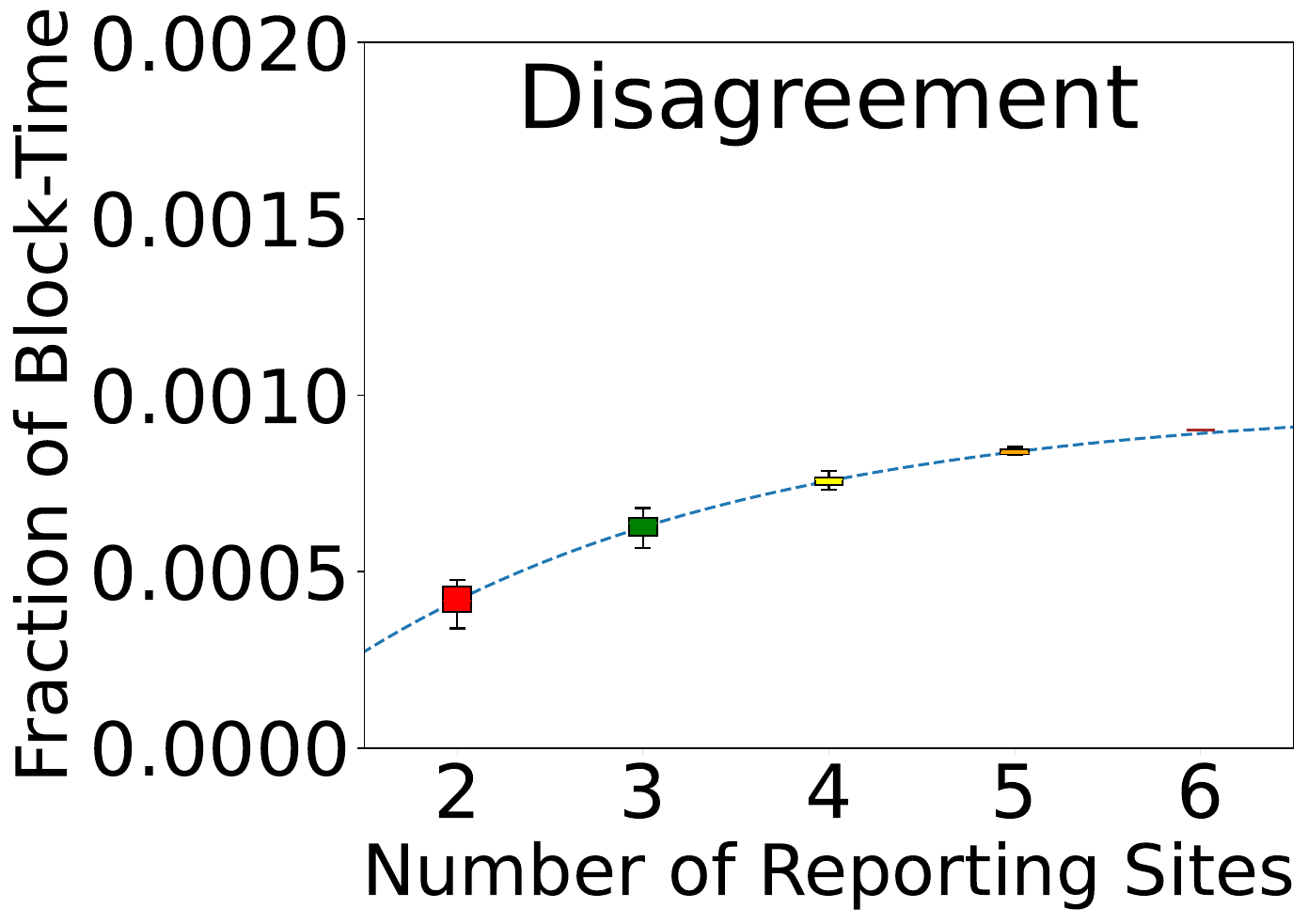}
\end{minipage}}\quad
\adjustbox{valign=b}{\begin{minipage}[b]{.4\linewidth}
    \includegraphics[width=.95\linewidth]{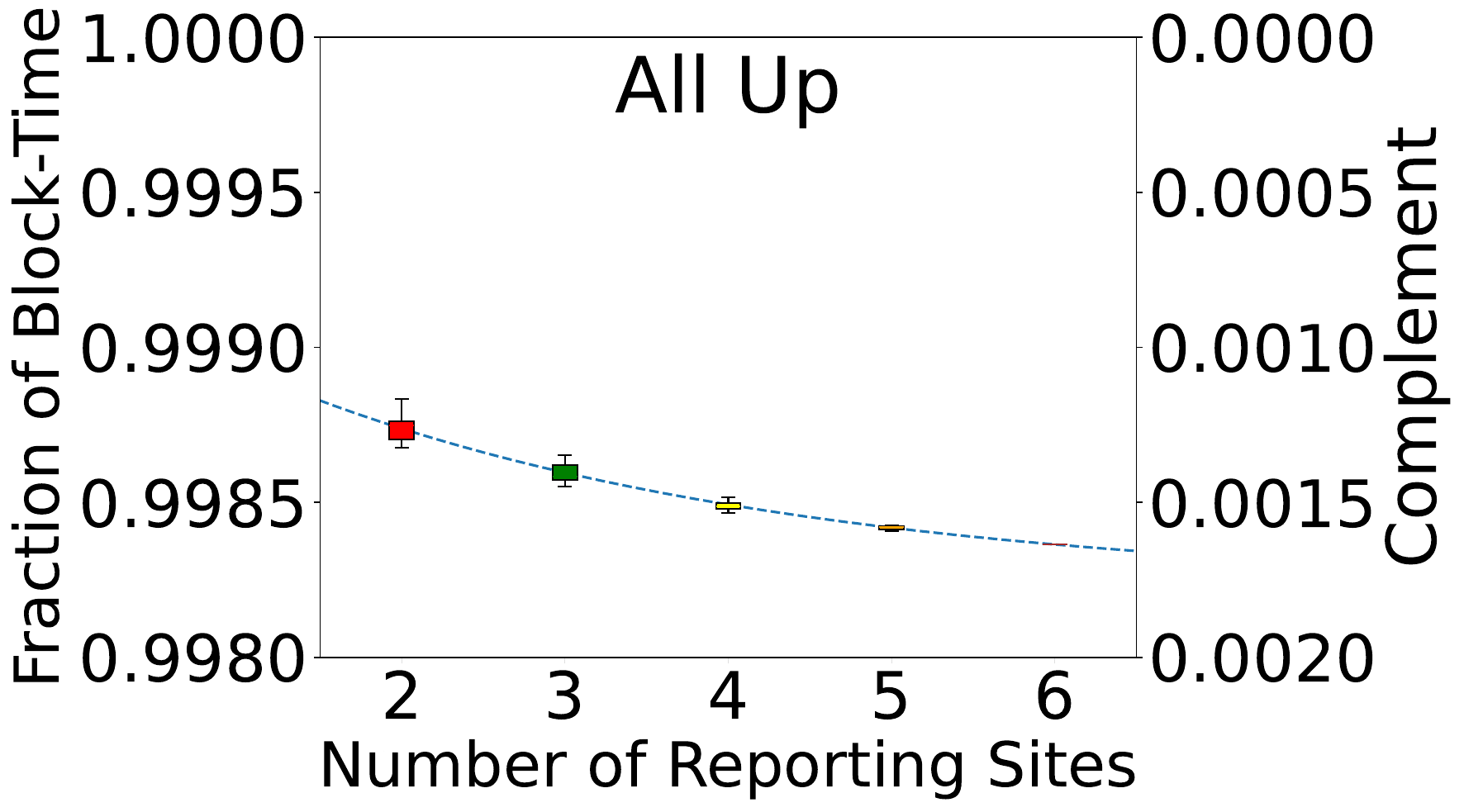}
\end{minipage}}
\caption{Distribution of block-time fraction: all-down
        (left), disagreement (center), and all-up (right), events $\ge 1$ hour.
        Data: 3.7M blocks, 2017-10-06 to -11-16, A30.}
\label{fig:a30all_peninsulas_duration_oct_nov}
\end{figure*}

We have computed peninsula-time
  by evaluating Taitao over Trinocular data for 2017q4~\cite{LANDER14d}.
\autoref{fig:a30all_peninsulas_duration_oct_nov} shows the distribution of peninsulas
measured as a fraction of block-time for an increasing number of sites.
We consider all possible combinations of the six sites.

First we examine the data with all 6 VPs (the rightmost points).
We see that %
  peninsulas (the middle, disagreement graph)
  are visible about 0.00075 of the time.
This data suggests \emph{peninsulas regularly occur,
  appearing at least 0.075\% of the time}.
Fortunately, large peninsulas are rare from many locations---our 6 VPs almost always see the same targets.

\textbf{Convergence:}
While more \acp{VP} provide a better view of the Internet core's overall state,
  but the \emph{global fraction} of affected networks will show diminishing returns
  after major problems are found.
That is previously inferred outages (all unreachable) should have been peninsulas,
  with partial reachability.
All-down (left) decreases from an average of 0.00082 with 2 \acp{VP} to 0.00074 for 6
\acp{VP}. All-up (right) goes down a relative 47\% from 0.9988 to 0.9984, while
disagreements (center) increase from 0.0029 to 0.00045.
Outages (left) converge after 3 sites,
  as shown by the fitted curve and decreasing variance.
Peninsulas and all-up converge more slowly.
We conclude that \emph{a few, independent sites (3 or 4)
  converge on a good estimate of the fraction of true islands
  and peninsulas}.

We support this claim by comparing
  all non-overlapping combinations of 3 sites.
If all combinations are equivalent,
  then a fourth site will not add new information.
Six \acp{VP} yield 10 possible sets of 3 sites;
  we examine those combinations for each of 21 quarters, from 2017q2 to 2020q1.
When we compare the one-sample Student $t$-test
  to evaluate if the difference of each pair of combinations of those 21 quarters
  is greater than zero,
  none of the combinations are rejected at confidence level 99.75\%,
  suggesting that any combination of three sites is statistically equivalent
  and confirm our claim that a few sites are sufficient for estimation.

\textbf{Relative impact:}
Finally, comparing outages (the left graph) with peninsulas (the middle graph),
  we see both occur about the same fraction of time (around 0.00075).
This comparison shows that \emph{peninsulas are about as common as outages},
  suggesting they deserve more attention.

\textbf{Generalizing:}
We confirm that each of these results holds in a subsequent year in
        \autoref{sec:2020},
  suggesting the result is not unique to this quarter.
While we reach a slightly different limit (in that case,
  peninsulas and outages appear about in 0.002 of data),
  we still see good convergence after 4 VPs.

While this data demonstrates convergence on the \emph{rate} of
  peninsulas and islands,
  we confirm the rate and show a larger absolute \emph{number} of peninsulas
  with DNSmon's 12k \acp{VP}.

\subsection{How Long Do Peninsulas Last?}
	\label{sec:peninsula_duration}

Peninsulas have multiple root causes:
  some are short-lived routing misconfigurations
  while others reflect long-term disagreements in routing policy.
In this section we determine the distribution of peninsulas in terms of their duration
  to determine the prevalence of persistent peninsulas.
We will show that there are millions of brief peninsulas,
  likely due to transient routing problems,
  but that 90\% of peninsula-time is in long-lived events (5\,h or more, following~\autoref{sec:taitao_validation}).

We use Taitao to see peninsula duration for all detected in 2017q4:
   some 23.6M peninsulas affecting 3.8M unique blocks.
If instead we look at \emph{long-lived} peninsulas (at least 5\,h),
  we see 4.5M peninsulas in 338k unique blocks.

\autoref{fig:a30_partial_outages_duration_cdf} examines peninsula
  duration
  in three ways:
  a cumulative distribution (CDF) counting all peninsula events
  (left, solid, purple line),
  the CDF of the number of peninsulas for VP-down events
  longer than 5 hours (middle, solid green line),
  and the cumulative size of peninsulas
  for VP down events longer than 5 hours (right, green dashes).

We see that there are many very brief peninsulas (purple line):
  about 65\% last only 20--60 minutes ($\sim$2--6 observations).
With two or more observations, these events are not just
  one-off measurement loss.
These results suggest that while the Internet core is robust,
there are many small connectivity glitches (7.8M events).
Events that are two rounds (20 minutes) or shorter
  may be due to transient BGP blackholes~\cite{bush2009internet}.

The number of day-long or multi-day peninsulas is small,
  only 1.7M events (2\%, the purple line).
However, about 57\% of all peninsula-time is in such longer-lived events
  (the right, dashed line),
  and 20\% of time is in events lasting 10 days or more,
  even when longer than 5 hours events are less numerous (compare the middle, green line to the left, purple line).
Day-long events persist long enough for human network operators to respond,
  and events lasting longer than a week suggest potential
  policy disputes and \emph{intentional} unreachability.
Together, these long-lived events suggest that
  there is benefit to identifying non-transient peninsulas
  and addressing the underlying routing problem.

\subsection{What Sizes Are Peninsulas?}
	\label{sec:peninsula_size}

When network issues cause connectivity problems like peninsulas,
  the \emph{size} of those problems may vary,
  from country-size%
  (see \autoref{sec:country_peninsulas}),
to \ac{AS}-size,
and also for routable prefixes or fractions of prefixes.
We next examine peninsula sizes.

We begin with Taitao peninsula detection at a /24 block level.
We match peninsulas across blocks within the same prefix by start time and
duration, both measured in one hour timebins.
This match implies that the Trinocular \acp{VP} observing the blocks as up are
also the same.

We compare peninsulas to routable prefixes from Routeviews \cite{routeviews},
  using
  longest prefix matches with /24 blocks.

Routable prefixes consist of many blocks, some of which may not be measurable.
We therefore define the \emph{peninsula-prefix fraction}
  for each routed prefix as fraction of blocks in the peninsula
  that are Trinocular-measurable blocks.
To reduce noise provided by single block peninsulas,
  we only consider peninsulas covering 2 or more blocks in a prefix.

\begin{figure*}
\begin{center}
  \subfloat[Number of Peninsulas]{
    \includegraphics[width=0.4\columnwidth]{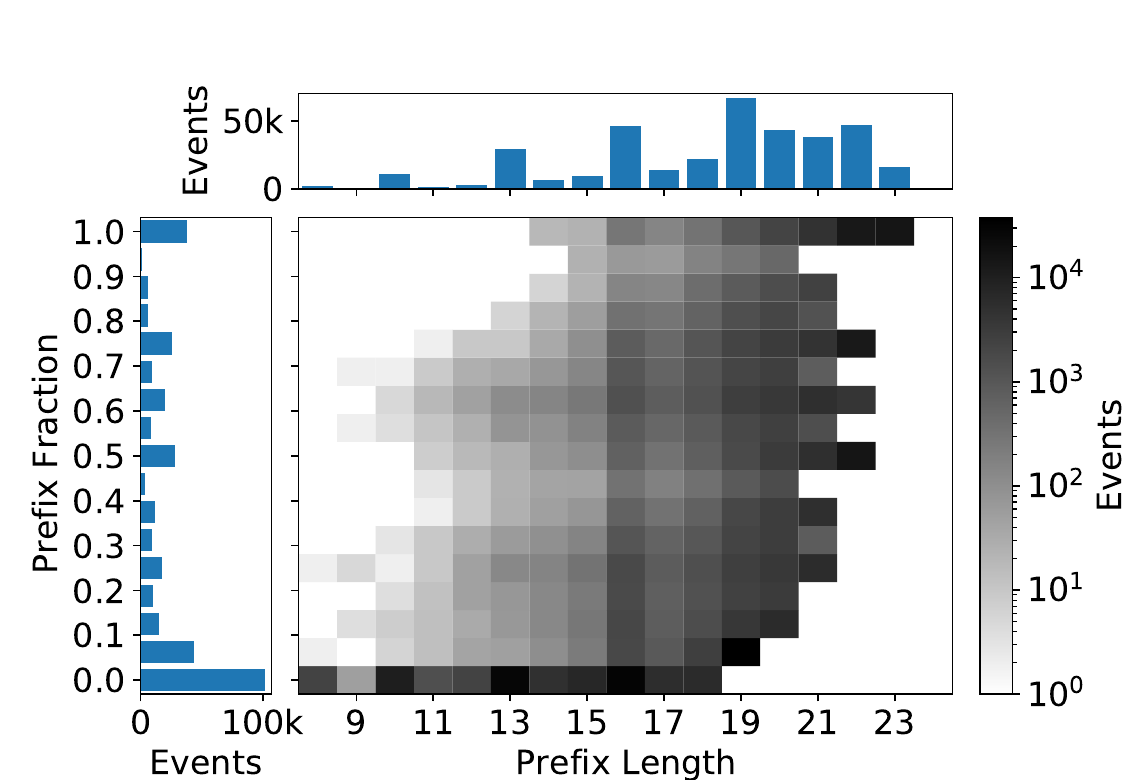}
    \label{fig:a30all_blocks_in_prefix_prefix_fraction_heatmap}
  }
\quad
  \subfloat[Duration fraction]{
    \includegraphics[width=0.4\columnwidth]{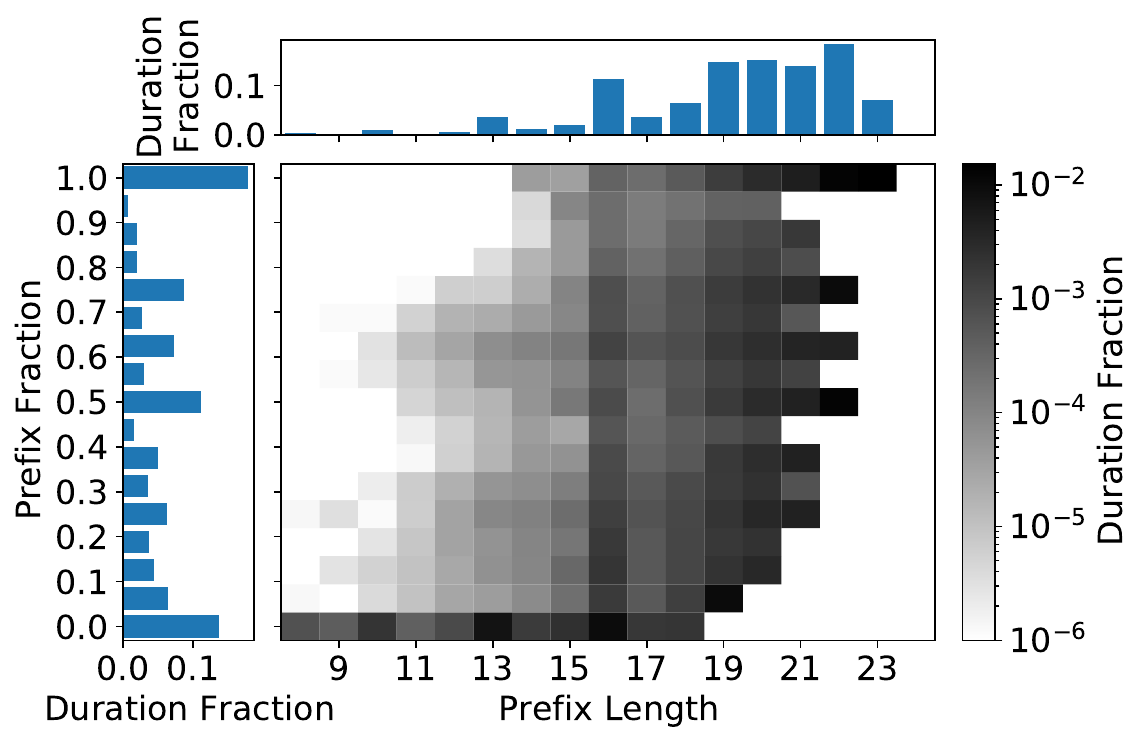}
    \label{fig:a30all_blocks_in_prefix_prefix_fraction_heatmap_duration}
  }
\end{center}
\caption{Peninsulas measured with per-site down events longer than 5 hours. Dataset A30, 2017q4.}
\end{figure*}

\autoref{fig:a30all_blocks_in_prefix_prefix_fraction_heatmap} shows the number
of peninsulas for different prefix lengths and the fraction of the prefix
affected by the peninsula
  as a heat-map,
  where we group them into bins.

We see that about 10\% of peninsulas
  are likely due to
  routing problems or policies,
  since 40k peninsulas affect the whole routable prefix.
However, a third of peninsulas
  (101k, at the bottom of the plot)
  affect
  only a very small fraction of the prefix.
These low prefix-fraction peninsulas suggest
  that they happen \emph{inside} an ISP and
  are not due to interdomain routing.

Finally, we show that \emph{long-lived peninsulas are likely due to routing or policy choices}.
\autoref{fig:a30all_blocks_in_prefix_prefix_fraction_heatmap_duration}
  shows the same data source,
  but weighted by fraction of time each peninsula
  contributes to the total peninsula time during 2017q4.
Here the larger fraction of weight are peninsulas covering
  full routable prefixes---20\% of all peninsula time during the
quarter (see left margin).

\subsection{Where Do Peninsulas Occur?}
	\label{sec:peninsula_locations}

Firewalls, link failures, and routing problems cause peninsulas on the Internet,
  and can occur at AS boundaries or inside an AS.
We next show that \emph{many peninsulas occur at AS boundaries,
  consistent with policies as a cause} for long-lived events.
(Short-lived events at AS boundaries may be routing transients
  or operator error that is quickly corrected.)

To detect \emph{where} the Internet breaks into peninsulas,
  we look at traceroutes that failed to reach their target address,
  either due to a loop or an ICMP unreachable message.
Then, we examine if the traceroute
  halts \emph{at} the target AS and target prefix,
  or \emph{before} the target AS and prefix.

For our experiment
  we run Taitao to detect peninsulas at target blocks over Trinocular VPs,
  we use Ark's traceroutes~\cite{ark_data_2020} to find last IP address before halt, and
  we get target and halting ASNs and prefixes using RouteViews.

\begin{table*}
\begin{minipage}[b]{.58\linewidth}
    \centering
    \footnotesize
    \resizebox{1\textwidth}{!}{

    \begin{tabular}{c r r | r r}
      & \multicolumn{2}{c | }{\textbf{Target AS}}
      & \multicolumn{2}{c}{\textbf{Target Prefix}} \\
      Sites Up & At & Before & At & Before \\
      \midrule
      0	& 21,765	    & 32,489	    & 1,775	    & 52,479 \\
      \rowcolor[HTML]{DCDCDC}
      1	& 587	    & 1,197	    & 113	    & 1,671 \\
      \rowcolor[HTML]{DCDCDC}
      2	& 2,981	    & 4,199	    & 316	    & 6,864 \\
      \rowcolor[HTML]{DCDCDC}
      3	& 12,709	    & 11,802	    & 2,454	    & 22,057 \\
      \rowcolor[HTML]{DCDCDC}
      4	& 117,377	& 62,881	    & 31,211	    & 149,047 \\
      \rowcolor[HTML]{DCDCDC}
      5	& 101,516	& 53,649	    & 27,298	    & 127,867 \\
  	\cline{2-5}
      \rowcolor[HTML]{DCDCDC}
      \textbf{1-5} & \cellcolor[HTML]{99ee77} \textbf{235,170} & \textbf{133,728} & \textbf{61,392} & \cellcolor[HTML]{99ee77}\textbf{307,506} \\
      6	& 967,888	& 812,430	& 238,182	& 1,542,136 \\
  \end{tabular}}
    \caption{Halt location of failed traceroutes for peninsulas longer than 5
    hours. Dataset A41, 2020q3.}
    \label{tab:peninsula_root_cause}
\end{minipage}
\hspace{2mm}
\begin{minipage}[b]{.4\linewidth}
    \centering
    	\footnotesize
        \resizebox{1\textwidth}{!}{
            \includegraphics[trim=25 0 25 0,clip,width=0.99\textwidth]{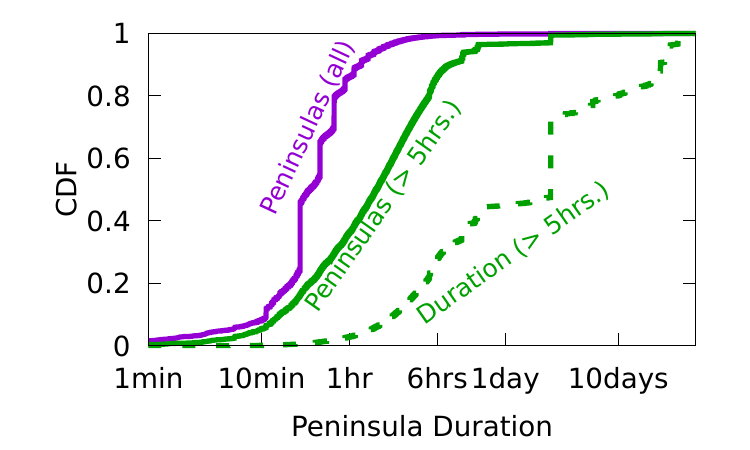}}
        \vspace*{-6mm}
    \captionsetup{type=figure}
    \captionof{figure}{Cumulative peninsulas and peninsula duration. Dataset A30, 2017q4.}
    \label{fig:a30_partial_outages_duration_cdf}
\end{minipage}
\end{table*}

In~\autoref{tab:peninsula_root_cause} we show how many traces halt
\emph{at} or \emph{before} the target network.
The center, gray rows show peninsulas (disagreement between \acp{VP})
  with their total sum in bold.
For all peninsulas (the bold row),
  more traceroutes halt at or inside the target AS (235k vs.~134k, the left columns),
  but they more often terminate before reaching the target prefix (308k vs.~61k, the right columns).
(While traceroutes are imperfect, these large differences ($2\times$ or more)
  suggest a robust qualitative conclusion.)
This difference suggests policy is implemented at or inside ASes, but not at routable prefixes.
By contrast, outages (agreement with 0 sites up)
  more often terminate before reaching the target AS.
Because peninsulas are more often at or in an AS,
  while outages occur in many places,
  it suggests that long-lived peninsulas are policy choices
  consistent with public operator reports~\cite{ipv6peeringdisputes,Leber09a,Anderson10a,Rayburn16a,ThinkBroadband24a,CartwrightCox24a}.

\subsection{How Common Are Islands?}
	\label{sec:how_common_are_islands}

Multiple groups have shown that there are many network outages in the Internet~\cite{Schulman11a,quan2013trinocular,Shah17a,richter2018advancing,guillot2019internet}.
We have described (\autoref{sec:problem}) two kinds of outages:
  full outages where all computers at a site are down (perhaps due to a loss of power),
  and islands, where the site is cut off from the Internet core, but computers
    at the site can talk between themselves.
We next use Chiloe to determine how often islands occur.
We study islands in two systems with 6 \acp{VP} for 3 years
  and 13k \acp{VP} for 3 months.

\textbf{Trinocular:}
We first consider three years of Trinocular data (\autoref{tab:data_types}),
  from 2017-04-01 to 2020-04-01.
We run Chiloe across each VP for this period.

\Cref{tab:island_summary} shows the number of islands per VP
  over this period.
Over the 3 years, all six \acp{VP} see from 1 to 5 islands.
In addition,
  we report as islands some cases even though
  not the \emph{entire} Internet core
  is unreachable.
This apparent discrepancy from our definition
  reflects the limitations of our necessarily
  non-instantaneous measurement of the Internet.
We expect such cases, and perhaps other 12 non-islands where 20\% to 50\% is inaccessible,
  are \emph{short-lived} true islands,
  that are incompletely measured because
  the island recovers before we complete
  an 11~minute-long evaluation of all 5M networks for a full Internet scan
  (see \autoref{sec:island_trinocular_threshold} for details).

\textbf{RIPE Atlas:}
For broader coverage we next consider RIPE Atlas'
  13k \acp{VP} for all of 2021q3~\cite{ripe_ping}.
While Atlas does not scan the whole Internet core,
  they do scan most root DNS servers every 240\,s.
Chiloe would like to observe the whole Internet core, and
  while Trinocular scans 5M /24s,
  it does so with only 6 VPs.
To use RIPE Atlas' VPs,
  we approximate a full scan
  with probes to 12 of the DNS root server systems (G-Root was unavailable in 2021q3).
Although far fewer than 5M networks,
  these targets provide a very sparse sample
  of usually independent destinations since each is independently operated.
Thus we have complementary datasets
  with sparse VPs and dense probing, and
  many VPs but sparse probing.
In other words, to get many VP locations
  we relax our conceptual definition by decreasing our target list.

\begin{figure*}
  \hspace*{-5ex}
  \captionsetup{justification=centering}
  \subfloat [Number of islands]{
    \includegraphics[trim=25 0 25 0,clip,width=0.33\linewidth]{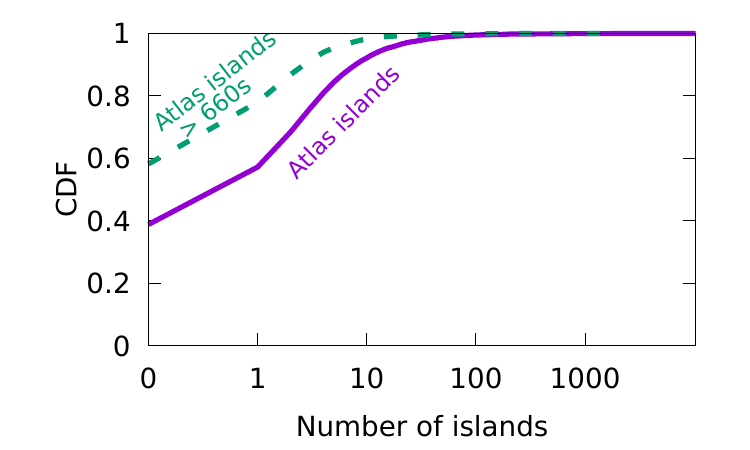}
    \label{fig:islands_per_node}
  }
  \subfloat[Duration of islands]{
    \includegraphics[trim=25 0 25 0,clip,width=0.33\textwidth]{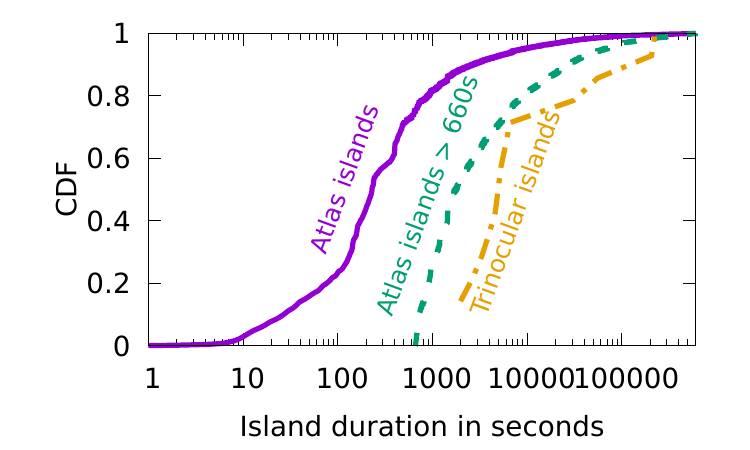}
    \label{fig:island_duration}
  }
  \subfloat[Size of islands]{
    \includegraphics[trim=25 0 25 0,clip,width=0.33\textwidth]{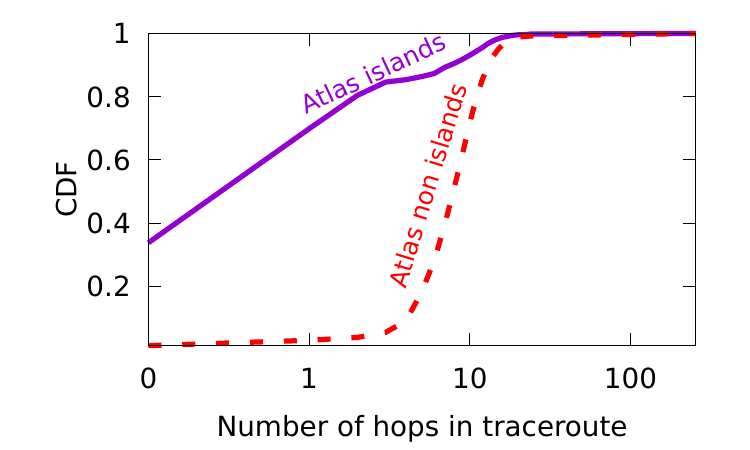}
    \label{fig:island_path_length}
  }
    \captionsetup{type=figure}
    \captionsetup{width=0.9\textwidth}
    \captionof{figure}{CDF of islands detected by Chiloe for data from Trinocular (3 years, Datasets A28-A39) and %
  Atlas (2021q3).}
        \label{fig:a30all_peninsulas_fig7}
\end{figure*}

\autoref{fig:islands_per_node} shows the CDF of the number of islands detected
per RIPE Atlas \ac{VP} during 2021q3.
During this period, 55\% of \acp{VP} observed one or no islands (the solid line).
We compare to Trinocular with only events longer than 660\,s (the dashed line).
We see that
  60\% of \acp{VP} have no islands; 19\%, one; with 21\% seeing more.
The annualized rate of the stable \acp{VP} that see 2 or fewer islands
  is 1.75 islands per year (a lower bound, since we exclude less stable \acp{VP}),
  compared to 1.28 for Trinocular (\Cref{tab:island_summary}).
We see islands are more common in Atlas, perhaps because it includes
  many VPs in homes.

We conclude that islands \emph{do} happen,
  but rarely, and occur at
  at irregular times.
This finding is consistent with importance of the Internet
  at the locations where we run VPs.

\subsection{How Long Do Islands Last?}
\label{sec:islands_duration}

Islands causes range from brief connectivity loss to
  long-term policy differences,
  so
  we next evaluate island duration.

We compare the distributions of island durations observed from
  RIPE Atlas (the left line) and Trinocular (right) in
  \autoref{fig:island_duration}.
Since Atlas' frequent polling means it detects islands lasting seconds,
  while Trinocular sees only islands of 660\,s or longer,
  we split out Atlas events lasting at least 660\,s
  (middle line).
All measurements follow a similar S-shaped curve,
  but for Trinocular, the curve is truncated at 660\,s.
With only 6 VPs, Trinocular sees far fewer events (23 in 3 years compared to 235k in a yearly quarter with Atlas),
  so the Trinocular data is quantized.
In both cases, about 70\% of islands are between 1000 and 6000\,s.
This graph shows that Trinocular's curve is similar in shape to Atlas-660\,s,
  but about $2\times$ longer.
All Trinocular observers are in datacenters,
  while Atlas devices are often at homes,
  so this difference may indicate that datacenter islands are rarer, but harder to resolve.

\subsection{What Sizes Are Islands?}
	\label{sec:islands_sizes}

\subsubsection{Island Size via Traceroute}
	\label{sec:traceroute_island_size}

First we evaluate island sizes,
  comparing traceroutes before and during an island.
We use traceroutes from RIPE Atlas \acp{VP} sent to 12 root DNS servers
  for 2021q3~\cite{ripe_traceroute}.
\autoref{fig:island_path_length} shows
  the distribution of number of traceroute hops reaching target (green),
  and \emph{not} reaching their target (purple),
  for VPs in islands (\autoref{sec:how_common_are_islands}).

Most islands are small, with 70\% at~0 or 1~hop. %
We believe huge islands (10 or more hops) are likely false positives.

\subsubsection{Country-sized Islands}
	\label{sec:country_sized_islands}

We have some evidence of country-sized islands:
In 2017q3, on 8 occasions it appears that most or all of China
  stopped responding to external pings
  (visualized in \autoref{fig:a29all_outagedownup_4096} in \autoref{sec:chinese_islands}).
We found no problem reports on network operator mailing lists,
  so we believe these outages were ICMP-specific and likely did not affect
  web traffic.
Since there were no public reports,
  we assume the millions of computers inside China continued to operate,
  suggesting that China was briefly a country-wide
  \emph{ICMP-island}.
Such large examples have not re-occurred.

\section{Applying These Tools}
        \label{sec:applications}

\subsection{Can the Internet Core Partition?}
	\label{sec:internet_partition}

In \autoref{sec:other_applications}
  we discussed secession and expulsion qualitatively.
Here we ask: Does any
  country or group have enough addresses to secede and claim to be
  ``the Internet core'' with a majority of addresses?
Alternatively,
  if a country were to exert control over their allocated addresses,
  would they become
  a country-sized island or peninsula?
We next use our reachability definition of more than 50\%
  to quantify control of the IP address space.

To evaluate the power of countries and \acp{RIR} over the Internet core,
  \Cref{tab:rir_allocation} reports the number of active IPv4
  addresses as determined by Internet censuses~\cite{Heidemann08c}
  for \acp{RIR} and selected countries.
Since estimating active IPv6 addresses is an open problem,
  we provide allocated addresses for both v4 and v6~\cite{iana_v4,
  iana_v6}.
(IPv4 has been fully allocated since 2011~\cite{ICANN11a}).

\Cref{tab:rir_allocation} shows that \emph{no individual \ac{RIR} or country can secede and take the Internet core},
  because none controls the majority of IPv4 addresses.
ARIN has the largest share with 1673M allocated (45.2\%).
Of countries, U.S. has the largest share of allocated IPv4 (1617M, 43.7\%).
Active addresses are more evenly distributed
  with APNIC (223M, 33\%) and the U.S.~(40M, 21\%) the largest \ac{RIR} and country.

\emph{IPv6 is also an international collaboration},
  since no \ac{RIR} or country surpasses a 50\% allocation for control.
RIPE (an RIR) is close with 46.7\%,
  and China and the U.S.~have large
  allocations;
  with most v6 unallocated, this balance may change.

IPv4 reflects a first-mover bias, where early adopters acquired
  many addresses,
  but this factor is smaller in IPv6.
Our definition's use of active addresses also reduces this bias,
  since
  numbers of \emph{active} IPv4 addresses
  is similar to allocated IPv6 addresses
  (legacy IPv4 addresses are less used).

\subsection{Other Applications of the Definition}
	\label{sec:other_applications}
	\label{sec:outage_detection_short}

We next examine how a clear definition
  of the Internet core can inform policy tussles~\cite{Clark02a}.
Our hope is that our conceptual definition can make
  sometimes amorphous concepts like ``Internet fragmentation''
  more concrete,
  and an operational definition can quantify impacts
  and identify thresholds.

\textbf{Secession and Sovereignty:}
The U.S.~\cite{cybersecurity_act_2010}, China~\cite{Anonymous12a,Anonymous14a},
and Russia~\cite{russian_internet} have all proposed unplugging from
the Internet.
Egypt did in 2011~\cite{Cowie11a},
  and several countries have during exams~\cite{Gibbs16a,Dhaka18a,Henley18a,Economist18a}.
When the Internet partitions,
  which part is still ``the Internet core''?
Departure of an ISP or small country do not change the Internet core much,
  but what if a large country, or group of countries, leave together?
Our definition (\autoref{sec:definition}) resolves this question,
  since requiring a majority defines an Internet core
  that can end (\autoref{sec:internet_partition})
  if multiple partitions leave none with a majority.

\textbf{Sanction:}
An opposite of secession is expulsion.
Economic sanctions are one method of asserting international influence,
  and events such as the 2022 war in Ukraine prompted
  several large ISPs to discontinue service to Russia~\cite{Reuters22a}.
De-peering does not affect reachability for ISPs that purchase transit,
  but Tier-1 ISPs that de-peer create peninsulas for their users.
As described below in \autoref{sec:internet_partition},
  \emph{no single country can eject another by de-peering with it}.
However, a coalition of multiple countries could
  de-peer and eject a country from the Internet core
  if they, together, control
  more than half of the address space.

\textbf{Repurposing Addresses:}
Given full allocation of IPv4,
  multiple parties proposed re-purposing currently allocated or reserved IPv4 space,
  such 0/8 (``this'' network), 127/8 (loopback), and 240/4 (reserved)~\cite{Fuller08a}.
New use of these long-reserved addresses is challenged
  by assumptions in widely-deployed, difficult to change, existing software
  and hardware.
Our definition demonstrates
  that an RFC re-assigning this space for public traffic
  cannot make it a truly effective part of the Internet core until
  implementations used by a majority of active addresses
  can route to it.

\textbf{IPv4 Squat Space:}
IP squatting is when an organization
  requiring private address space beyond RFC1918
  takes over allocated but currently unrouted IPv4 space~\cite{Aronson15a}.
Several IPv4 /8s allocated to the U.S.~DoD have been used this way~\cite{Richter16c}
  (they were only publicly routed in 2021~\cite{Timberg21a}).
By our definition, such space is not part of the Internet core without
  public routes,
  and if more than half of the Internet is squatting on it,
  reclamation may be challenging.

\textbf{The IPv4/v6 Transition:}
We have defined two Internet cores: IPv4 and IPv6.
Our definition can determine when one supersedes the other.
After more than half of all IPv4 hosts are dual-homed,
  IPv6 will supersede IPv4 when
  a majority of hosts on IPv6 can no longer reach IPv4.
Current limits on IPv6 measurement mean evaluation
  here is future work,
  and show the strength and limits of our definition:
  since IPv6 is already economically important,
  a definition seems unnecessary.
But providing a sharp threshold that makes the maturity of IPv6 definitive may
  help motivate late-movers.

\textbf{Outage Detection:}
Prior outage detection systems have struggled
  with conflicting observations~\cite{Schulman11a,quan2013trinocular,Shah17a,richter2018advancing,guillot2019internet}.
We instead
  recognize such cases as peninsulas in a normal Internet, not measurement error.
(We expand in \autoref{sec:local_outage_eval}.)

\begin{figure*}
\begin{minipage}[b]{.49\linewidth}
    \centering
    \footnotesize
    \resizebox{1\textwidth}{!}{
    \includegraphics[trim=0 0 0 5,clip,width=1\textwidth]{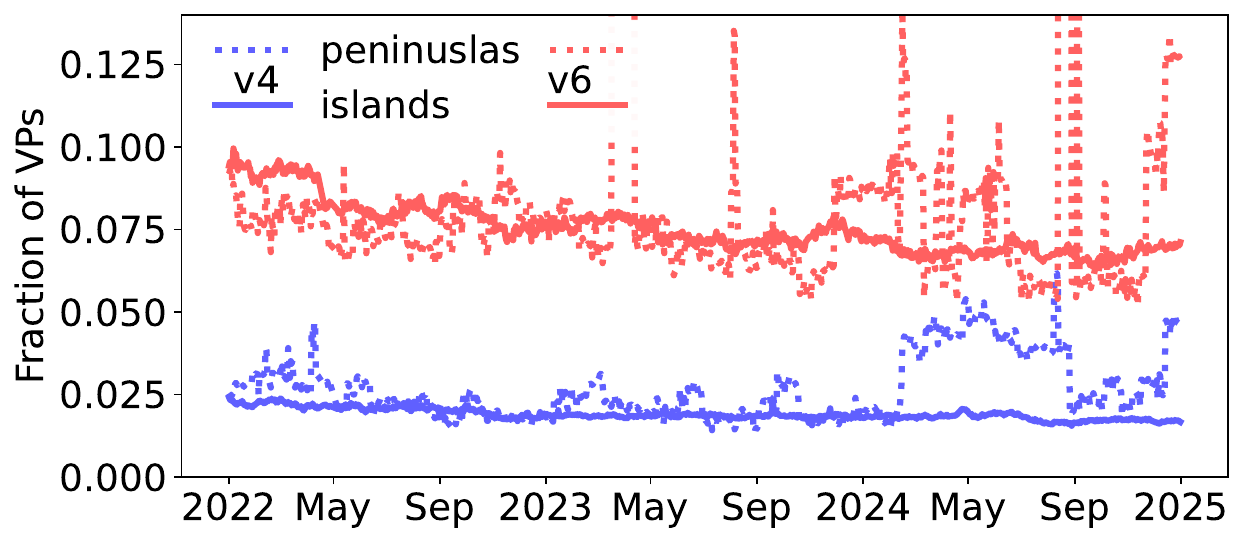}}
        \vspace*{-4mm}
    \captionsetup{type=figure}
    \captionsetup{width=0.9\textwidth}
    \captionof{figure}{Fraction of VPs observing islands and peninsulas for IPv4 and IPv6, 2022--2025.}
  \label{fig:tsaluja_longitudinal_partial_outages}
\end{minipage}
\hspace{2mm}
\begin{minipage}[b]{.49\linewidth}
    \centering
    	\footnotesize
        \resizebox{1\textwidth}{!}{
        \includegraphics[trim=0 0 0 5,clip,width=1\textwidth]{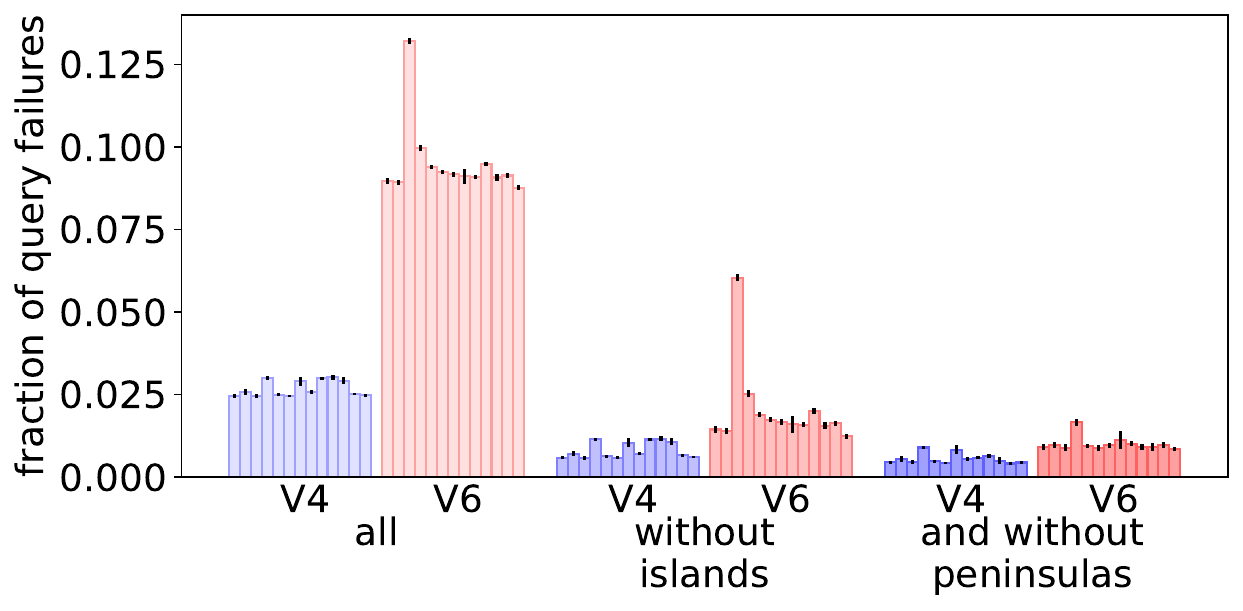}}
        \vspace*{-6mm}
    \captionsetup{type=figure}
    \captionof{figure}{Atlas queries from all available VPs to 13
    Root Servers for IPv4 and IPv6 on 2022-07-23.}
    \label{fig:atlas_revisited}
\end{minipage}
\end{figure*}

\subsection{Improving DNSmon Sensitivity}
	\label{sec:dnsmon}

DNSmon~\cite{Amin15a}
  monitors the Root Server System~\cite{RootServers16a},
  with each RIPE Atlas \ac{VP} measuring its anycast-determined neighbor~\cite{Ripe15c}.
For years, DNSmon has often reported IPv6 loss rates of 4-10\%.
Since the DNS root  is well provisioned and anycast,
  we expect minimal or no loss.

RIPE Atlas operators are aware of problems with some Atlas VPs.
Some VPs support IPv6 on their LAN, but not to the global IPv6 Internet---such VPs
  are IPv6 islands.
Atlas periodically tags and culls these VPs
  from DNSmon.
However, our study of DNSmon
  for islands and peninsulas
  improves their results.
Using concepts pioneered here (\autoref{sec:problem} and \autoref{sec:design}),
  we give full analysis in a
    workshop paper~\cite{Saluja22a};
Here we add new data
  showing these results persist for 3~years (\autoref{fig:tsaluja_longitudinal_partial_outages}).

Groups of bars in
  \autoref{fig:atlas_revisited} show query loss
  for each of the 13 root service identifiers,
  as observed
  from all available Atlas VPs (10,082 IPv4, and 5,173 IPv6)
  on 2022-07-23.
(We are similar to DNSmon, but it uses only about 100 well-connected ``anchors'',
  so our analysis is wider.)
The first two groups show loss rates for IPv4 (light blue, left most) and IPv6 (light red),
  showing IPv4 losses around 2\%, and IPv6 from 9 to 13\%.

We apply Chiloe to these VPs, detecting as islands those VPs that
  cannot see \emph{any} of the 13 root identifiers over 24~hours.
(This definition is stricter than regular Chiloe because these VPs attempt only 13 targets,
  and we apply it over a full day to consider only long-term trends.)
The middle two groups of bars show IPv4 and IPv6 loss rates
  after removing 188 v4 and 388 v6 \acp{VP} that
  are islands.
Without islands,
  v4 loss drops to 0.005 from 0.01, and v6 to 0.01 from 0.06.
These rates represent a more
  meaningful estimate of DNS reliability.
Users of VPs that are IPv6 islands
  will not expect global IPv6,
  and such VPs should not be used for IPv6 in DNSmon.

The third bar in each red cluster of IPv6 is an outlier:
  that root identifier shows 13\% IPv6 loss with all VPs,
  and 6\% loss after islands are removed.
This result is explained by
  persistent routing disputes between Cogent (the operator of C-Root) and Hurricane Electric~\cite{ipv6peeringdisputes}.
Omitting islands (the middle bars) makes this difference much clearer.

Applying Taitao to detect peninsulas, we
  find 14 to 57 v4 peninsulas and 266 (Cogent) and 19 to 49 (others) v6 peninsulas.
Peninsulas suggest persistent routing problems
  meriting attention from ISPs and root operators.
The darker, right two groups show loss remaining (after removal of
  islands and peninsulas),
  representing \emph{underlying events worth root operator attention}.
These bars show all letters see similar events rates,
  \emph{after} we remove persistent problems.

This example shows how \emph{understanding partial reachability
  can improve the sensitivity of existing measurement systems}.
Removing islands makes it easy to identify persistent routing problems.
Removing peninsulas makes
  transient changes (perhaps from failure, DDoS, routing)
  more visible.
Each layer of these problems can be interesting,
  but considering each separately,
  the interesting ``signal'' of routing changes
  (appearing in the right two groups in \autoref{fig:atlas_revisited}),
  is hidden under the
  $5\times$ or $9.7\times$ times larger peninsulas and islands (the left two groups).
Improved sensitivity also \emph{shows a need to improve
  IPv6 provisioning},
  since
  IPv6 loss is statistically higher than
  IPv4 loss (compare the right blue and red groups),
  even accounting for known problems.
After sharing the results with root operators and RIPE Atlas,
  two operators adopted them in regular operation.

\subsection{Outages Given Partial Reachability}
	\label{sec:local_outage_eval}
	\label{sec:representing_the_internet}

\begin{figure}
  \centering
  \includegraphics[width=6cm]{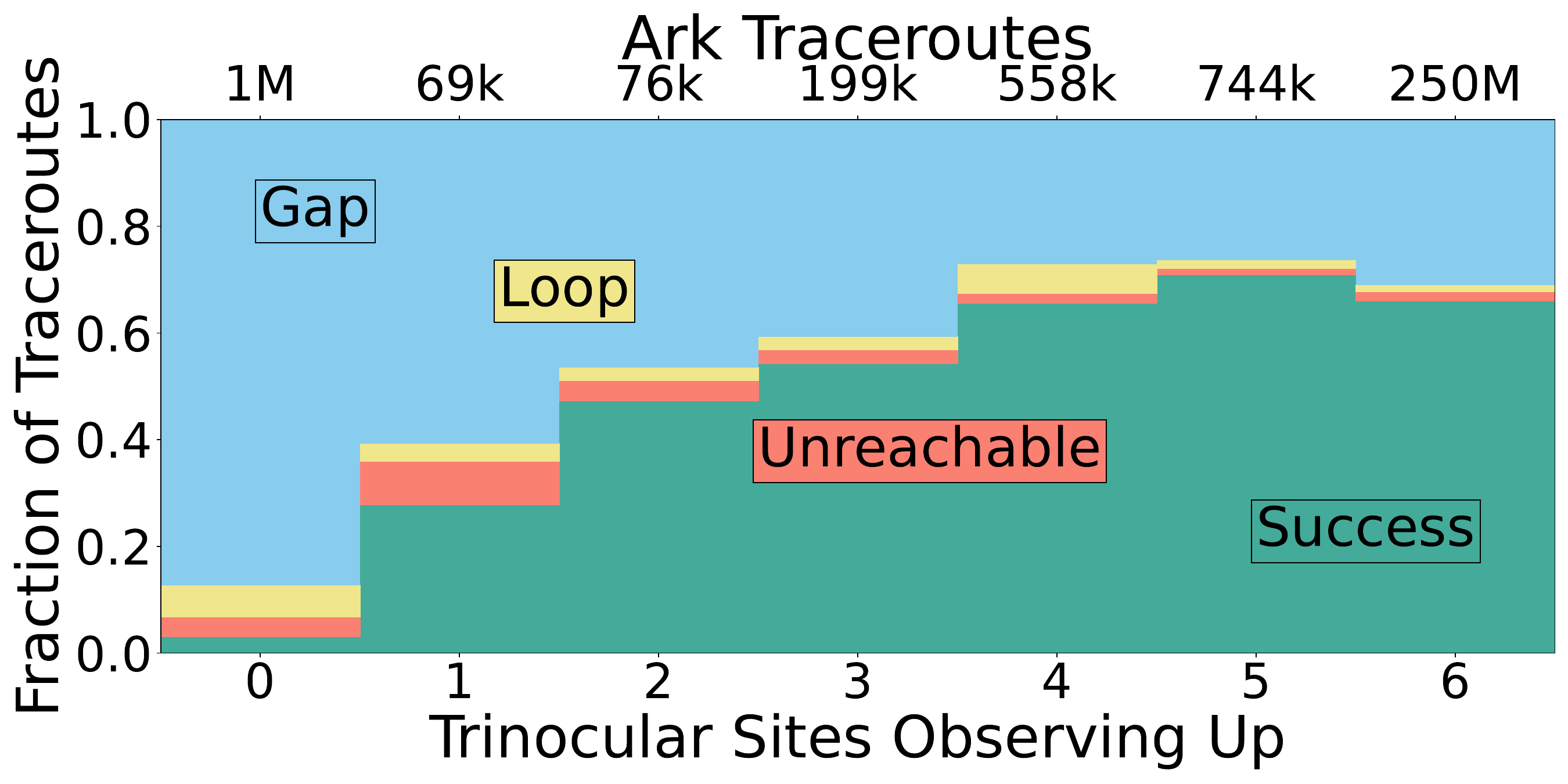}
        \caption{Ark traceroutes sent to targets under partial outages
        (2017-10-10 to -31). Dataset A30.}
  \label{fig:a30all_reach_fraction}
\end{figure}

We next re-evaluate reports from existing outage detection systems,
  considering how to resolve conflicting information in light of
  our new algorithms.
We compare findings to external information in traceroutes from CAIDA Ark.

\autoref{fig:a30all_reach_fraction} compares Trinocular with
  21 days of Ark topology data, from 2017-10-10 to -31 from all 3 probing teams.
For each Trinocular outage we classify the Ark result as success
  or three types of failure: unreachable, loop, or gap.

Trinocular's 6-site-up case suggests a working network,
  and we consider this case as typical.
However, we see that about 25\% of Ark traceroutes are ``gap'',
  where several hops fail to reply.
We also see about 2\% of traceroutes are unreachable
  (after we discard traceroutes to never reachable addresses).
Ark probes a random address in each block;
  many addresses are non-responsive,
  explaining these.

With 1 to 11 sites up, Trinocular is reporting disagreement.
We see that the number of Ark success cases (the green, lower portion of each bar)
  falls roughly linearly with the number of
  successful observers.
This consistency suggests that Trinocular and Ark are seeing similar behavior,
  and that there is partial reachability---these events
  with only partial Trinocular positive results
  are peninsulas.

Since 5 sites give the same results as all 6,
  single-\ac{VP} failures likely represent problems local to that \ac{VP}.
This data suggests that all-but-one voting will track true outages.

With only partial reachability, with 1 to 4 VPs (of 6),
  we see likely peninsulas.
These cases confirm that partial connectivity is common:
  while there are 1M traceroutes sent to outages where no \ac{VP} can see the target
  (the number of events is shown on the 0 bar),
  there are 1.6M traceroutes sent to partial outages
  (bars 1 to 5),
  and 850k traceroutes sent to definite peninsulas (bars 1 to 4).
This result is consistent with the convergence we see in
\autoref{fig:a30all_peninsulas_duration_oct_nov}.

\section{Related Work}
	\label{sec:related}

Prior definitions of the
    Internet exist at the IP-layer~\cite{Cerf74a,Postel80b,nitrd,huawei2020}
    of their time, or the AS-level~\cite{Gao01b,Luckie13a}.
We consider the IP-layer,
  and seek to address today's challenges (see \autoref{sec:problem}).

Cannon explored legal definitions of the Internet~\cite{Cannon02a},
  recognizing limitations of early definitions %
  and need to be application-independent. %
Like us, he considers connectivity and addressing important,
  but he questions if a firm legal definition is possible.
While we do not comment legalities,
  we suggest our technical definition
  may address his questions.

Several systems mitigate partial outages.
RON provides alternate-path routing
  around failures for a mesh of sites~\cite{andersen2001resilient}.
Hubble monitors in real-time reachability problems
  when working physical paths exist~\cite{katz2008studying}.
LIFEGUARD, remediates route failures
  by rerouting traffic using BGP to select a
  working path~\cite{katz2012lifeguard}.
While addressing the problem of partial outages,
  these systems do not quantify their duration or scope.

Prior work studied partial reachability, showing
  it is a common transient occurrence
  during routing convergence~\cite{bush2009internet}.
They reproduced partial connectivity with controlled experiments;
  we study it from Internet-wide \acp{VP}.

Internet scanners have examined bias by location~\cite{Heidemann08c},
  more recently looking for policy-based filtering~\cite{wan2020origin}.
  We measure policies with our country specific algorithm, and
  we extend those ideas to defining the Internet core.

Active outage detection systems have encountered partial outages.
Thunderping's ``hosed'' state recognizes mixed replies,
  but its study is future work~\cite{Schulman11a}.
Trinocular discards partial outages by
  reporting the target block ``up'' if any VP can reach
  it~\cite{quan2013trinocular}.
Disco identifies partial connectivity as future work~\cite{Shah17a}.
None of these systems
  consistently report partial outages in the Internet core,
  nor study their extent.

We use the idea of majority to define the Internet core in the face of secession.
That idea is fundamental in many algorithms for distributed consensus~\cite{Lamport82a,Lamport98a,Nakamoto09a},
  for example, with applications to certificate authorities~\cite{birge2018bamboozling}.

Recent work considered policies about Internet fragmentation~\cite{Drake16a,Drake22a}, but do not define it---a need we hope to meet.

\section{Conclusions}

Our new definition of the Internet core
  leads to new algorithms: Taitao, to find peninsulas of partial connectivity,
  and Chiloe, to find islands.
We validate these algorithms and show
  partial reachability is as common as simple outages.
They have important applications about Internet sovereignty
  and to improve outage and DNSmon measurement systems.

\label{page:last_body}

  \printbibliography

\appendix

\section{Discussion of Research Ethics}
	\label{sec:research_ethics}

Our work poses no ethical concerns as described in \autoref{sec:introduction}.
We elaborate here.

First, we collect no additional data, but instead
  reanalyze data from several existing sources
  (see \autoref{sec:data_sources_list}).
Our work therefore poses no additional load on the Internet,
  nor any new risk from data collection.

Our analysis poses no risk to individuals
  because our subject is network topology and connectivity.
There is a slight risk to individuals in that we
  examine responsiveness of individual IP addresses.
With external information, IP addresses can sometimes be traced to individuals,
  particularly when combined with external data sources like DHCP logs.
We avoid this risk in three ways.
First, we do not have DHCP logs for any networks
  (and in fact, most are unavailable outside of specific ISPs).
Second, we commit, as research policy, to not combine
  IP addresses with external data sources
  that might de-anonymize them to individuals.
Finally, except for analysis of specific cases as part of validation,
  all of our analysis is done in bulk over the whole dataset.

We do observe data about organizations such as ISPs,
  and about the geolocation of blocks of IP addresses.
Because we do not map IP addresses to individuals,
  this analysis poses no individual privacy risk.

Finally, we suggest that while our work poses minimal privacy risks
  to individuals,
  to also provides substantial benefit to the community and to individuals.
For reasons given in the introduction
  it is important to improve network reliability and understand
  now networks fail.
Our work contributes to that goal.

Our work was reviewed by the
  Institutional Review Board at our university
  and because it poses no risk to individual privacy,
  it was identified as non-human subjects research
  (USC IRB IIR00001648).

\section{Proof of Majority Enforcing One or No Internet}
    \label{sec:half_proof}

Our definition in \autoref{sec:definition} is complete,
  and Bitcoin provides an example of majority forcing consensus.
However, here we provide a proof and discuss
  scenarios that, at first glance, may appear challenging.

Our conceptual definition is ``the strongly-connected component
  of more than 50\% of active, public IP addresses that can initiate communication with each other'',
  is chosen to ensure there can be only one Internet in each address space (IPv4 and IPv6).
We next prove this definition yields one result,
  both with and without peninsulas.

The reasoning for this choice in \autoref{sec:definition} is straightforward:
  if a connected component has some fraction $A$, where $1 > A > 0.5$,
  than this component \emph{must} be larger than any other component $B$.
One can prove this by contradiction:
  (i) assume some $B'$ exists, such that $B' > A$.
  (ii) Since $A > 0.5$, then (i) implies $B' > 0.5$.
  (iii) We then must conclude that $A+B' > 1$,
    but by definition, we measure only the whole address space,
    so it is also required that $A+B' \le 1$.
Therefore $B' < A$ and A forces a single clear component.
Q.E.D.

\textbf{Resolving competing ``cores'':}
This definition handles cases with multiple overlapping
  but incompletely communicating groups.
If members of those groups can reach half the active addresses,
  they are part of the Internet even if some are on peninsulas relative to each other.
Consider a simplified version of \autoref{fig:term_concept}
  with only three with three pluralities
  of connectivity, $A$, $B$, and $C$, each representing one third of the addresses,
  where $A$ and $B$ are strongly and directly connected,
  and $A$ and $C$ are strongly and directly connected,
  but $B$ and $C$ cannot directly reach each other.
(Recall that strong connections in graph theory
  means bi-directional connectivity,
  but it does not require \emph{direct}
  and allows connections through multiple hops.)
In this example, $B$ and $C$ can reach each other,
  but only through $A$, so they are strongly connected
  but not directly connected.
Our Internet core requires strong connections,
  but if it required direct connections, it would become a clique
  (a fully connected graph), forbidding peninsulas.

In this example
  there are two, partially overlapping, large, components that are
  both strongly and directly connected:
  $A \cup B$ and $A \cup C$.
Here \emph{all} ($A \cup B \cup C$) are part of the Internet,
  because any address can directly reach more than half of the active addresses:
  address $b \in B$ can reach $A \cup B$,
  $c \in C$ can reach $B \cup C$,
  and $a \in A$ can reach anyone.
While all addresses are in one Internet,
  $B$ and $C$ are on peninsulas.
The example in \autoref{fig:term_concept} is similar to this thought experiment.
In practice, we know that peninsulas occur in less then 1\% of block-time
  (\autoref{sec:peninsula_frequency}),
  so typically $A \ge 0.98$, with other components $B, C < 0.01$,
  quite different from this theoretical case where $A=B=C= 0.33$,
  or an asymmetric case where $A=0.49$ and $B = C = 0.02$.
However, the definition applies whenever $A \cup B \cup C > 0.5$.

\textbf{Resolving disagreements with incomplete knowledge:}
In the above discussion we apply our conceptual definition
  assuming an  omniscience view of connectivity.
All parties must agree that $A$ directly reaches both $B$ and $C$,
  but $B$ and $C$ can reach each other only indirectly through $A$.
An omniscient observer must recognize they are all part of the same core,
  in spite of the peninsula.

In practice, no real-world system will have omniscient knowledge of
  connectivity.
However, this scenario works even with incomplete knowledge.
Imagine observers only in $B$ and $C$
  both might assert they are ``the'' core,
  since both can observe direct, strong connectivity to more than half
  of the active, public addresses.

When faced with seemingly conflicting claims of what the core is,
  all parties must share their observations with each other
  to make their case.
In this case, $B$ and $C$ will recognize they are both reporting
  $A$ as part of their core, and that $A$ overlaps---they must
  therefore recognize the reachable core is $A \cup B \cup C$,
  even though they cannot directly reach each other.

This seeming disagreement highlights the requirement that $B$ and $C$ recognize
  that the $A$ they each measure is the same $A$.
This requirement is met by our definition
  of what a public, global address space is---we assume some authority
  allocated addresses.
In today's Internet, this authority is IANA\@.
Note that IANA is not saying who is in our out of the Internet,
  but only who is responsible for a given fraction of the address space.

If all parties cannot agree on a shared address space,
  then our definition cannot be used.
For example, if one party asserts the entire 0/0 IPv4 address space is
  theirs to reallocate, then one cannot use address to resolve disputes.
Fortunately, address assignment has historically been coordinated
  to avoid overlaps.
(One exception is DISA's 4 /8 prefixes.
These were clearly allocated to DISA, but lack of global routing
  prompted multiple organizations to squat on them,
  using them as additional private address space.
Fortunately this variance is not a practical problem
for several reasons:
Since 2021 DISA has announced routes for these blocks on the public Internet.
Their actual allocation has never been disputed.
And even if they were disputed, this 4/256ths of the address space
  is not enough to change control of a majority.)

\section{Additional Results about Islands}

We define islands and give examples in \autoref{sec:island}.
Here we supplement those results with  examples of country-sided islands (\autoref{sec:country_sized_islands}).
We also show the raw data we use to justify our choice of 50\% unreachability
  to define islands in Trinocular (\autoref{sec:island_trinocular_threshold}).

\subsection{Visualizing Potential 2017q3 Islands}
	\label{sec:chinese_islands}

\begin{figure}
\centering
        \includegraphics[width=0.6\linewidth]{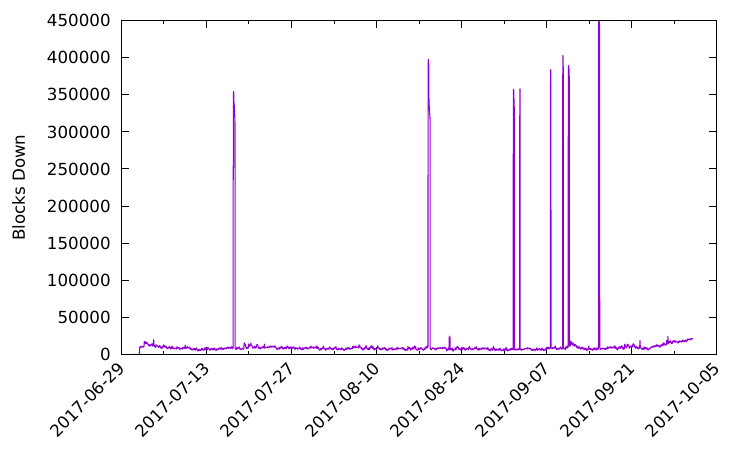}
	\caption{Unreachable blocks over time.  Large spikes are unreachability to Chinese-allocated IPv4 addresses. Dataset: A29, 2017q3.}
  \label{fig:a29all_outagedownup_4096}
\end{figure}

In \autoref{sec:country_sized_islands} we discuss evidence for country-sized
  islands.
In 2017q3, on 8 occasions it appears that most or all of China
  stopped responding to external pings.
\autoref{fig:a29all_outagedownup_4096} shows
  the number of /24 blocks that were down over time,
  each spike more than 200k /24s,
  between two to eight hours long.

\subsection{Longitudinal View Of Islands}
\label{sec:island_trinocular_threshold}

\begin{figure*}
        \begin{center}
          \includegraphics[width=0.95\linewidth]{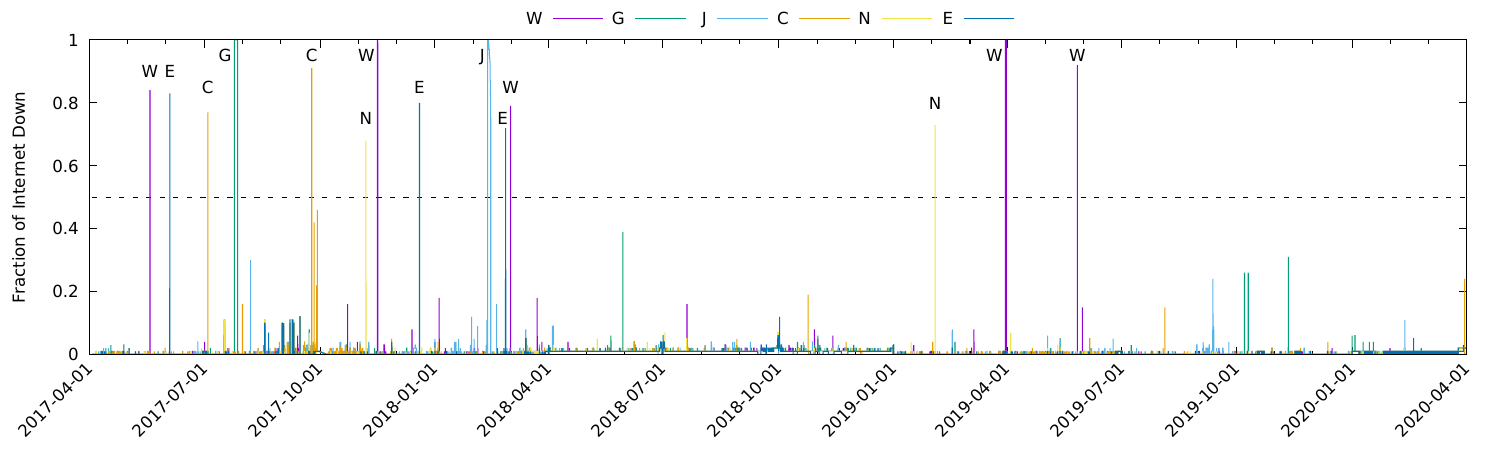}
        \end{center}
        \caption{Islands detected across 3 years using six \acp{VP}.
        Datasets A28-A39.}
  \label{fig:islands_plot_down_fraction}
\end{figure*}

We first consider three years of Trinocular data (described in
        \autoref{sec:data_sources_list}),
  from 2017-04-01 to 2020-04-01.
\autoref{fig:islands_plot_down_fraction}
  shows the fraction of the Internet that is
  reachable
  as a dotted line at the 50\% threshold that Chiloe uses to detect an island (\autoref{sec:chiloe}).
We run Chiloe across each VP for this period.

\section{Summary of Key Claims and Limitations}

\subsection{Data Sources and Key Claims}
	\label{sec:data_sources_list}
	\label{sec:key_claims_list}

\begin{table*}
  \resizebox{\textwidth}{!}{%
	\begin{tabular}{llllp{6.5cm}}
	\textbf{Dataset Name} & \textbf{Source} & \textbf{Start Date} & \textbf{Duration} & \textbf{Where Used} \\
	\hline
	\rowcolor[HTML]{DCDCDC}
	internet\_outage\_adaptive\_a28w-20170403 & Trinocular~\cite{trinoculardatasets} & 2017-04-03 & 90 days &  \\
	\rowcolor[HTML]{DCDCDC}
	\quad Polish peninsula subset & & 2017-06-03 & 12 hours & \autoref{sec:island}, \autoref{sec:polish_peninsula_validation} \\
	internet\_outage\_adaptive\_a28c-20170403 & Trinocular & 2017-04-03 & 90 days & \\
	\quad Polish peninsula subset & & 2017-06-03 & 12 hours & \autoref{sec:polish_peninsula_validation}\\
	\rowcolor[HTML]{DCDCDC}
	internet\_outage\_adaptive\_a28j-20170403 & Trinocular & 2017-04-03 & 90 days & \\
	\rowcolor[HTML]{DCDCDC}
	\quad Polish peninsula subset & & 2017-06-03 & 12 hours & \autoref{sec:polish_peninsula_validation}\\
	internet\_outage\_adaptive\_a28g-20170403 & Trinocular & 2017-04-03 & 90 days & \\
	\quad Polish peninsula subset & & 2017-06-03 & 12 hours & \autoref{sec:polish_peninsula_validation}\\
	\rowcolor[HTML]{DCDCDC}
	internet\_outage\_adaptive\_a28e-20170403 & Trinocular & 2017-04-03 & 90 days & \\
	\rowcolor[HTML]{DCDCDC}
	\quad Polish peninsula subset & & 2017-06-03 & 12 hours & \autoref{sec:island}, \autoref{sec:polish_peninsula_validation} \\
	internet\_outage\_adaptive\_a28n-20170403 & Trinocular & 2017-04-03 & 90 days & \\
	\quad Polish peninsula subset & & 2017-06-03 & 12 hours & \autoref{sec:island}, \autoref{sec:polish_peninsula_validation} \\
	\rowcolor[HTML]{DCDCDC}
	internet\_outage\_adaptive\_a28all-20170403 & Trinocular & 2017-04-03 & 89 days & \autoref{sec:chiloe_validation},
  											   \autoref{sec:how_common_are_islands},
    											   \autoref{sec:islands_duration},
  											   \autoref{sec:island_trinocular_threshold} \\
	internet\_outage\_adaptive\_a29all-20170702 & Trinocular & 2017-07-02 & 94 days  & \autoref{sec:island},
											   \autoref{sec:chiloe_validation},
  											   \autoref{sec:how_common_are_islands},
    											   \autoref{sec:islands_duration},
  											   \autoref{sec:island_trinocular_threshold} \\
	\rowcolor[HTML]{DCDCDC}
	internet\_outage\_adaptive\_a30w-20171006 & Trinocular & 2017-10-06 & 85 days & \\
	\rowcolor[HTML]{DCDCDC}
	\quad Site E Island & & 2017-10-23 & 36 hours & \autoref{sec:peninsula_definition}, \autoref{sec:polish_peninsula_validation} \\
	internet\_outage\_adaptive\_a30c-20171006 & Trinocular & 2017-10-06 & 85 days & \\
	\quad Site E Island & & 2017-10-23 & 36 hours & \autoref{sec:polish_peninsula_validation} \\
	\rowcolor[HTML]{DCDCDC}
	internet\_outage\_adaptive\_a30j-20171006 & Trinocular & 2017-10-06 & 85 days & \\
	\rowcolor[HTML]{DCDCDC}
	\quad Site E Island & & 2017-10-23 & 36 hours & \autoref{sec:polish_peninsula_validation} \\
	internet\_outage\_adaptive\_a30g-20171006 & Trinocular & 2017-10-06 & 85 days & \\
	\quad Site E Island & & 2017-10-23 & 36 hours & \autoref{sec:polish_peninsula_validation} \\
	\rowcolor[HTML]{DCDCDC}
	internet\_outage\_adaptive\_a30e-20171006 & Trinocular & 2017-10-06 & 85 days & \\
	\rowcolor[HTML]{DCDCDC}
	\quad Site E Island & & 2017-10-23 & 36 hours & \autoref{sec:peninsula_definition}, \autoref{sec:polish_peninsula_validation} \\
	internet\_outage\_adaptive\_a30n-20171006 & Trinocular & 2017-10-06 & 85 days & \\
	\quad Site E Island & & 2017-10-23 & 36 hours & \autoref{sec:peninsula_definition}, \autoref{sec:polish_peninsula_validation} \\
	\rowcolor[HTML]{DCDCDC}
	internet\_outage\_adaptive\_a30all-20171006 & Trinocular & 2017-10-06 & 85 days  & \autoref{sec:chiloe_validation},
  											   \autoref{sec:how_common_are_islands},
    											   \autoref{sec:islands_duration},
    											   \autoref{sec:site_correlation},
  											   \autoref{sec:island_trinocular_threshold} \\
	\rowcolor[HTML]{DCDCDC}
	\quad   Oct. Nov. subset & & 2017-10-06 & 40 days  &
	\autoref{sec:country_validation},
											   \autoref{sec:peninsula_duration},
											   \autoref{sec:peninsula_size}\\
	\rowcolor[HTML]{DCDCDC}
	\quad	Oct. subset & & 2017-10-10 & 21 days  & \autoref{sec:taitao_validation},
  											   \autoref{sec:representing_the_internet} \\
	internet\_outage\_adaptive\_a31all-20180101 & Trinocular & 2018-01-01 & 90 days  & \autoref{sec:chiloe_validation},
  											   \autoref{sec:how_common_are_islands},
    											   \autoref{sec:islands_duration},
  											   \autoref{sec:island_trinocular_threshold} \\
	\rowcolor[HTML]{DCDCDC}
	internet\_outage\_adaptive\_a32all-20180401 & Trinocular & 2018-04-01 & 90 days  & \autoref{sec:chiloe_validation},
  											   \autoref{sec:how_common_are_islands},
    											   \autoref{sec:islands_duration},
  											   \autoref{sec:island_trinocular_threshold}\\
	internet\_outage\_adaptive\_a33all-20180701 & Trinocular & 2018-07-01 & 90 days  & \autoref{sec:chiloe_validation},
  											   \autoref{sec:how_common_are_islands},
    											   \autoref{sec:islands_duration},
  											   \autoref{sec:island_trinocular_threshold}\\
	\rowcolor[HTML]{DCDCDC}
	internet\_outage\_adaptive\_a34all-20181001 & Trinocular & 2018-10-01 & 90 days  & \autoref{sec:chiloe_validation},
  											   \autoref{sec:how_common_are_islands},
    											   \autoref{sec:islands_duration},
											   \autoref{sec:additional_confirmation},
  											   \autoref{sec:island_trinocular_threshold}\\
	internet\_outage\_adaptive\_a35all-20190101 & Trinocular & 2019-01-01 & 90 days  & \autoref{sec:chiloe_validation},
  											   \autoref{sec:how_common_are_islands},
    											   \autoref{sec:islands_duration},
  											   \autoref{sec:island_trinocular_threshold}\\
	\rowcolor[HTML]{DCDCDC}
	internet\_outage\_adaptive\_a36all-20190401 & Trinocular & 2019-01-01 & 90 days  & \autoref{sec:chiloe_validation},
  											   \autoref{sec:how_common_are_islands},
    											   \autoref{sec:islands_duration},
  											   \autoref{sec:island_trinocular_threshold}\\
	internet\_outage\_adaptive\_a37all-20190701 & Trinocular & 2019-01-01 & 90 days  & \autoref{sec:chiloe_validation},
  											   \autoref{sec:how_common_are_islands},
    											   \autoref{sec:islands_duration},
  											   \autoref{sec:island_trinocular_threshold}\\
	\rowcolor[HTML]{DCDCDC}
	internet\_outage\_adaptive\_a38all-20191001 & Trinocular & 2019-01-01 & 90 days  & \autoref{sec:chiloe_validation},
  											   \autoref{sec:how_common_are_islands},
    											   \autoref{sec:islands_duration},
  											   \autoref{sec:island_trinocular_threshold}\\
	internet\_outage\_adaptive\_a39all-20200101 & Trinocular & 2020-01-01 & 90 days  & \autoref{sec:chiloe_validation},
  											   \autoref{sec:how_common_are_islands},
    											   \autoref{sec:islands_duration},
  											   \autoref{sec:island_trinocular_threshold}\\
	\rowcolor[HTML]{DCDCDC}
	internet\_outage\_adaptive\_a41all-20200701 & Trinocular & 2020-07-01 & 90 days  & \autoref{sec:peninsula_locations} \\
	\hline
	prefix-probing & Ark~\cite{CAIDA07b} \\
	\quad Oct. 2017 subset & & 2017-10-10 & 21 days & \autoref{sec:taitao_validation}, \autoref{sec:representing_the_internet} \\
	\quad 2020q3 subset & & 2020-07-01 & 90 days & \autoref{sec:peninsula_locations}\\
	\rowcolor[HTML]{DCDCDC}
	probe-data     & Ark & & & \\
	\rowcolor[HTML]{DCDCDC}
	\quad Oct 2017 subset     & & 2017-10-10 & 21 days & \autoref{sec:taitao_validation}, \autoref{sec:representing_the_internet} \\
	\rowcolor[HTML]{DCDCDC}
	\quad 2020q3 subset     & & 2020-07-01 & 90 days & \autoref{sec:peninsula_locations}\\
	\hline
	routeviews.org/bgpdata & Routeviews~\cite{routeviews} & 2017-10-06 & 40 days
    &
      \autoref{sec:country_validation},
  								      \autoref{sec:polish_peninsula_validation} \\
	\hline
	\rowcolor[HTML]{DCDCDC}
	Atlas Recurring Root Pings (id: 1001 to 1016) & Atlas~\cite{ripe_ping} & 2021-07-01 & 90 days & \autoref{sec:peninsula_frequency},
													\autoref{sec:islands_duration} \\
	\hline
	nro-extended-stats & NRO~\cite{iana_v4,iana_v6} & 1984 & 41 years & \autoref{sec:internet_partition} \\

	\end{tabular}
  }
   \caption{All datasets used in this paper.}
   \label{tab:datasets}
\end{table*}

\autoref{tab:datasets} provides a full list of datasets used in this paper
  and where they may be obtained.

\begin{table*}
\resizebox{\textwidth}{!}{%
\begin{tabular}{lll}
 & \textbf{claim} & \textbf{support} \\
\multirow{3}{*}{\rotatebox[origin=c]{90}{\footnotesize examples}}
 & IPv4 islands exist (\autoref{sec:island}) & example Trinocular/2017q2 \\
 & IPv6 peninsulas exist (\autoref{sec:peninsula_definition}) & public news~\cite{ipv6peeringdisputes,google_cogent,cloudflare_he}, DNSmon~\cite{dnsmon}, looking glass~\cite{he_looking_glass,cogent_looking_glass} \\
 & IPv4 peninsulas exist (\autoref{sec:peninsula_definition}) & example Trinocular/2017q4, Ark~\cite{CAIDA07b}, traceroutes and Routeviews (\autoref{sec:polish_peninsula_validation}) \\
  \hline
\multirow{3}{*}{\rotatebox[origin=c]{90}{\footnotesize validation}}
 & Taitao correctness (\autoref{sec:taitao_validation}) & Trinocular/2017q4, validated with Ark \\
 & Chiloe correctness (\autoref{sec:chiloe_validation}) & Trinocular/2017q1 to 2020q1 \\
 & 6 Trinocular sites are independent (\autoref{sec:site_correlation}) & Trinocular/2017q4 \\
  \hline
\multirow{5}{*}{\rotatebox[origin=c]{90}{\small observations}}
 & Peninsulas are common (\autoref{sec:peninsula_frequency}) & Trinocular/2017q1, 2018q4, 2020q3 \\
 & Some peninsulas are long-lived (\autoref{sec:peninsula_duration}) & Trinocular/2017q3 to 2020q1 and RIPE Atlas (2021q3) \\
 & Islands are common (\autoref{sec:how_common_are_islands}) &  Trinocular/2017q3 to 2020q1 and RIPE Atlas (2021q3) \\
 & Most island are hours or less (\autoref{sec:islands_duration}) &  Trinocular/2017q3 to 2020q1 and RIPE Atlas (2021q3) \\
 & Most islands are small (\autoref{sec:islands_sizes}) & RIPE Atlas/2021q3 \\
   \hline
\multirow{4}{*}{\rotatebox[origin=c]{90}{\small applications}}
 & IPv4 and IPv6 cannot partition (\autoref{sec:internet_partition}) & ICANN IPv4 allocations, Routeviews \\
 & Policy applications (\autoref{sec:other_applications}) & ICANN IPv4 allocations \\
 & DNSmon sensitivity can improve (\autoref{sec:dnsmon}) & DNSmon/2022 (uses RIPE Atlas) \\
 & peninsulas can clarify outages (\autoref{sec:local_outage_eval}) & Trinocular/2017q4 with Ark (21 days) \\
\end{tabular}
}
\caption{Key observations made in the paper and which datasets support each.}
  \label{tab:claims}
\end{table*}

\autoref{tab:claims} summarizes the key
  observations this paper makes about the Internet,
  and what datasets support each.
The paper body provides all of our key claims
  and validates them with multiple data sources
  and broad, Internet-wide data as observed
  from 6 (Trinocular), about 150 (Ark),  and about 13k (RIPE Atlas)
  locations.
We emphasize that \emph{all} our key results
  use data from multiple data sources.
Some graphs emphasize data source from Trinocular,
  since it provides very broad coverage,
  all key validation and observational results are supported
  with data from either Ark or RIPE Atlas.
All of our trends are verified with
  Trinocular data from 6 sites scanning millions of networks,
  and confirmed by data from many sites scanning less frequently
    (Ark, with 150 sites scanning millions of networks daily)
  or less completely (RIPE Atlas, with 13k sites scanning 13 destinations).

Our core quantitative results use Trinocular data,
  because it is the only currently available data source
  that provides very broad coverage (5M IPv4 blocks)
  with sufficient frequency (updates every 11 minutes)
  to approach an Internet wide view.
Each of these conclusions are validated with Ark,
  as described in \autoref{sec:validation}.
These results strongly confirm that
  Taitao and Chiloe true positives and true negatives are correct,
  and suggest there are relatively few true positives.
With only 6 \acp{VP}, we expect that
  Trinocular's observations
  underrepresent peninsulas,
  and that Taitao thus has an unknown number of false negatives.
Finally, in many cases, we validate observations with multiple quarters
  of Trinocular data to show that the results are consistent over time.

DNSmon provides a second system where we apply our algorithms,
  and the overall DNSmon analysis provides trends consistent with
  what we see from Trinocular (\autoref{sec:dnsmon}).
DNSmon provides observations from about 13k VPs,
  although they contact 13 sites (the DNS root servers)
  and therefore require a relaxed version of the algorithms.
These results show peninsulas and islands have profound effects
  on DNSmon reports about root-server-system reliability,
  confirming the Trinocular Internet-wide data.

\subsection{Limitations}
	\label{sec:limitations_summary}

Our work is a first step in understanding partial reachability.
We recognize the its limitations,
  as in any system spanning theory and practical implementation.
All of these limitations were described in context in the body of the paper,
  but we summarize them here so, as with our key results, there is a central list.

Our \emph{conceptual definition can never be completely realized}.
Like most conceptual definitions,
  we imagine reachability from between all Internet addresses.
It is clearly impossible to actually carry out that $O(n^2)$ measurement.
However, we suggest that this conceptual definition can help us reason
  about partial reachability,
  particularly in providing a politically neutral discussion about ``the Internet'',
  and that it can help clarify
  measurement error.
We discussion this limitation in \autoref{sec:definition_motivations}
  and provide two realizations with Trinocular and RIPE Atlas (\autoref{sec:deployment_status}).

\emph{Any realization of reachability will be protocol specific.}
We explore two protocols for measurement, ICMP echo request with Trinocular,
  and DNS, with RIPE Atlas.
Other options are possible (see \autoref{sec:definition_motivations}),
  and while we recognize the limitations of any specific choice,
  we believe the flexibility of the conceptual definition leaves this option open.

As a data source, \emph{Trinocular is measured from a few \acp{VP}}.
With only 6 locations, Trinocular provides only a few perspectives on the Internet.
We use it because it provides excellent coverage of destinations
  (about 5 million destination networks),
  and provides years of 24x7 historical data we can examine.
Having few locations limits this source's ability to be on two sides of peninsula
  (as required by Taitao)
  or to be inside an island (as required by Chiloe).
We use Trinocular because its long duration and wide destination coverage
  is important to detect peninsulas, which, although more frequent than outages,
  are still relatively rare.
In \autoref{sec:site_correlation}
  we provide experimental data that shows \emph{Trinocular's 6 sites are independent}
  and that as \emph{few as 3 sites converge on a stable estimate} of
  the fraction of peninsula,
  strongly suggesting 6 independent observers can provide consistent results.
Finally, Trinocular makes available public outage results,
  and comparison of the fraction of outages to peninsulas
  from the same data source suggests that \emph{peninsulas are more widespread than outages},
  and the Taitao's ``two side'' limitations suggests that more observers
  would strengthen this claim.

As a data source, \emph{Trinocular does not provide IPv6}.
To our knowledge there is no public source for IPv6 outage data today.
However, we provide IPv6 results using RIPE Atlas as a data source.

As a data source, \emph{RIPE Atlas measures few destinations}.
Our analysis of RIPE Atlas uses traffic to the DNS roots,
  13 IPv4 and 13 IPv6 destinations that are anycast.
We use RIPE Atlas because it provide a large number of \acp{VP}
  (about 13k, with global distribution),
  because it provides years of 24x7 historical data we can examine,
  and because to provides IPv6 data suitable for analysis.
Since RIPE Atlas has few destinations,
  we chose to be stricter in applying Chiloe
  to insure any error results in an undercount of islands.

As a data source, \emph{RIPE Atlas destinations are anycast}.
Since the destinations of Atlas DNS queries are anycast,
  we expect each \ac{VP} to measure only its 26 nearest anycast destinations,
  not all of the nearly 2000 actual root server sites that exist.
Having anycast targets means that RIPE is best at measuring the network
  relatively near each \ac{VP} to the public Internet.
However, since many Atlas \acp{VP} are in eyeball networks
  and most root DNS sites are in commercial hosting facilities,
  DNS queries from Atlas typically traverse several routers
  and more than one AS\@.

As data sources, \emph{neither Trinocular nor RIPE Atlas
  provides both many \acp{VP} and destinations}.
It is important to recognize that \emph{no} tractable data source
  can have both many \acp{VP} and destinations because of the quadratic
  cost in traffic.
This limitation motivates our use of both data sources to evaluate
  nearly all research questions (\autoref{sec:key_claims_list})),
  and an data sources (CAIDA Archipelago and Routeviews) for validation.

\emph{Is data from 2017 too old?}
Our results come from many years, with many results comparing 2017 and 2020,
  and RIPE Atlas results covering 3 years (\autoref{tab:claims}).
We believe our results from 2017 demonstrate our claims,
  and the additional data shows they are relevant today.
We discuss specific technical reasons we use data from specific years
  in \autoref{sec:2020}.

\emph{Comparisons of Taitao with CAIDA Ark are difficult.}
We summarize how we validate Taitao with Ark data
  in \autoref{sec:taitao_validation},
  and expand on that discussion in \autoref{sec:false_negative_details}.
We agree this comparison is difficult because neither data source
  was designed to compare to the other,
  particularly with Ark's relatively sparse probing and random destination.
To account for this challenge, we use the comparison to establish lower and upper
  bounds, rather than a single estimate.
We consider this challenge and limitation the reality
  when evaluating a new method that detects relatively rare events.
We suggest that these bounds establish the plausibility of our approach
  and motivate potential future work to identify better methods to validate,
  and to tighten these claims,

\section{Additional Details About Examples}

\subsection{Additional Details About Example Islands}
	\label{sec:additional_details_island_example}
\begin{figure}
\begin{center}
    \mbox{\includegraphics[width=0.7\textwidth,trim=20 90 42 35,clip]{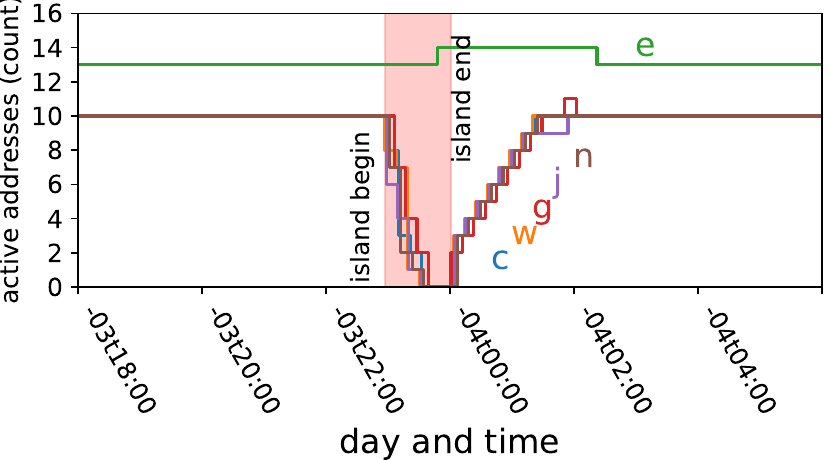}}
\end{center}
    \caption{Estimates of an island in Trinocular data
      starting 2017-06-03t23:06Z and lasting just longer than one hour.}
        \label{fig:a28all_417bca00_accum}
\end{figure}

\textbf{Observation with IP Address Accumulation:}
In \autoref{fig:a28all_417bca00_accum} shows an example island we observed.
In this graph, each line shows the count of ping-responsive
  addresses observed from a different \ac{VP}
  over 12~hours.
In this graph, VP E always sees 13 or 14 active addresses,
  while the other 5 VPs (C, W, G, J, and N)
  all start and end seeing 10 addresses,
  but their estimates of active addresses drop during the island.

The island is shown as a red shaded bar, starting at
  2017-06-03t23:06Z.
During the island \ac{VP} E is disconnected from the Internet,
  but it continues to see 13 or 14 active addresses.
By contrast, the other 5 \acp{VP}
  estimates of active addresses gradually drop when the island begins,
  and gradually recover then the island ends.
(These estimates drop and recover gradually
  because Trinocular only probes a few addresses every 11 minutes.)
We confirmed this island was a network outage that disconnected
  the office hosting \ac{VP} E from the Internet
  from network operators at that office.

\textbf{Raw Trinocular Observations:}
During the time of the event, E was able to successfully probe addresses within
the same block, however, unable to reach external addresses.
This event started at 2017-06-03t23:06Z, and
can be observed in ~\autoref{fig:islands_plot_down_fraction}.

\begin{figure}
  \includegraphics[width=1\columnwidth]{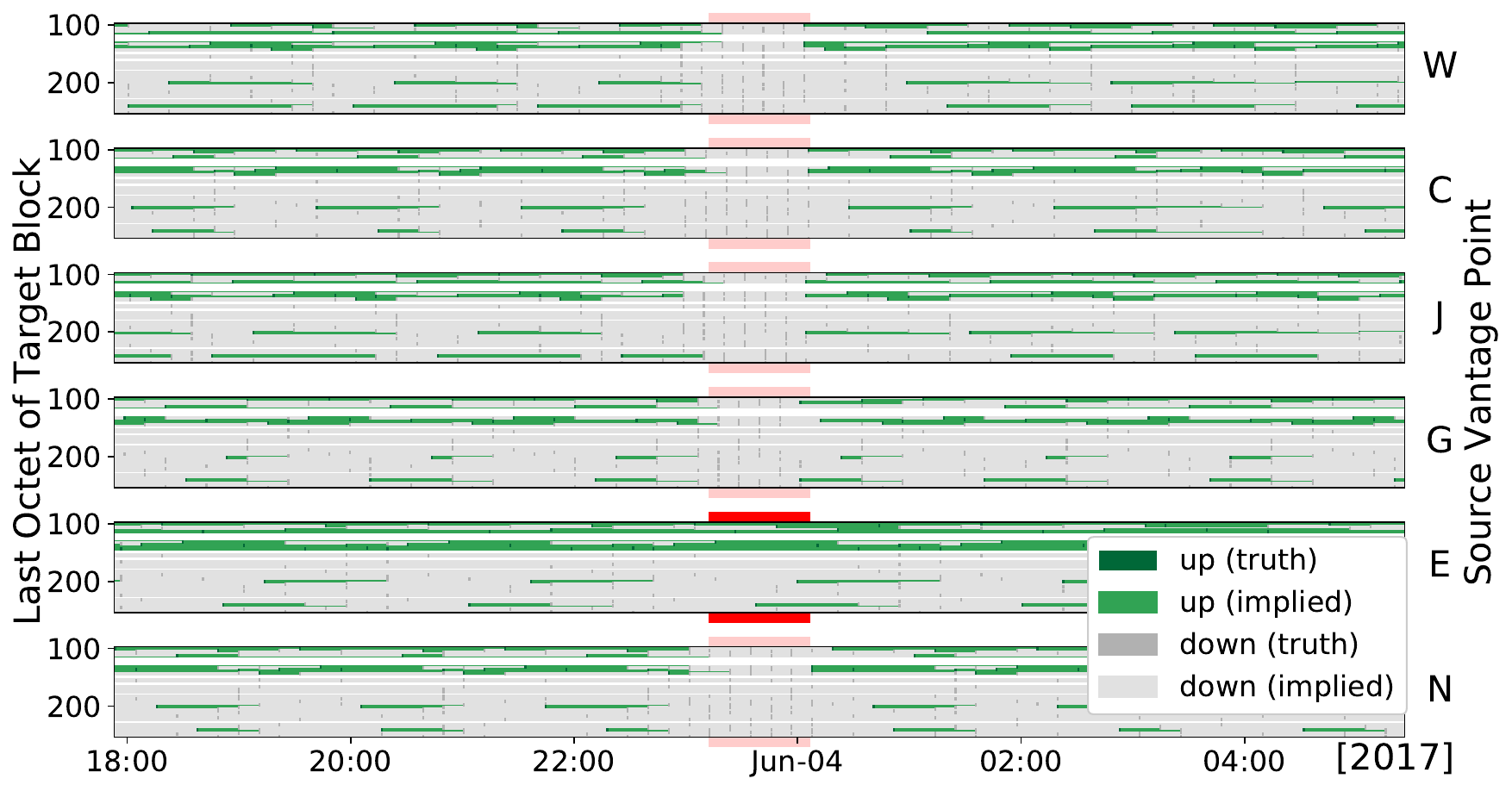}
  \caption{A block
    showing a 1-hour island for this block and \ac{VP} E, while other five
    VPs cannot reach it.}
  \label{fig:a28all_raw_417bca00_6sites}
\end{figure}

Furthermore, no other \ac{VP} was able to reach the affected block for the
time of the island as shown in \autoref{fig:a28all_raw_417bca00_6sites}.

\subsection{Additional Details About Example Peninsulas}
	\label{sec:additional_details_peninsula_example}
	\label{sec:polish_peninsula_validation}

In \autoref{sec:peninsula_definition} we defined peninsulas and discussed
  the 2017-10-23 Multimedia Polksa peninsula.
On 2017-10-23, for a period of 3 hours starting at 22:02Z,
  five Polish \acp{AS}
  had 1716 blocks that were unreachable from five \acp{VP}
  while the same blocks remained reachable from a sixth \ac{VP}.

\begin{figure}
\begin{center}
    \mbox{\includegraphics[width=.7\textwidth,trim=5 5 45 0,clip]{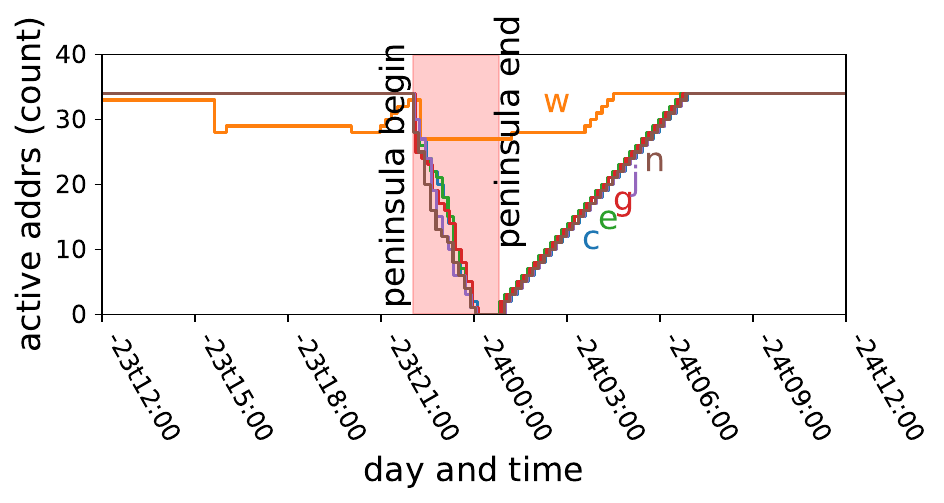}}
\end{center}
 \caption{Estimates of reachable addresses during a peninsula found in Trinocular data
  starting 2017-10-23t22:02Z for 3\,h.}
    \label{fig:a30all_50f5b000_accum}
\end{figure}

\textbf{Observation with IP Address Accumulation:}
To show how our algorithms detect this event,
  \autoref{fig:a30all_50f5b000_accum} shows how many addresses are visible
  from each observer.
Observer w, the top line, always sees about 30 addresses,
  but all other observers overlap and lose track of addresses
  as the peninsula begins and re-detects them when it heals.

\begin{figure}
  \includegraphics[width=0.9\columnwidth]{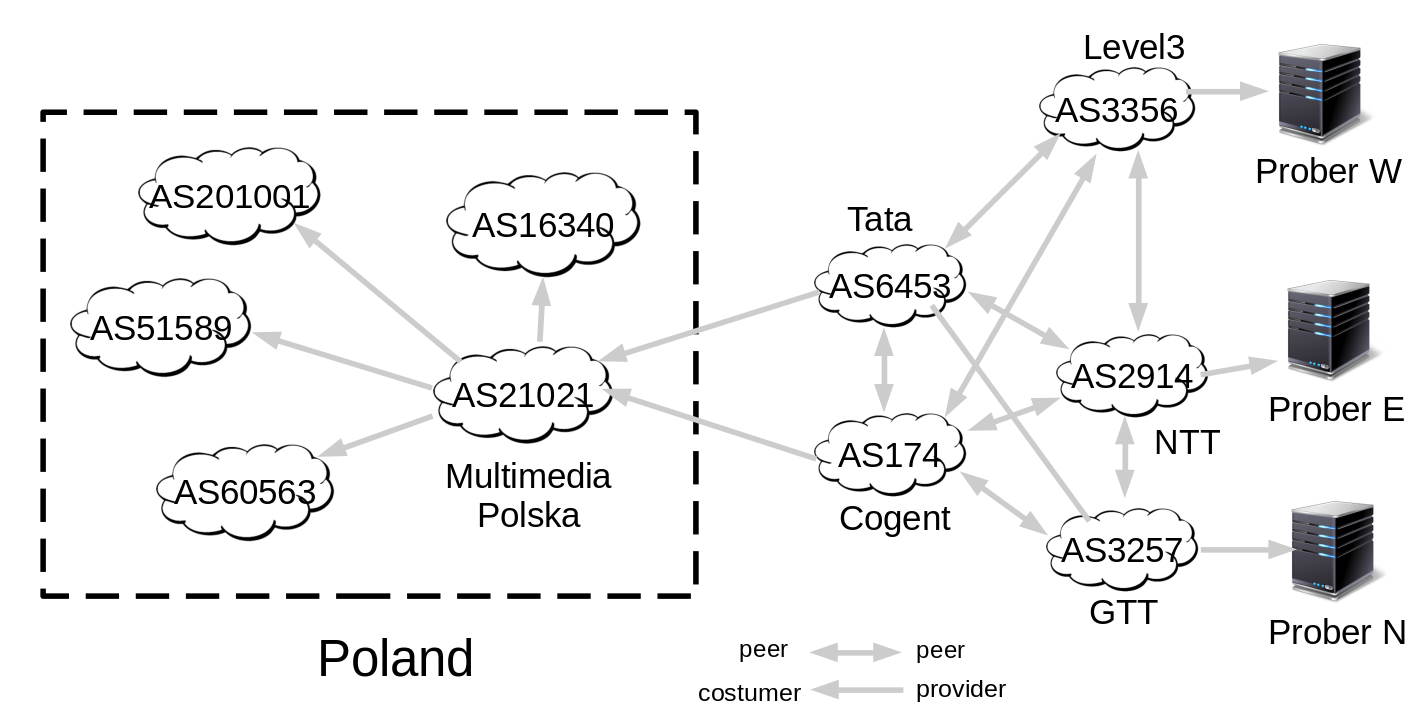}
  \caption{AS level topology during the Polish peninsula.}
  \label{fig:polish_peninsula}
\end{figure}

\begin{figure}
  \includegraphics[width=1\columnwidth]{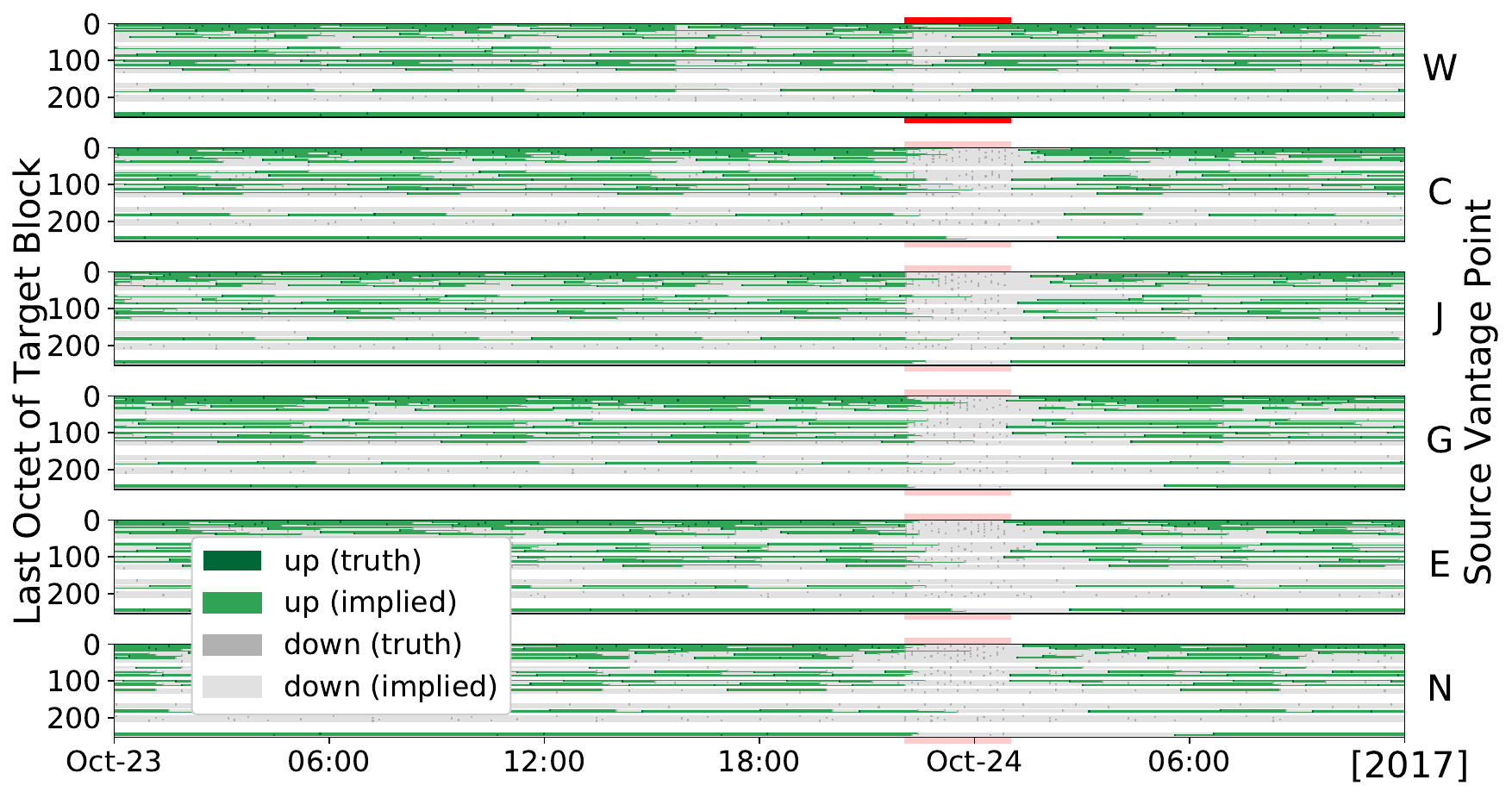}
  \caption{A block
            (80.245.176.0/24)
      showing a 3-hour peninsula accessible only from \ac{VP} W (top bar)
      and not from the other five \acp{VP}.  Dataset: A30.}
  \label{fig:a30all_raw_50f5b000_6sites}
\end{figure}

\textbf{Raw Trinocular Observations:}
In \autoref{fig:a30all_raw_50f5b000_6sites} we provide data from 6 Trinocular \acp{VP},
where W is uniquely capable of reaching the target block, thus living in the
same peninsula.
The address accumulation data represents counts of the green lines in this graph.

\textbf{Observations in BGP:}
\autoref{fig:polish_peninsula} shows the AS-level relationships
  at the time of the peninsula.
Multimedia Polska (AS21021, or \emph{MP}) provides service to the other 4 ISPs.
MP has two Tier-1 providers:
  Cogent (AS174) and Tata (AS6453).
Before the peninsula, our VPs see MP
  through Cogent.

\begin{figure}
  \includegraphics[width=0.70\columnwidth]{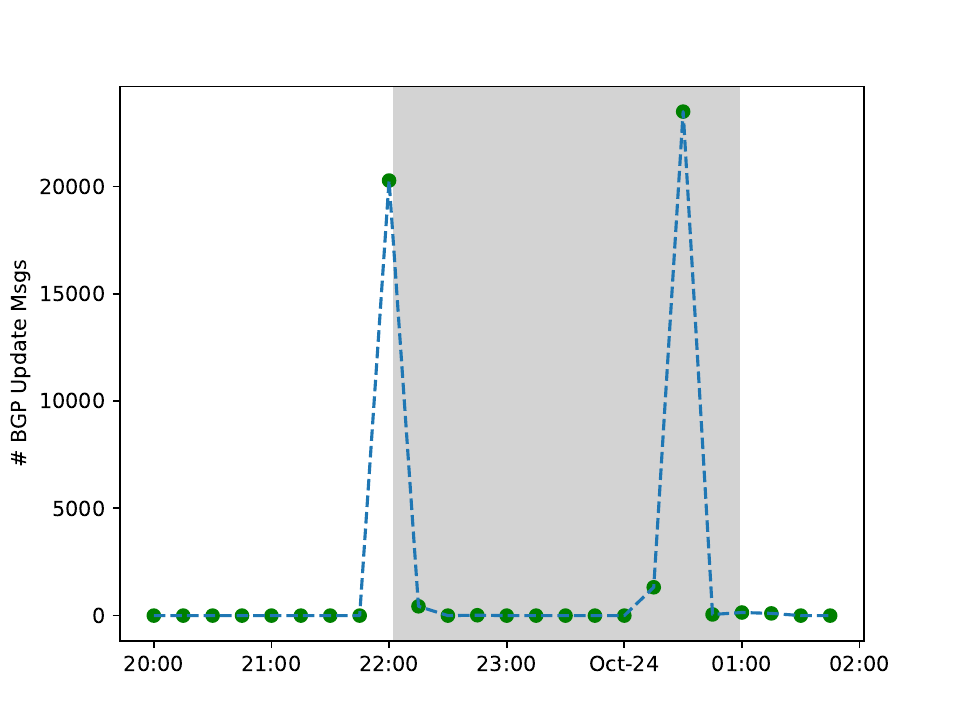}
	\caption{ BGP update messages sent for affected Polish blocks starting
    2017-10-23t20:00Z. Data source: RouteViews. }
  \label{fig:rv_time_plot}
\end{figure}

At event start, we observe many BGP updates (20,275) announcing
and withdrawing routes to the affected blocks(see~\autoref{fig:rv_time_plot}).
These updates correspond to Tata announcing MP's prefixes.
Perhaps MP changed its peering
  to prefer Tata over Cogent,
  or the MP-Cogent link failed.

Initially, traffic from most VPs continued through Cogent
  and was lost; it did not shift to Tata.
One \ac{VP} (W) could reach MP
  through  Tata for the entire event,
  proving MP was connected.
After 3 hours, we see another burst of BGP updates (23,487 this time),
  making MP reachable again from all VPs.

\begin{table*}
\resizebox{\textwidth}{!}{%
    \begin{tabular}{c  c  c  >{\raggedright\arraybackslash}m{120mm}}
            \toprule
src block  &    dst block   &   time        &   traces \\
\hline
c85eb700   &    50f5b000    &   1508630032  & q, 148.245.170.161, 189.209.17.197, 189.209.17.197, 38.104.245.9, 154.24.19.41, 154.54.47.33, 154.54.28.69, 154.54.7.157, 154.54.40.105, 154.54.40.61, 154.54.43.17, 154.54.44.161, 154.54.77.245, 154.54.38.206, 154.54.60.254, 154.54.59.38, 149.6.71.162, 89.228.6.33, 89.228.2.32, 176.221.98.194
\\
\hline
c85eb700   &     50f5b000   &     1508802877 &
q, 148.245.170.161, 200.38.245.45, 148.240.221.29 \\
\bottomrule
    \end{tabular}
}
\caption{Traces from the same Ark VPs (mty-mx) to the same destination
            before and during the event block}
    \label{tab:traceroutes}
\end{table*}

\textbf{Observations in CAIDA Ark:}
We further verify this event by looking at traceroutes.
During the event we see 94 unique Ark VPs attempted 345 traceroutes to the affected blocks.
Of the 94 VPs, 21 VPs (22\%) have their last responsive
  traceroute hop in the same \ac{AS} as the
  target address, and 68 probes (73\%) stopped before reaching that \ac{AS}.
\autoref{tab:traceroutes} shows traceroute data from a single CAIDA Ark VP
  before and during the peninsula described in \autoref{sec:peninsula_definition} and
  \autoref{fig:a30all_50f5b000_accum}.
This data confirms the block was reachable from some locations and not others.
During the event, this trace breaks at the last hop within the source \ac{AS}.

\section{Additional Validation}

\subsection{Additional Analysis of Potential Missed Peninsulas}
	\label{sec:false_negative_details}

In \autoref{sec:taitao_validation}
  we suggest that it is difficult to use Ark
  to predict a peninsula because its methodology
  causes it to usually probe a non-responsive address.
This result follows from Ark's design as a topology discovery system,
  not partition detection---as a result,
  the majority of Ark probes fail to reach a target,
  even for a block that is reachable.
(We do not mean this statement as a negative comment about Ark.
It is well suited for topology discovery and building router-level
  network topologies.
This problem occurs when we do our best to reuse Ark data do
  validate network partitions.)

\textbf{Comparing Ark and Trinocular data:}
We reuse Ark to validate network partitions for two reasons:
  first, each traceroute targets one address (prefix probing, the .1, and team probing a random address),
  not multiple addresses.
Second, Ark visits each block on average every 36 minutes,
  compared to Trinocular's observation every 2 minutes (6 observers every 11 minutes).

The biggest challenge is that Ark probes either only one destination address.
By contrast, Trinocular probes up to 15 addresses, stopping on success,
  and those addresses are drawn from addresses that have previously responded.
Predicted addresses respond 49\% of the time,
  while a random address responds in less than 1\% of blocks (\cite{Fan10a}, Figure 7).
Probing 15 addresses is successful more than 90\% of time
  for 5/6ths of responsive blocks (\cite{quan2013trinocular}, Figure 6),
  while probing .1 is responsive only in 1/6th of responsive blocks,
  and other addresses respond even less frequently.
As a result, we expect 4/6ths of Ark targets to be non-responsive,
  even when the block is reachable.
While this choice does not limit Ark for building router-level topologies,
  prediction of peninsulas from multiple Ark attempts
  with mixed results, \emph{almost always reflects Ark targeting
  a non-responsive address, not an actual peninsula}.

Ark's less frequent probing is a second factor.
In aggregate, Ark probes a target block every 36 minutes (40 teams,
  each trying each block once per day),
  while Trinocular probes 18x more frequently (6 observers,
  each probing every block every 11 minutes).
As a result, combinations of Ark observations are often from different times,
  blurring outages and peninsulas.

A ``positive'' is Taitao detecting a peninsula that Ark confirms,
  and a negative is Taitao not detecting a peninsula
  (showing all positive or all negative)
  that Ark has mixed results about.
However, since 90\% of Ark results are negative because
  it chooses a target address that doesn't respond,
  \emph{most} Ark results are mixed and are false indications of peninsulas.
For this reason, we consider Ark reports about false negatives untrustworthy,
  and we discount Ark's evaluation of metrics that depend on the false negative rate
    (such as recall, but not precision).
In other words, Ark can confirm what Taitao finds,
  but it is insufficient to evaluate what Taitao misses.

\textbf{Implications in comparing Ark and Taitao:}
We observe that it is quite challenging to compare Taitao and Ark.
We would like to take Ark as ``ground truth''
  from more VPs (150, not just 6),
  but it is designed for topology discovery, not peninsula detection,
  makes it an imperfect tool for this application.

However, we suggest it does provide useful validation
  for true positives, true negatives, and false positives,
  as we describe next.

\textbf{Implications of potential false negatives:}
While we cannot trust Ark's judgment of false negatives,
  it is likely that \emph{some} are correct---Ark likely sees
  some partitions that Taitao misses when using only Trinocular as a data source.
We expect that there are many very small peninsulas (micro-peninsulas),
  and that adding
  more VPs will increase the number of these micro-peninsulas.
As a thought experiment, \emph{every} computer that
  can route to a LAN using public IP addresses,
  but lacks a global route, is a peninsula.

We align our claims with this potential:
  we claim there are many peninsulas, at least more than there are outages,
  as supported by \autoref{fig:a30all_peninsulas_duration_oct_nov}.
Although we think there are not many large peninsulas,
  we recognize there are many,
  and some are long-lasting (\autoref{fig:a30_partial_outages_duration_cdf}).
If some or many of the false negatives suggested by Ark
  are actual peninsulas,
  implies that understanding peninsulas is even more important
  than we predict.

\textbf{Implications of true positives, true negatives, and false positives:}
While care must be taken when using Ark as ``ground truth''
  to judge false negatives (peninsulas missed by Taitao),
  it is much more promising to test other conditions.

True positives are
  peninsulas identified by Taitao because it sees conflicting VPs.
For these cases,
  presence of conflicting Ark data is consistent with Taitao.
Even if Ark has excessive last-hop failures,
  its large number of observers suggest that some may be correctly unreachable,
  and the presence of reachability from some Ark VPs confirms
  partial reachability.

True negatives last a long time, so in these cases
  all Taitao VPs reach the target,
  and the long duration of ``all Taitao VPs reach''
  makes it very likely some Ark VP reaches the target.

Finally, the very small number when Taitao reports all down
but Ark shows reachability (6 cases) confirms that in a few cases,
  6 sites are not enough to see all peninsulas.

\subsection{Stability of Results over Time}
\label{sec:2020}

Our paper body uses Trinocular measurements for 2017q4 because
  this time period had six active VPs,
  allowing us to make strong statements about how multiple perspectives help.
Those three months of data provide evidence of result stability,
  but those observations are now several years old.
Because IP address allocation and partial reachability
  are associated with organizational policy (of ICANN and the \acp{RIR})
  and business practices of thousands of ISPs,
  we expect them to change relatively slowly.
Here we verify this assumption,
  showing our results from 2017 hold in 2018 and 2020.
They do---we find quantitatively similar results between 2017
  for number and sizes of peninsulas in 2018q4 in \autoref{sec:additional_confirmation},
  and duration in 2020q3 in \autoref{sec:additional_peninsula_duration},
  confirming these results in \autoref{sec:evaluation} hold.

\section{Additional Results about Peninsulas}
	\label{sec:additional_peninsula_appendix}

Multiple well-known examples of peering disputes
  document one root-cause of peninsulas.
The paper body lists examples in \autoref{sec:introduction}
  and \autoref{sec:peninsula_definition},
  and shows evidence found with Taitao in DNSmon in \autoref{sec:dnsmon}.

Here we supplement those results with additional details
  to evaluate peninsula sizes (\autoref{sec:peninsula_size})
   and locations (\autoref{sec:peninsula_locations}).

\subsection{Detecting Country-Level Peninsulas}
\label{sec:detecting_country_peninsulas}

Taitao detects peninsulas based on differences in observations.
Long-lived peninsulas are likely intentional, from policy choices.
One policy is filtering based on national boundaries,
  possibly to implement legal requirements about data sovereignty
  or economic boycotts.

We identify country-specific peninsulas as a special case of Taitao
  where a given destination block is reachable (or unreachable) from only one country,
  persistently for an extended period of time.
(In practice,
  the ability to detect country-level peninsulas is somewhat limited because
  the only country with multiple VPs in our data is the United States.
However, we augment non-U.S.~observers with data from other non-U.S.~sites
  such as Ark or RIPE Atlas.)

A country level peninsula occurs when
  \emph{all} available \acp{VP} from the same country as the target block
  successfully reach the target block and all available \acp{VP} from different countries
  fail.
Formally, we say there is a country peninsula when the set of observers claiming
block $b$ is up at time $i$ is equal to $O_{i,b}^c \subset O_{i,b}$
the set of all available observers with
valid observations at country $c$.
\begin{align}
  O_{i,b}^\Vs{up} = O_{i,b}^c
\end{align}

\subsection{Results of Country-Level Peninsulas}
	\label{sec:country_peninsulas}

Country-specific filtering is a routing policy made by networks to
restrict traffic they receive.
We next look into  what type of organizations actively block overseas
traffic.
For example, good candidates to restrain who can reach them for security purposes
  are government related organizations.

\begin{table}
    \centering
    \footnotesize
    \begin{tabular}{l c c}
            Industry           & ASes  & Blocks   \\
            \midrule
            ISP               & 23  & 138 \\
            Education         & 21  & 167 \\
            Communications    & 14  & 44  \\
            Healthcare        & 8   & 18  \\
            Government        & 7   & 31  \\
            Datacenter        & 6   & 11  \\
            IT Services       & 6   & 8   \\
            Finance           & 4   & 6   \\
            Other (6 types)
            & \multicolumn{2}{c}{6 (1 per type)}   \\
    \end{tabular}
    \caption{U.S. only blocks. Dataset A30, 2017q4}
    \label{tab:industry}
\end{table}

We test for country-specific filtering (\autoref{sec:detecting_country_peninsulas}) over 2017q4 and find 429
unique U.S.-only blocks in 95 distinct ASes.
We then manually verify each AS categorized by industry
   in \autoref{tab:industry}.
It is surprising how many universities filter by country.
While not common, country specific blocks do occur.

\subsection{Verifying Country-Level Peninsulas}
	\label{sec:country_validation}

Next, we verify detection of country-level peninsulas
(\autoref{sec:detecting_country_peninsulas}).
We expect that legal requirements sometimes result in long-term
  network unreachability.
For example, blocking access from Europe
  is a crude way to comply with the EU's GDPR~\cite{eu_gdpr}.

Identifying country-level peninsulas requires
  multiple VPs in the same country.
Unfortunately the source data we use only has multiple \acp{VP} for the United States.
We therefore look for U.S.-specific peninsulas
  where only these \acp{VP} can reach the target and the non-U.S.-\acp{VP} cannot,
  or vice versa.

We first consider the 501 cases where Taitao reports that only U.S.~\acp{VP}
  can see the target, and compare to how Ark \acp{VP} respond.
For Ark, we follow
\autoref{sec:taitao_validation},
  except retaining blocks with less than 85\% uptime.
We only consider Ark VPs that are able
  to reach the destination (that halt with ``success'').
We note blocks that can only be reached by Ark VPs within the same
country as domestic, and blocks that can be reached from VPs located in other
countries as foreign.

In \autoref{tab:taitao_countries_validation_table}
  we show the number of blocks that uniquely
  responded to all U.S.~\ac{VP} combinations during the quarter.
We contrast these results against Ark reachability.

True positives are when  Taitao shows a peninsula
  responsive only to U.S.~\acp{VP}
  and nearly all Ark \acp{VP} confirm this result.
We see 211 targets are U.S.-only, and another 171 are available to only a few
  non-U.S.~countries.
The specific combinations vary: sometimes allowing access from the U.K.,
  or Mexico and Canada.
Together these make 382 true positives, most of the 501 cases.
Comparing all positive cases, we see a
  very high precision of 0.99 (382 green of 385 green and red reports)---our
  predictions are nearly all confirmed by Ark.

In yellow italics we show 47 cases of false positives
  where more than five non-U.S. countries are allowed access.
In many cases these include many European countries.
Our recall is therefore 0.89 (382 green of 429 green and yellow true country peninsulas).

In light green we show true negatives.
Here we include blocks that filter one or more U.S. \acp{VP},
and are reachable from Ark VPs in multiple
countries, amounting to a total of 69 blocks.
There are other categories involving non-U.S. sites,
  along with other millions of true
  negatives, however, we only concentrate in these few.

In red and bold we show three false negatives.
These three blocks seem to have strict filtering policies,
  since they were reachable only from one U.S.~site (W)
  and not the others (C and E) in the 21 days period.

\begin{table}
  \begin{minipage}[b]{.6\linewidth}
  	\footnotesize
  	\tabcolsep=0.1cm
  	\renewcommand{\arraystretch}{1.1}
\resizebox{\textwidth}{!}{
    \begin{tabular}{c c | c c c | c}
      & & \multicolumn{3}{c}{\normalsize \textbf{Ark}} & \\
      & U.S. VPs & Domestic Only & $\leq5$ Foreign & $>5$ Foreign & Total\\
  	\cline{2-6}
      \multirow{7}{2pt}{\rotatebox[origin=l]{90}{\parbox{50pt}{\centering \normalsize \textbf{Trinocular}}}}
      & WCE & \cellcolor[HTML]{99ee77}211  & \cellcolor[HTML]{99ee77}171  & \cellcolor[HTML]{FFF9C4}\emph{47} & 429  \\
      & WCe & \cellcolor[HTML]{F0ABAB}\textbf{0}  & \cellcolor[HTML]{CCFF99}5  & \cellcolor[HTML]{CCFF99}1 & 6  \\
      & WcE & \cellcolor[HTML]{F0ABAB}\textbf{0}  & \cellcolor[HTML]{CCFF99}1  & \cellcolor[HTML]{CCFF99}0 & 1   \\
      & wCE & \cellcolor[HTML]{F0ABAB}\textbf{0}  & \cellcolor[HTML]{CCFF99}0  & \cellcolor[HTML]{CCFF99}0 & 0  \\
      & Wce & \cellcolor[HTML]{F0ABAB}\textbf{3}  & \cellcolor[HTML]{CCFF99}40 & \cellcolor[HTML]{CCFF99}11 & 54\\
      & wcE & \cellcolor[HTML]{F0ABAB}\textbf{0} & \cellcolor[HTML]{CCFF99}4 & \cellcolor[HTML]{CCFF99}5    & 9     \\
      & wCe & \cellcolor[HTML]{F0ABAB}\textbf{0} & \cellcolor[HTML]{CCFF99}1 & \cellcolor[HTML]{CCFF99}1    & 2 \\
      \midrule
      & Marginal distr.  & 214 & 222 & 65 & 501 \\
  \end{tabular}}
    \caption{Trinocular U.S.-only blocks. Dataset: A30, 2017q4.}
    \label{tab:taitao_countries_validation_table}
  \end{minipage}
\end{table}

\subsection{Additional Confirmation of the Number of Peninsulas}
\label{sec:additional_confirmation}

\begin{figure*}
\adjustbox{valign=b}{\begin{minipage}[b]{.31\linewidth}
    \includegraphics[width=1\linewidth]{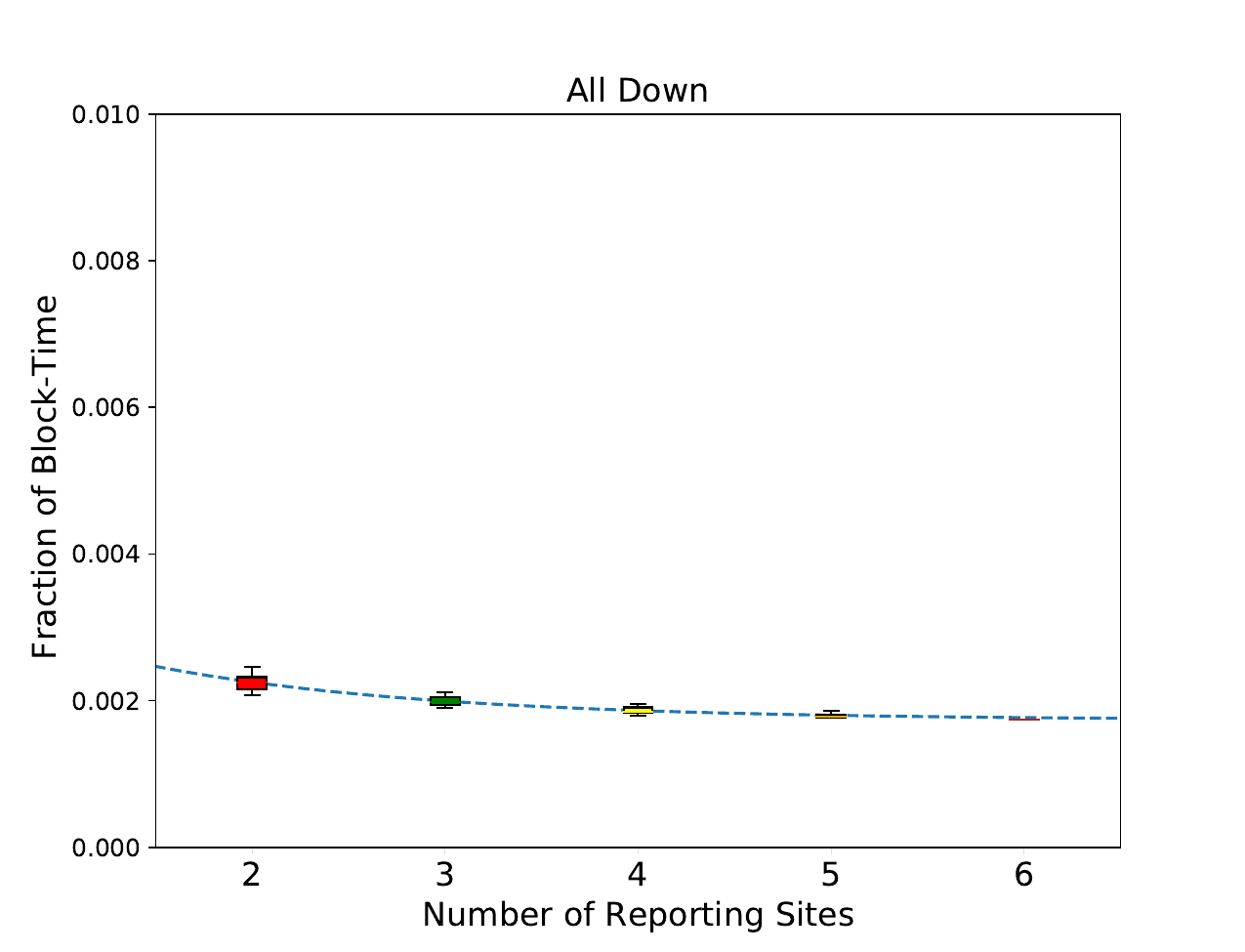}
\end{minipage}}
\adjustbox{valign=b}{\begin{minipage}[b]{.31\linewidth}
    \includegraphics[width=1\linewidth]{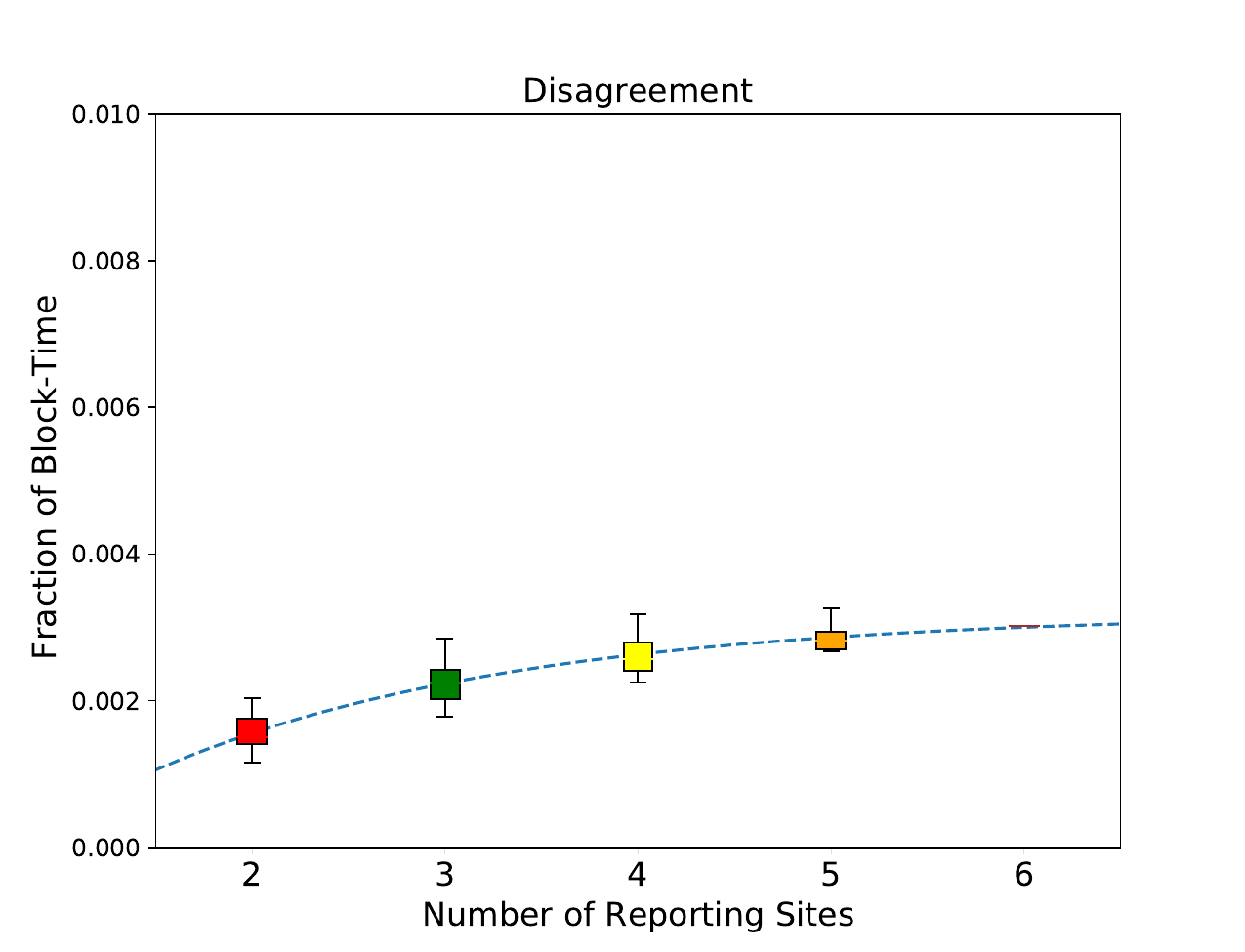}
\end{minipage}}
\adjustbox{valign=b}{\begin{minipage}[b]{.31\linewidth}
    \includegraphics[width=1\linewidth]{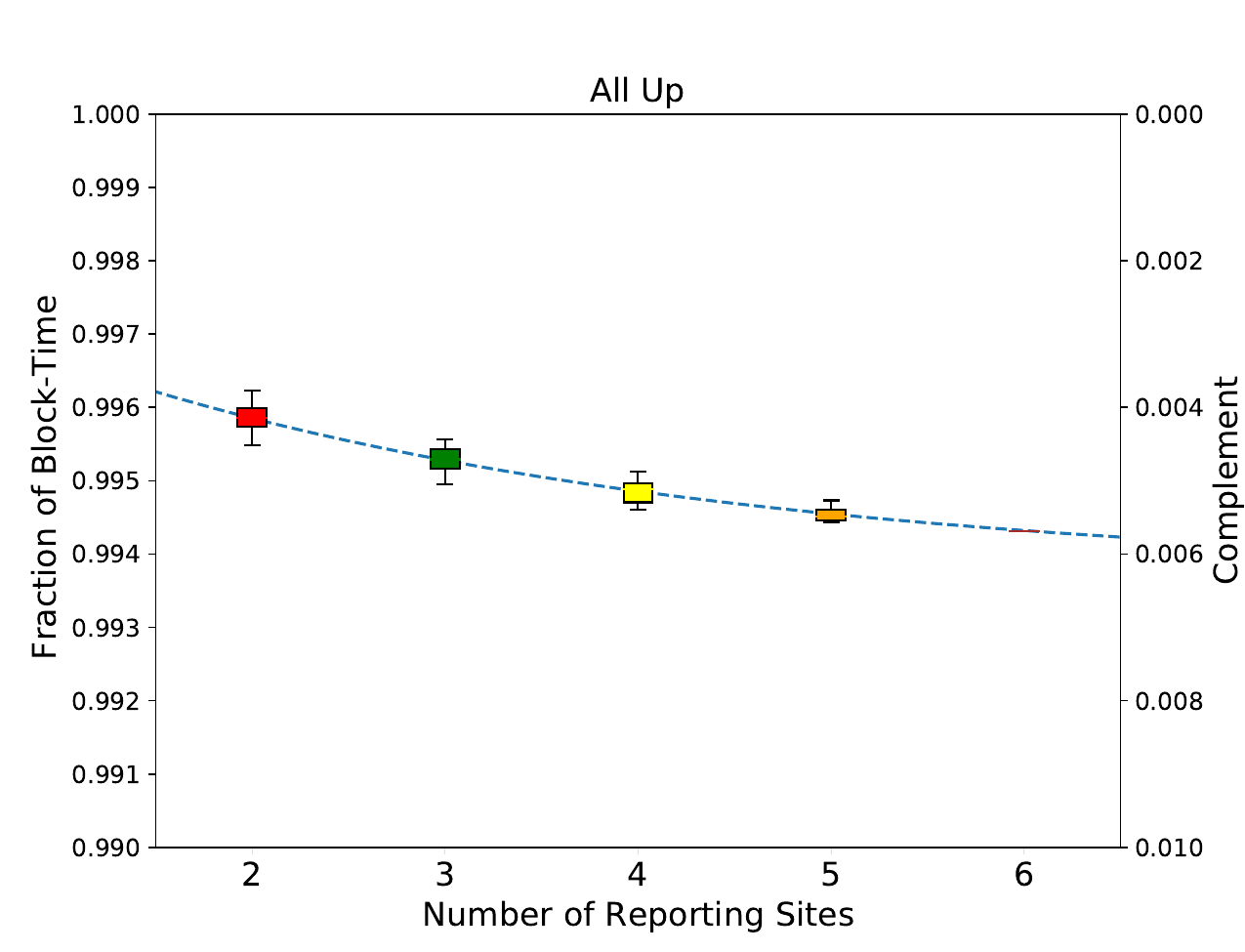}
\end{minipage}}
\caption{Distribution of block-time fraction over sites reporting all down
(left), disagreement (center), and all up (right), for events longer than five
hour. Dataset A34, 2018q4.}
\label{fig:a34all_peninsulas_duration_box}
\end{figure*}

In \autoref{sec:peninsula_frequency}
  we quantify how common peninsulas are.
Here we confirm we see qualitatively similar results,
  but in Trinocular from 2018q4 data.

In \autoref{fig:a34all_peninsulas_duration_box} we confirm,
  that with more \acp{VP} more peninsulas are discovered,
  providing a better view of the Internet's overall state.

\emph{Outages (left) converge after 3 sites},
  as shown by the fitted curve and decreasing variance.
Peninsulas and all-up converge more slowly.

At six \acp{VP}, here we find and even higher difference between all down and
disagreements.
Confirming that peninsulas are a more pervasive problem than outages.

\subsection{Additional Confirmation of Peninsula Duration}
	\label{sec:additional_peninsula_duration}

In \autoref{sec:peninsula_duration} we characterize peninsula duration for
2017q4,
  to determine peninsula root causes.
To confirm our results, we repeat the analysis, but with 2020q3 data.

\begin{figure*}
\begin{center}
  \subfloat [Cumulative events (solid) and duration (dashed)]{
    \includegraphics[width=.60\columnwidth]{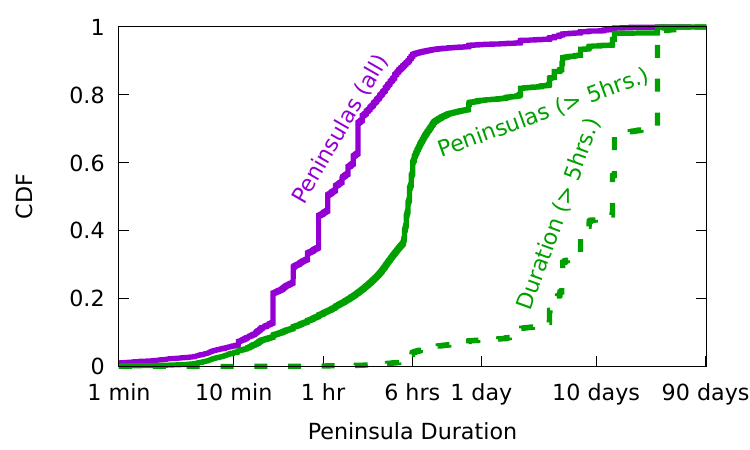}
    \label{fig:a41_partial_outages_duration_cdf}
  }
\hspace{1ex}
  \subfloat[Number of Peninsulas]{
    \includegraphics[width=0.48\columnwidth]{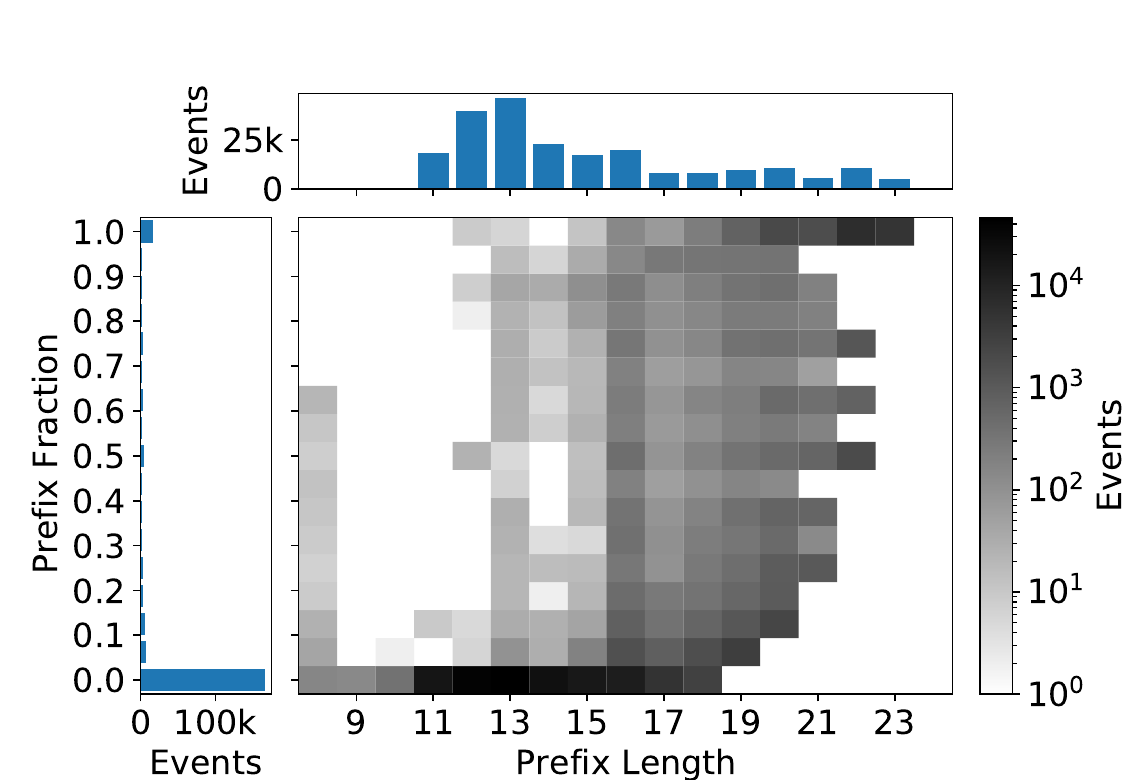}
    \label{fig:a41all_blocks_in_prefix_prefix_fraction_heatmap}
  }
\hspace{1ex}
  \subfloat[Duration fraction]{
    \includegraphics[width=0.48\columnwidth]{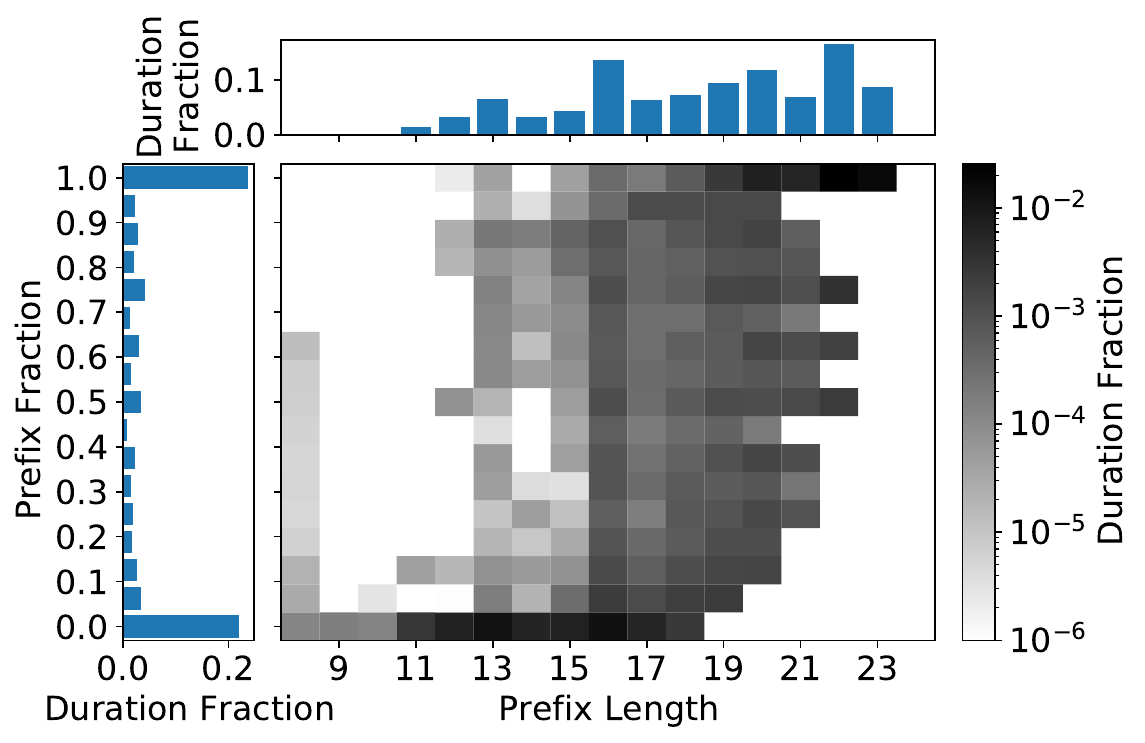}
    \label{fig:a41all_blocks_in_prefix_prefix_fraction_heatmap_duration}
  }
\end{center}
\caption{Peninsulas measured with per-site down events longer than 5 hours during 2020q3. Dataset A41.}
          \label{fig:sim_connection_rollup}
\end{figure*}

As \autoref{fig:a41_partial_outages_duration_cdf} shows,
similarly, as in our 2017q4 results,
  we see that there are many very brief peninsulas (from 20 to 60 minutes).
These results suggest that while the Internet is robust,
there are many small connectivity glitches.

Events shorter than two rounds (22 minutes),
  may represent BGP transients or failures due to random packet loss.

The number of multi-day peninsulas is small,
However, these represent about 90\% of all peninsula-time.
Events lasting a day are long-enough that can be debugged by human network operators,
  and events lasting longer than a week are long-enough that
    they may represent policy disputes.
Together, these long-lived events suggest that
  there is benefit to identifying non-transient peninsulas
  and addressing the underlying routing problem.

\subsection{Additional Confirmation of Size of Peninsulas}

In \autoref{sec:peninsula_size} we discussed the size of peninsulas measured as
a fraction of the affected routable prefix.
In the latter section, we use 2017q4 data.
Here we use 2020q3 to confirm our results.

\autoref{fig:a41all_blocks_in_prefix_prefix_fraction_heatmap} shows the peninsulas
per prefix fraction, and \autoref{fig:a41all_blocks_in_prefix_prefix_fraction_heatmap_duration}.
Similarly,
  we find that while small prefix fraction peninsulas are more in numbers,
  most of the peninsula time is spent in peninsulas covering the whole prefix.
This result is consistent with long lived peninsulas being caused by policy
choices.

\label{page:last_page}

\end{document}